\documentclass{article}
\usepackage{natbib,psfig,emulateapj}

\newcommand\hi{{\small H\thinspace \footnotesize\sc\romannumeral 1 }}

\newcommand\nv{{\small N\thinspace \footnotesize\sc\romannumeral 5 }}
\newcommand\nvi{{\small N\thinspace \footnotesize\sc\romannumeral 6 }}

\newcommand\civ{{\small C\thinspace \footnotesize\sc\romannumeral 4 }}
\newcommand\cv{{\small C\thinspace \footnotesize\sc\romannumeral 5 }}

\newcommand\oi{{\small O\thinspace \footnotesize\sc\romannumeral 1 }}
\newcommand\oiii{{\small O\thinspace \footnotesize\sc\romannumeral 3 }}
\newcommand\oiv{{\small O\thinspace \footnotesize\sc\romannumeral 4 }}
\newcommand\ov{{\small O\thinspace \footnotesize\sc\romannumeral 5 }}
\newcommand\ovi{{\small O\thinspace \footnotesize\sc\romannumeral 6 }}
\newcommand\ovii{{\small O\thinspace \footnotesize\sc\romannumeral 7 }}

\newcommand\neiv{{\small Ne\thinspace \footnotesize\sc\romannumeral 4 }}

\newcommand\nh{\hbox{{$N_{\rm H}$}}}
\newcommand\nhgal{\hbox{{$N_{\rm H}^{\rm Gal}$}}}

\newcommand\mabs{\hbox{{$M_{\rm abs}$}}}
\newcommand\mhot{\hbox{{$M_{\rm hot}$}}}
\newcommand\tacc{\hbox{{$t_{\rm acc}$}}}
\newcommand\msun{\hbox{{$M_{\sun}$}}}
\newcommand\mdot{\hbox{{$\dot{M}$}}}
\newcommand\tage{\hbox{{$t_{\rm age}$}}}

\newcommand\solar{\hbox{{$Z_{\sun}$\thinspace}}}

\newcommand\einstein{{\sl Einstein} }
\newcommand\rosat{{\sl ROSAT} }
\newcommand\asca{{\sl ASCA} }
\newcommand\chandra{{\sl Chandra} }
\newcommand\xmm{{\sl XMM} }

\newcommand\xspec{{\sc xspec} }

\newcommand\ergcms{{erg cm$^{-2}$ s$^{-1}$\thinspace}}

\newcommand\emin{\hbox{{$E_{\rm min}$}}}

\begin{document} 

\title{\rosat Evidence for Intrinsic Oxygen Absorption in Cooling Flow \\
Galaxies and Groups}

\author{David A. Buote\altaffilmark{1}}

\affil{UCO/Lick Observatory, University of California at Santa Cruz,
Santa Cruz, CA 95064; buote@ucolick.org}

\altaffiltext{1}{Chandra Fellow}

\slugcomment{Accepted for Publication in The Astrophysical Journal}

\begin{abstract}

The existence of large quantities of gas that have cooled and dropped
out of the hot phase in massive elliptical galaxies, groups, and
clusters is the key prediction of the inhomogeneous cooling flow
scenario. Using spatially resolved, deprojected \rosat PSPC spectra of
10 of the brightest cooling flow galaxies and groups with low Galactic
column densities we have detected intrinsic absorption over energies
$\sim 0.4-0.8$ keV at the $2\sigma/3\sigma$ level in half of the
sample. Since no intrinsic absorption is indicated for energies below
$\sim 0.4$ keV, the most reasonable model for the absorber is
collisionally ionized gas at temperatures $T=10^{5-6}$ K with most of
the absorption arising from ionized states of oxygen but with a
significant contribution from carbon and nitrogen. The soft X-ray
emission of this warm gas can also explain the sub-Galactic column
densities of cold gas inferred within the central regions of most of
the systems. (This could not be explained by an absorber composed only
of dust.)  Attributing the absorption to ionized gas reconciles the
large columns of cold H and He inferred from \einstein and \asca with
the lack of such columns inferred from \rosat.

Within the central $\sim 10-20$ kpc, where the constraints are most
secure, the mass of the ionized absorber is consistent with most
(perhaps all) of the matter deposited by a cooling flow over the
lifetime of the flow. Since the warm absorber produces no significant
H or He absorption the large absorber masses are consistent with the
negligible atomic and molecular H inferred from \hi\, and CO
observations of cooling flows. It is also found that if $T\ga 2\times
10^5$ K then the optical and FUV emission implied by the warm gas does
not violate published constraints. An important theoretical challenge
is to understand how the warm temperature is maintained and how the
gas is supported gravitationally, and we discuss possible solutions to
these problems that would require fundamental modification of the
standard cooling flow scenario.  Finally, we discuss how the
prediction of warm ionized gas as the product of mass drop-out in
these and other cooling flows can be verified with new \chandra and
\xmm observations.
\end{abstract}


\section{Introduction}
\label{intro}

The evolution of the hot gas in the centers of massive elliptical
galaxies, groups, and galaxy clusters has been most frequently
interpreted in terms of the cooling flow paradigm (e.g., Fabian
1994). However, the characteristic radially increasing temperature
profiles and centrally peaked X-ray surface brightness profiles
usually attributed to cooling flows can be successfully described by
only assuming a two-tier structure for the gravitational potential
(e.g., Ikebe et al 1996; Xu et al 1998); i.e., the cooler gas sits in
the shallower potential associated with the central galaxy whereas the
hotter gas sits in the deeper potential of the surrounding group or
cluster. The two-tier structure is typically incorporated into cooling
flow models (e.g., Thomas, Fabian, \& Nulsen 1987; Brighenti \&
Mathews 1998), but since cooling flows are not necessarily required to
explain the temperatures and surface brightness profiles of the hot
gas why should they still receive attention?

Unlike the empirical two-tier potential model, cooling flows attempt
to offer a nearly complete theoretical description of the time
evolution of the gas properties. A cooling flow describes the
dynamical evolution and the X-ray emission of the hot gas by
considering the following simple picture for the energy balance of a
parcel of gas. Since the parcel of gas emits X-rays it loses
energy. As the parcel radiates it sinks deeper (i.e., flows inward)
into the potential well of the system where it encounters regions of
higher density and therefore higher pressure. Consequently, the gas
parcel contracts but is then heated as the result of PdV work. This
balance between cooling and heating is expected to apply over much of
the cooling flow. Within the central regions of highest density this
balance is broken as cooling overwhelms the heating leading to the key
prediction of the inhomogeneous cooling flow scenario: in massive
elliptical galaxies, groups, and clusters large quantities of gas
should have cooled and dropped out of the flow and be distributed at
least over the central regions of the flow.

This prediction has inspired many searches in \hi\, and CO for cold
gas at the centers of cooling flows, and all such attempts have either
detected small gas masses or placed upper limits which are in
embarrassing disagreement with the large masses expected to have been
deposited in a cooling flow (e.g., Bregman, Hogg, \& Roberts 1992;
O'Dea et al 1994).  If instead the mass drop-out is in the form of
dust then current constraints on the infrared emission in cluster
cores are not inconsistent with cooling flow models (e.g., Voit \&
Donahue 1995; Allen et al 2000b). But cooling flows are not required
to explain the infrared data in clusters \citep{l95} and individual
elliptical galaxies \citep{tm96}.

The case for mass drop-out received a substantial boost with the
discovery of intrinsic soft X-ray absorption in the \einstein spectral
data of cooling flow clusters (White et al 1991; Johnstone et al
1992). The \einstein results have been verified with multitemperature
models of the \asca spectral data of clusters (Fabian et al 1994;
Allen et al 2000b), elliptical galaxies, and groups (Buote \& Fabian
1998; Buote 1999, 2000a; Allen et al 2000a). If the soft X-ray
absorption is interpreted as cold gas then the large intrinsic column
densities of cold H suggested by the \einstein and \asca observations
still suffer from the tremendous disagreement with the \hi\, and CO
observations noted above. The \einstein and \asca results appear even
more suspect when considering that in systems with low Galactic
columns (where any intrinsic absorption should be easier to detect) no
significant excess absorption from cold gas is ever found with the
\rosat PSPC which should be more sensitive to the absorption because
of its softer bandpass, 0.1-2.4 keV (e.g., David et al 1994; Jones et
al 1997; Briel \& Henry 1996).

We have re-examined the \rosat PSPC data of cooling flows to search
for evidence of intrinsic soft X-ray absorption and in particular have
allowed for the possibility that the absorber is not cold. Previously
in Buote (2000c; hereafter PAPER2) we have presented temperature and
metallicity profiles of the hot gas in 10 of the brightest cooling
flow galaxies and groups inferred from deprojection analysis of the
PSPC data. We refer the reader to that paper for details on the data
reduction and deprojection procedure.

In this paper we present the absorption profiles of the 10 galaxies
and groups, each of which have low Galactic column densities (see
Table 1 of PAPER2). Partial results for two systems analyzed in the
present paper (NGC 1399 and 5044) also appear with results for the
cluster A1795 in Buote (2000b; hereafter PAPER1). In \S \ref{models}
we describe the models used to parameterize the soft X-ray
absorption. We present the radial absorption profiles for a standard
absorber model with solar abundances in \S \ref{solar}. The effects of
partial covering and the sensitivity of the results to the bandpass
are discussed in \S \ref{partial} and \S \ref{bandpass}. The results
of modeling the absorption with an oxygen edge are presented in \S
\ref{edge}. In \S \ref{multi} we consider multiphase models such
as cooling flows (\S \ref{cf}). Evidence for emission from a warm
gaseous component is described in \S \ref{xray}.  We demonstrate the
consistency of the \rosat and \asca absorption measurements in \S
\ref{asca}. In \S \ref{warm} we discuss in detail the implications of
our absorption measurements for the physical state of the absorber,
the cooling flow scenario, observations at other wavelengths, and
theoretical models.  Finally, in \S \ref{conc} we present our
conclusions and discuss prospects for verifying our prediction of warm
ionized gas in cooling flows with future X-ray observations.

\section{Models}
\label{models}

\subsection{Hot Plasma}
\label{plasma}

As discussed in PAPER2 we use the MEKAL plasma code to represent the
emission from a single temperature component of hot gas. Because of
the limited energy resolution of \rosat\, we initially focus on a
``single phase'' representation of the hot gas such that a single
temperature component exists within each three-dimensional radial
annulus. Multiphase models are examined in \S \ref{multi}.

\subsection{Absorber}

It is standard practice to represent the soft X-ray absorption arising
from the Milky Way by material with solar abundances distributed as a
foreground screen at zero redshift. In this standard absorption model
the X-ray flux is diminished according to $A(E)=\exp(-\nh\sigma(E))$,
where \nh\, is the hydrogen column density and $\sigma(E)$ is the
energy-dependent photo-electric absorption cross section for an
absorber with solar abundances. We allow \nh\, to be a free parameter
in our fits to indicate any excess absorption intrinsic to a galaxy or
group and also to allow for any errors in the assumed Galactic value
for \nh\, and for any calibration uncertainties. Note that in this
standard model \nh\, is measured as a function of two-dimensional
radius, $R$, on the sky.

Intrinsic absorption is expressed more generally as,
\begin{displaymath}
A(E) = f\exp(-\nh\sigma[E(1+z)]) + (1-f),
\end{displaymath}
where $z$ is the source redshift and $f$ is the covering factor.
Since the redshifts are small for the objects in our sample any
absorption in excess of the Galactic value indicated by the standard
model is essentially that of an intrinsic absorber with $f=1$ placed
in front of the source. We discuss the effects of $f<1$ in \S
\ref{partial}.

As discussed in PAPER1 we consider oxygen absorption intrinsic to the
galaxy or group which we represent by the simple parameterization of
an edge, $\exp\left[-\tau(E/E_0)^{-3}\right]$ for $E\ge E_0$, where
$E_0$ is the energy of the edge in the rest frame of the galaxy or
group, and $\tau$ is the optical depth. To facilitate a consistent
comparison to the standard absorber model we place this edge in front
of the source, and thus $\tau$ is also measured as a function of
two-dimensional radius, $R$, on the sky. Models with $f<1$ behave
similarly to the solar-abundance absorber (\S \ref{partial}).

The photo-electric absorption cross sections used in this paper are
given by \citet{phabs}. Although \citet{ab} point out that the He
cross section at 0.15 keV is in error by 13\%, since we analyze $E>
0.2$ keV we find that our fits do not change when using the
\citet{wabs} cross sections which have the correct He value.  Further
details on the spectral models used in the analysis are given in
PAPER2.

\section{Radial Absorption Profiles}
\label{abs}

\begin{figure*}[t]

{\Large\boldmath \Large\bf 
\hskip 1.25cm $\emin = 0.2$ keV \hskip 2.1cm $\emin = 0.5$ keV \hskip 2.2cm
$\emin = 0.2$ keV 
}

\vskip 0.5cm

\centerline{\large\bf NGC 507} \vskip 0.1cm
\parbox{0.32\textwidth}{
\centerline{\psfig{figure=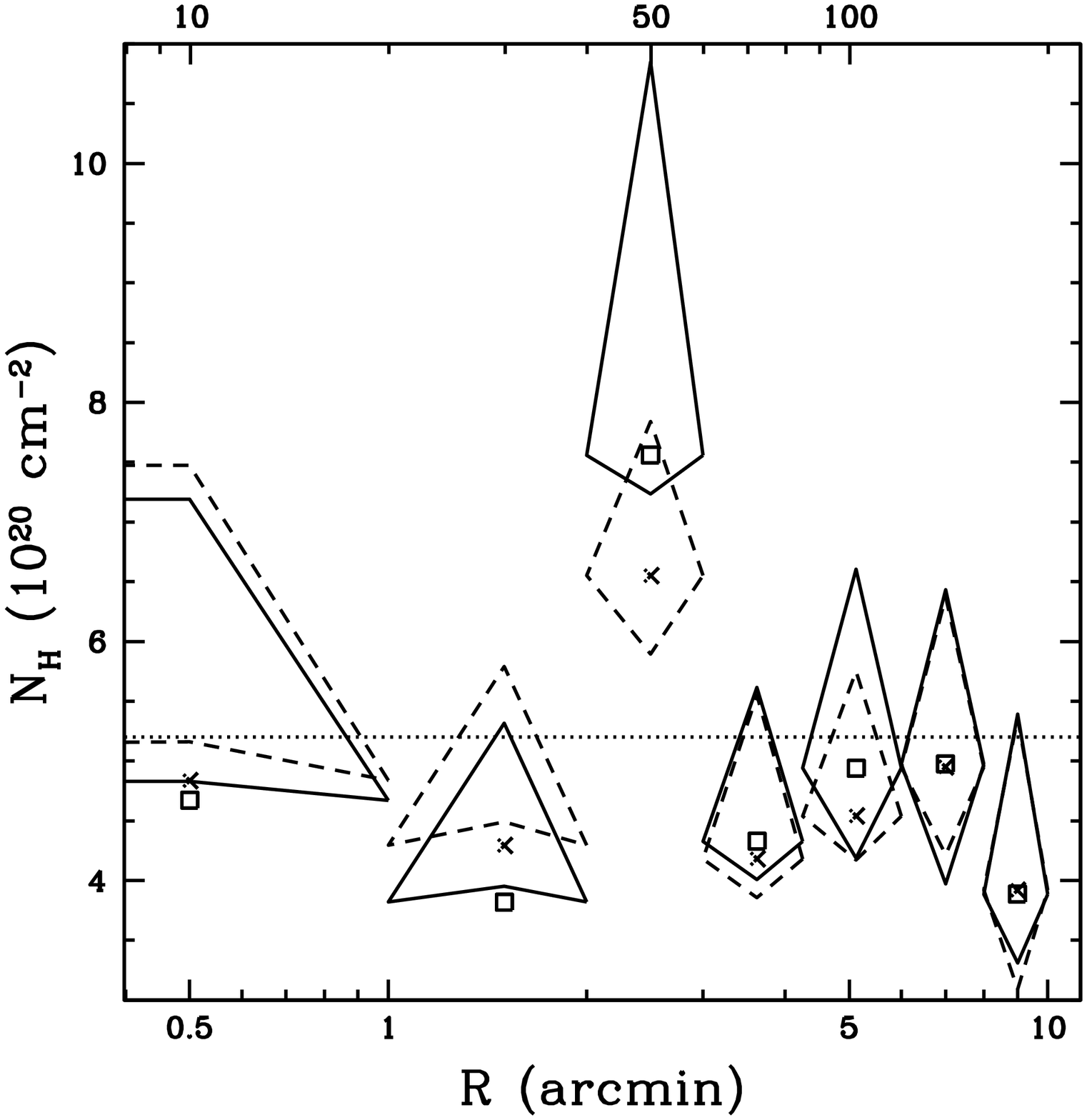,height=0.22\textheight}}
}
\parbox{0.32\textwidth}{
\centerline{\psfig{figure=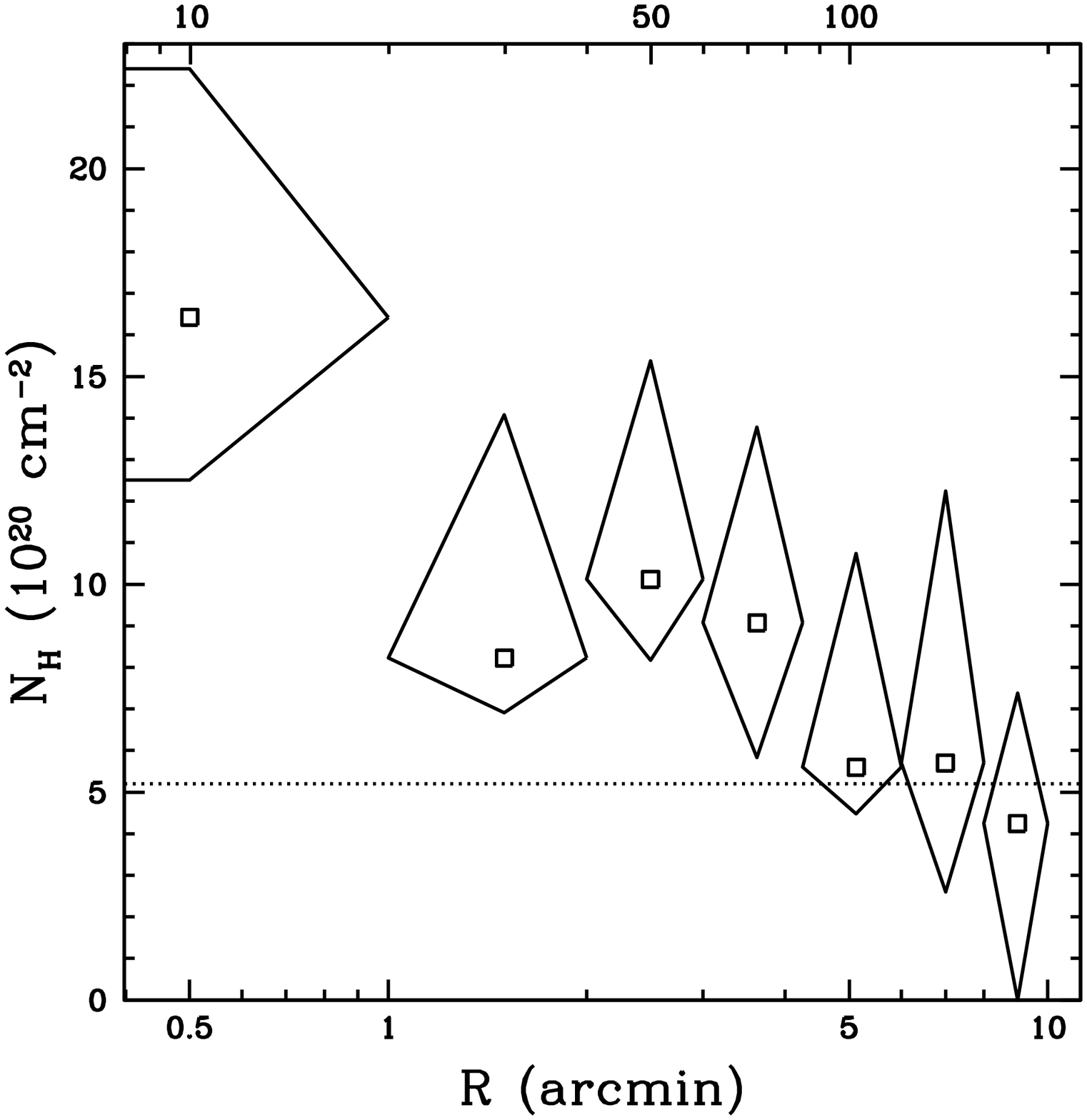,height=0.22\textheight}}
}
\parbox{0.32\textwidth}{
\centerline{\psfig{figure=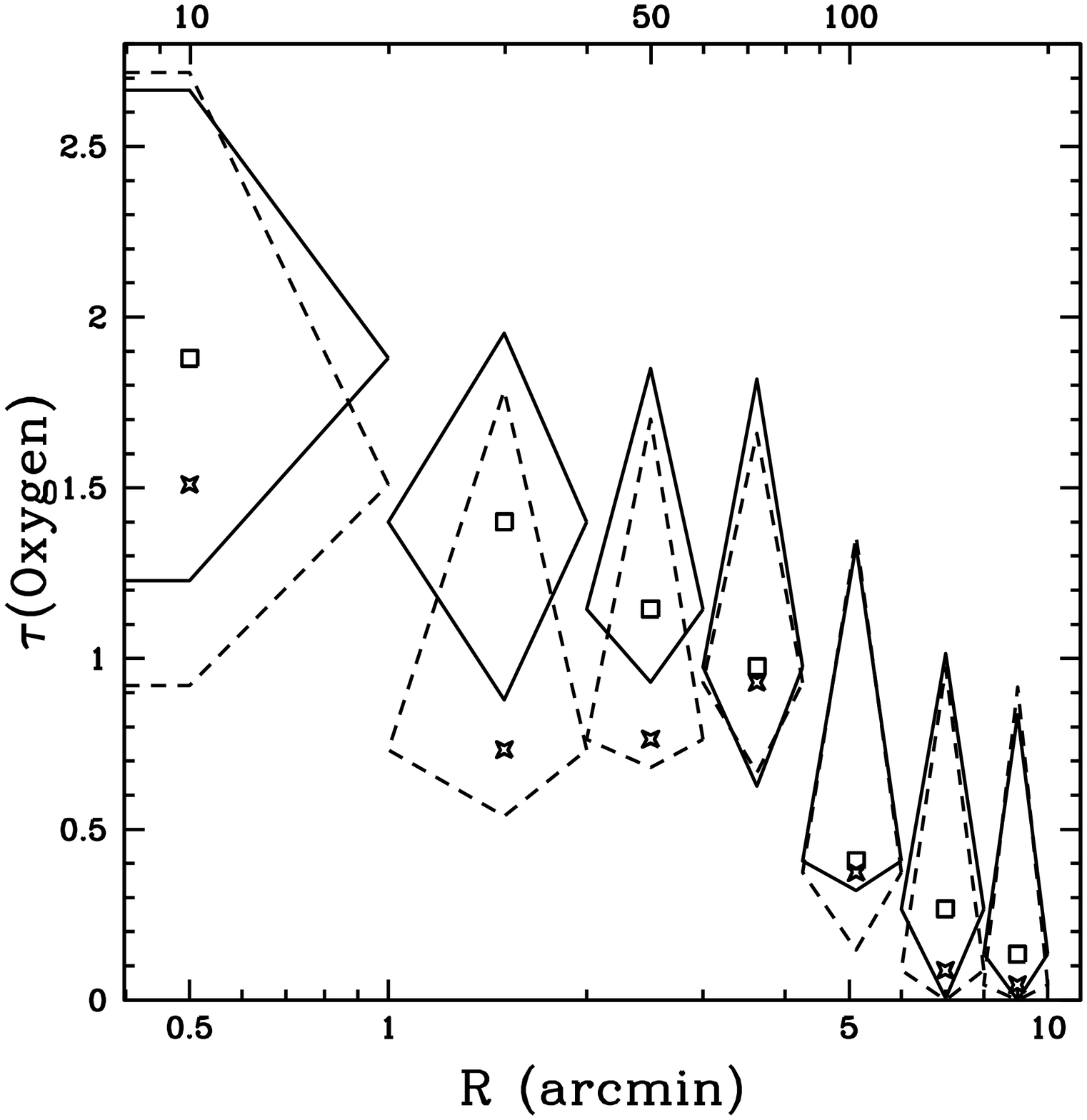,height=0.22\textheight}}
}
\vskip 0.25cm
\centerline{\large\bf NGC 1399} \vskip 0.1cm
\parbox{0.32\textwidth}{
\centerline{\psfig{figure=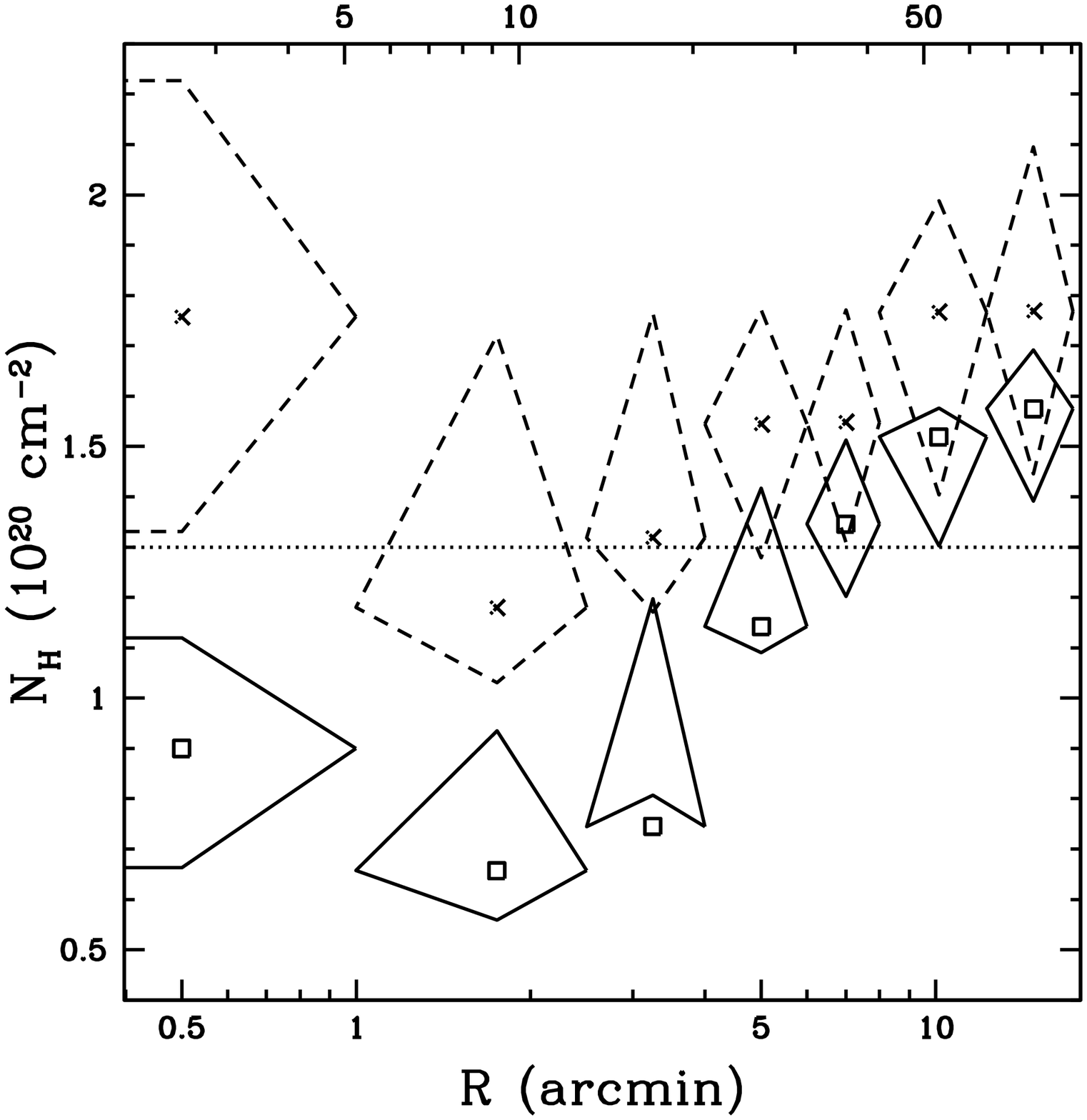,height=0.22\textheight}}
}
\parbox{0.32\textwidth}{
\centerline{\psfig{figure=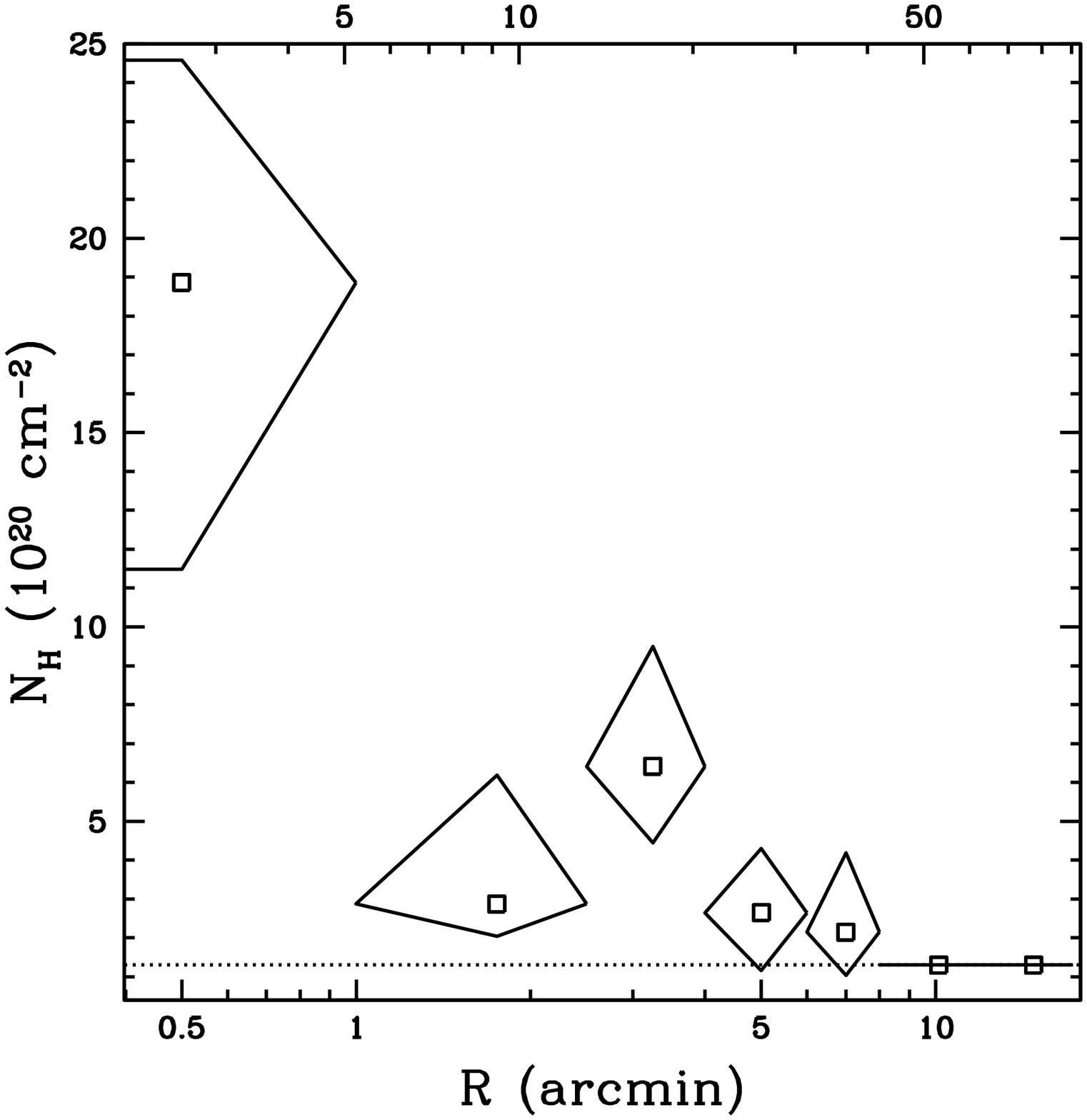,height=0.22\textheight}}
}
\parbox{0.32\textwidth}{
\centerline{\psfig{figure=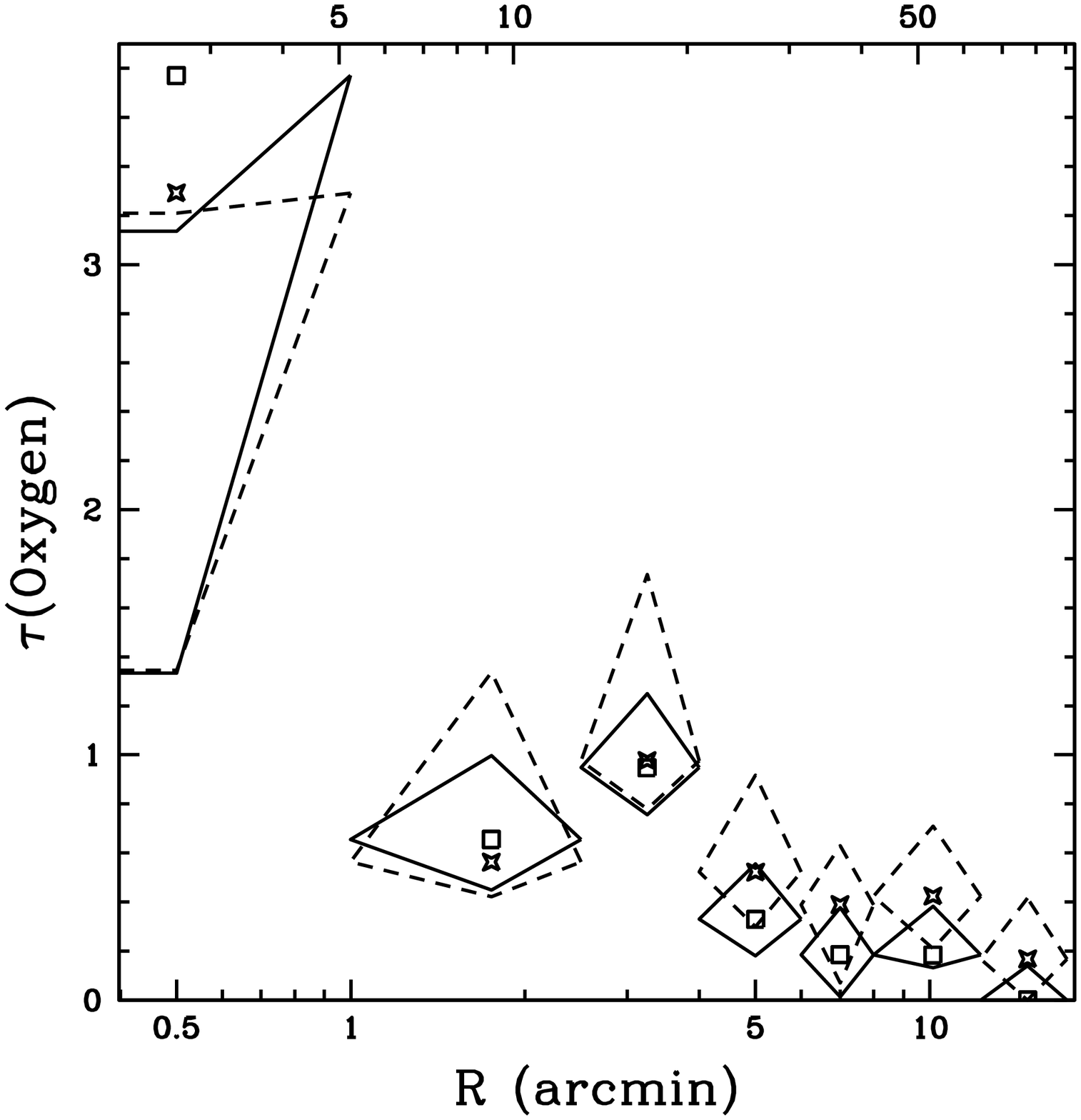,height=0.22\textheight}}
}
\vskip 0.25cm
\centerline{\large\bf NGC 5044} \vskip 0.1cm
\parbox{0.32\textwidth}{
\centerline{\psfig{figure=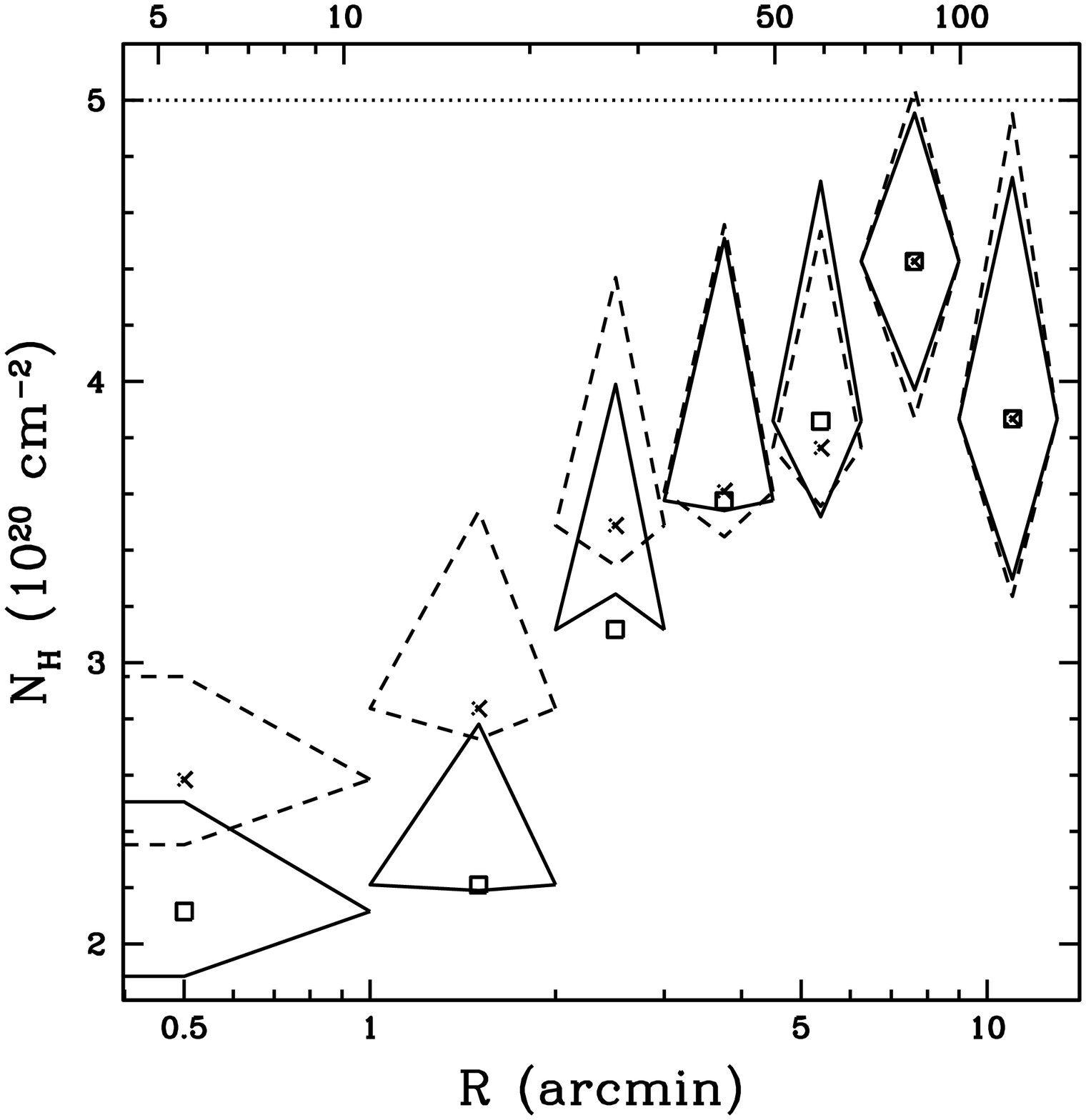,height=0.22\textheight}}
}
\parbox{0.32\textwidth}{
\centerline{\psfig{figure=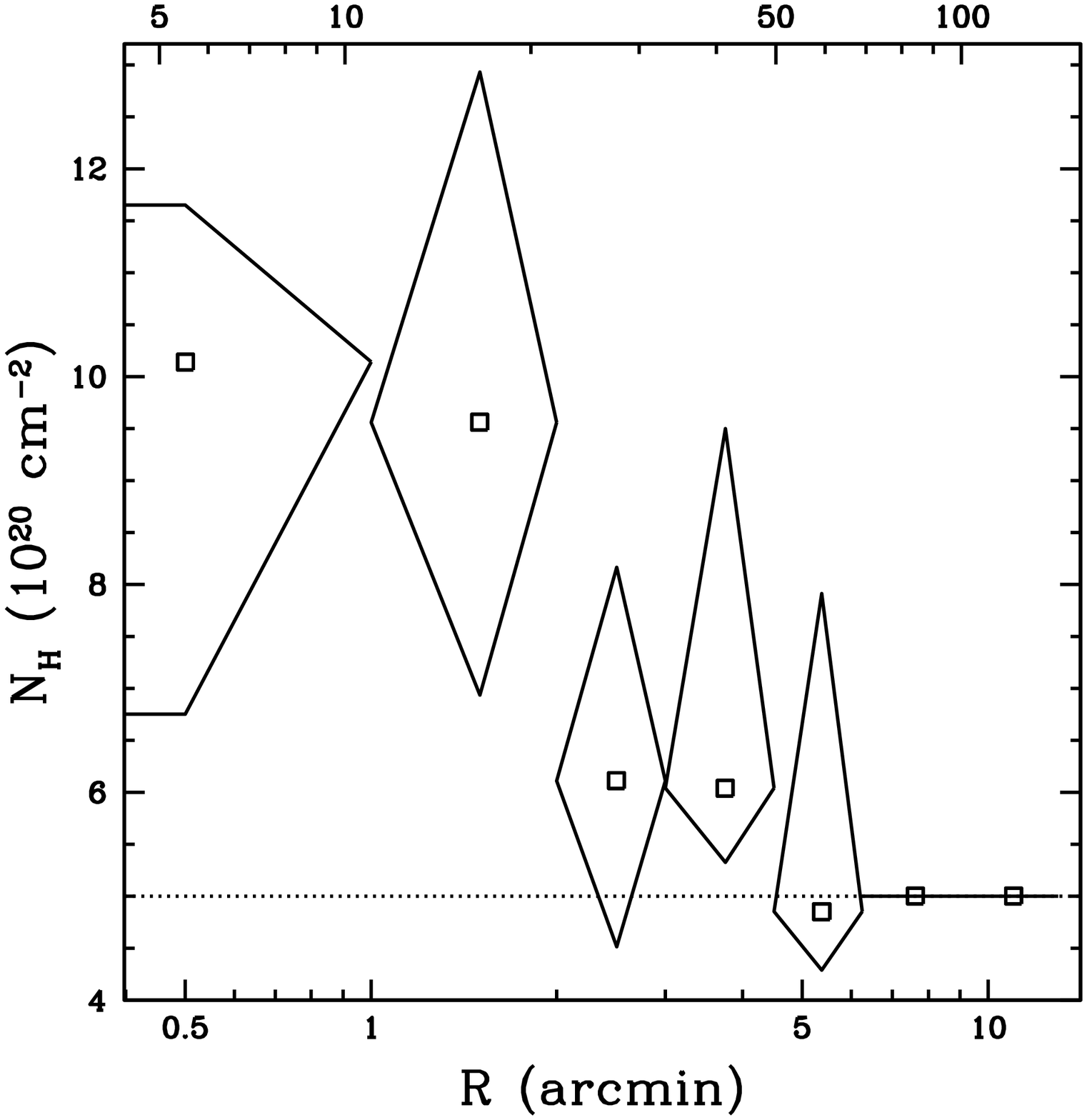,height=0.22\textheight}}
}
\parbox{0.32\textwidth}{
\centerline{\psfig{figure=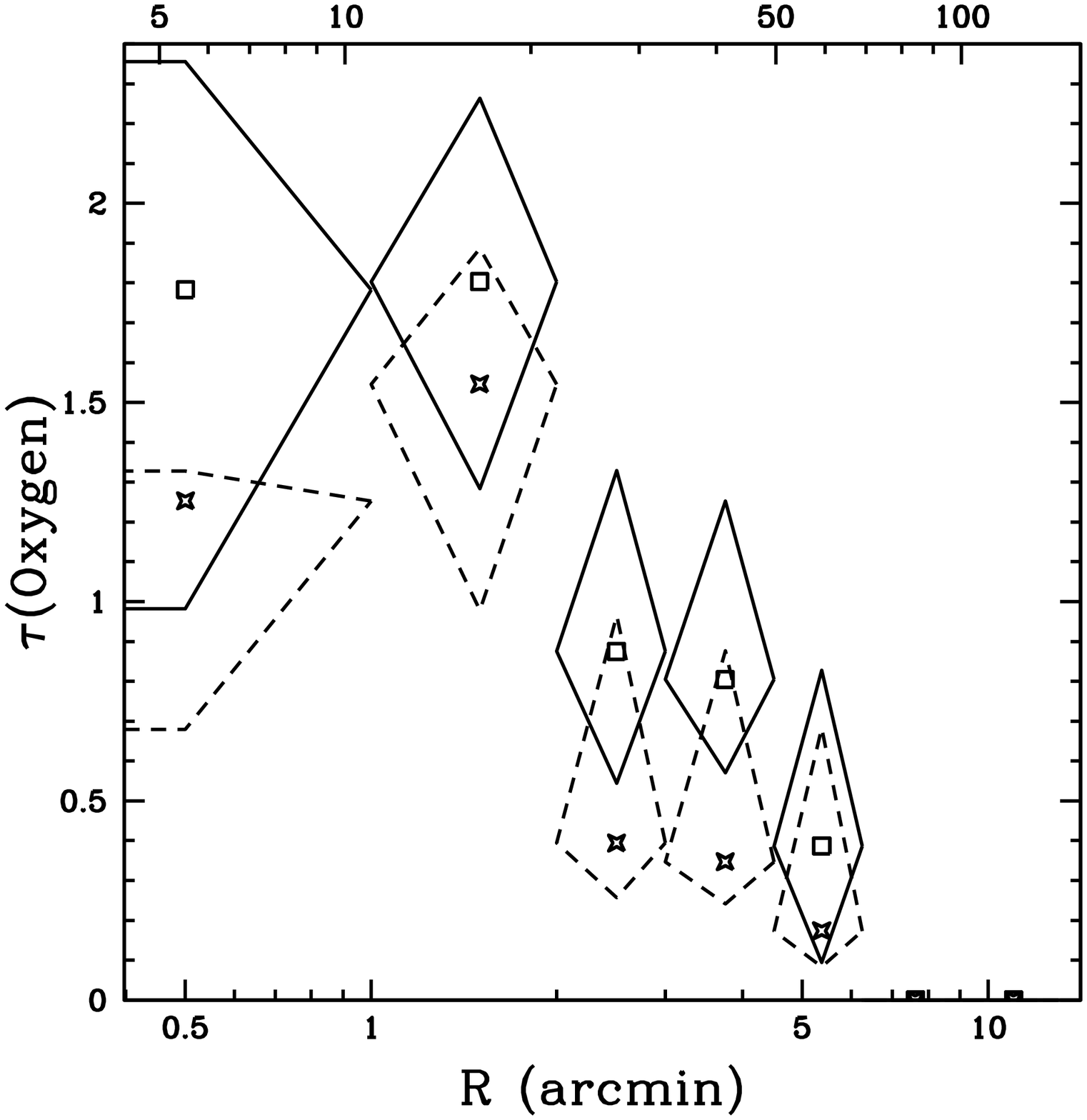,height=0.22\textheight}}
}
\caption{\label{fig.7ann} Results of the deprojection analysis for the
systems where interesting constraints were obtained in 7 annuli. (Left
panels) The column density profiles for a standard foreground absorber
with solar abundances obtained from fits over 0.2-2.2 keV (i.e.,
$\emin=0.2$ keV) are denoted by open squares for best fit and solid
diamonds for $1\sigma$ error bars; models which also have an intrinsic
oxygen edge are represented by crosses and dashed diamonds. The
Galactic hydrogen column density \citep{dl} is shown as a dotted
line. Radial units are arcminutes on the bottom axis and kpc on the
top. (Middle panels) The column density profiles for a standard
absorber obtained from fits over 0.5-2.2 keV (i.e., $\emin=0.5$
keV). (Right panels) Optical depths for an oxygen edge at 0.532 keV
(rest frame) obtained from fits with $\emin=0.2$ keV. The open squares
and solid diamonds are respectively the best fit and $1\sigma$ errors
for models where the standard absorber has \nh\, fixed to the Galactic
value; crosses and dashed diamonds refer to models with variable \nh\,
and thus correspond to similarly marked models in the left panels.}
\end{figure*}

\begin{figure*}[t]

{\Large\boldmath \Large\bf 
\hskip 1.25cm $\emin = 0.2$ keV \hskip 2.1cm $\emin = 0.5$ keV \hskip 2.2cm
$\emin = 0.2$ keV 
}

\vskip 0.5cm

\centerline{\large\bf NGC 2563} \vskip 0.1cm
\parbox{0.32\textwidth}{
\centerline{\psfig{figure=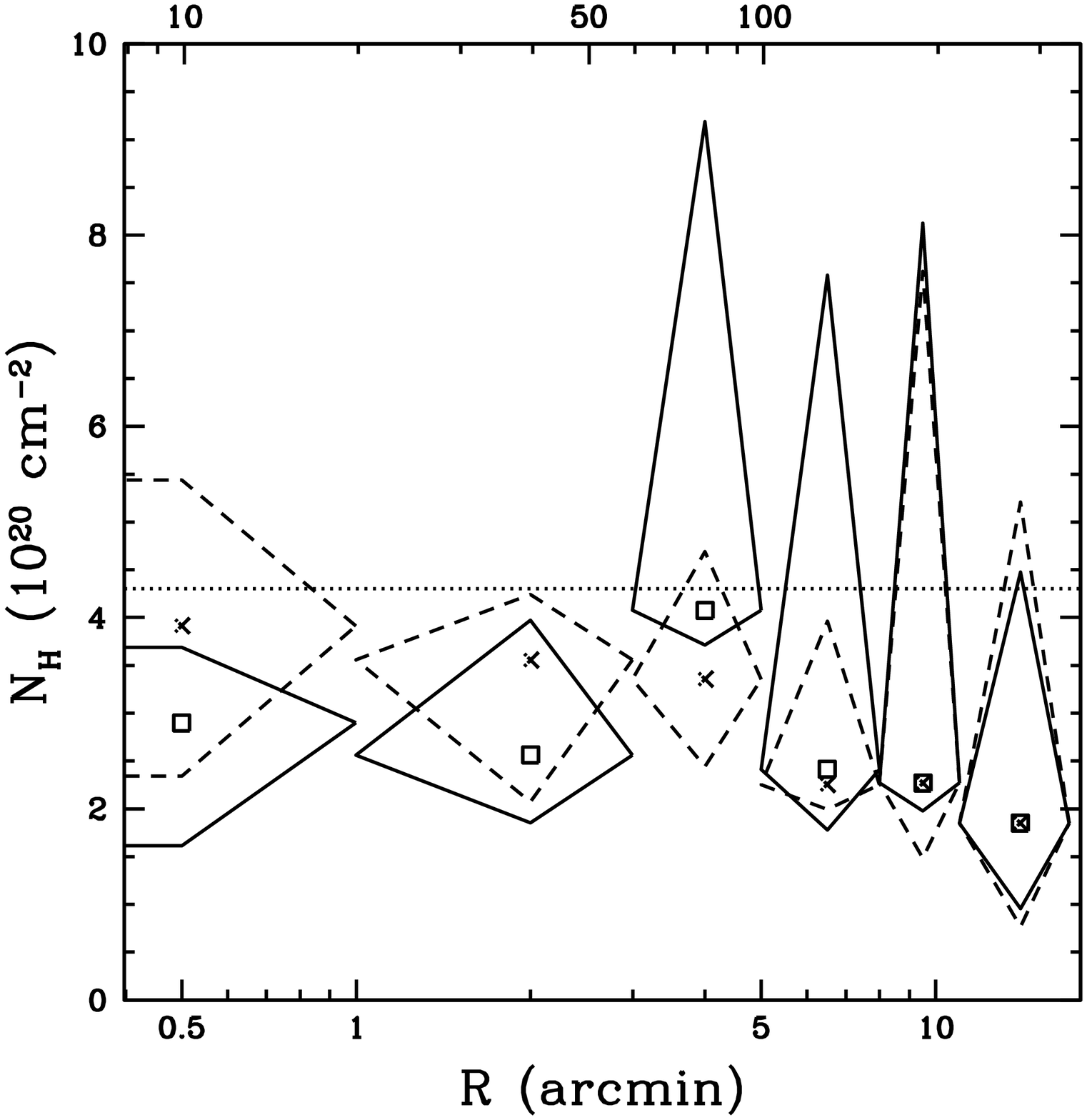,height=0.22\textheight}}
}
\parbox{0.32\textwidth}{
\centerline{\psfig{figure=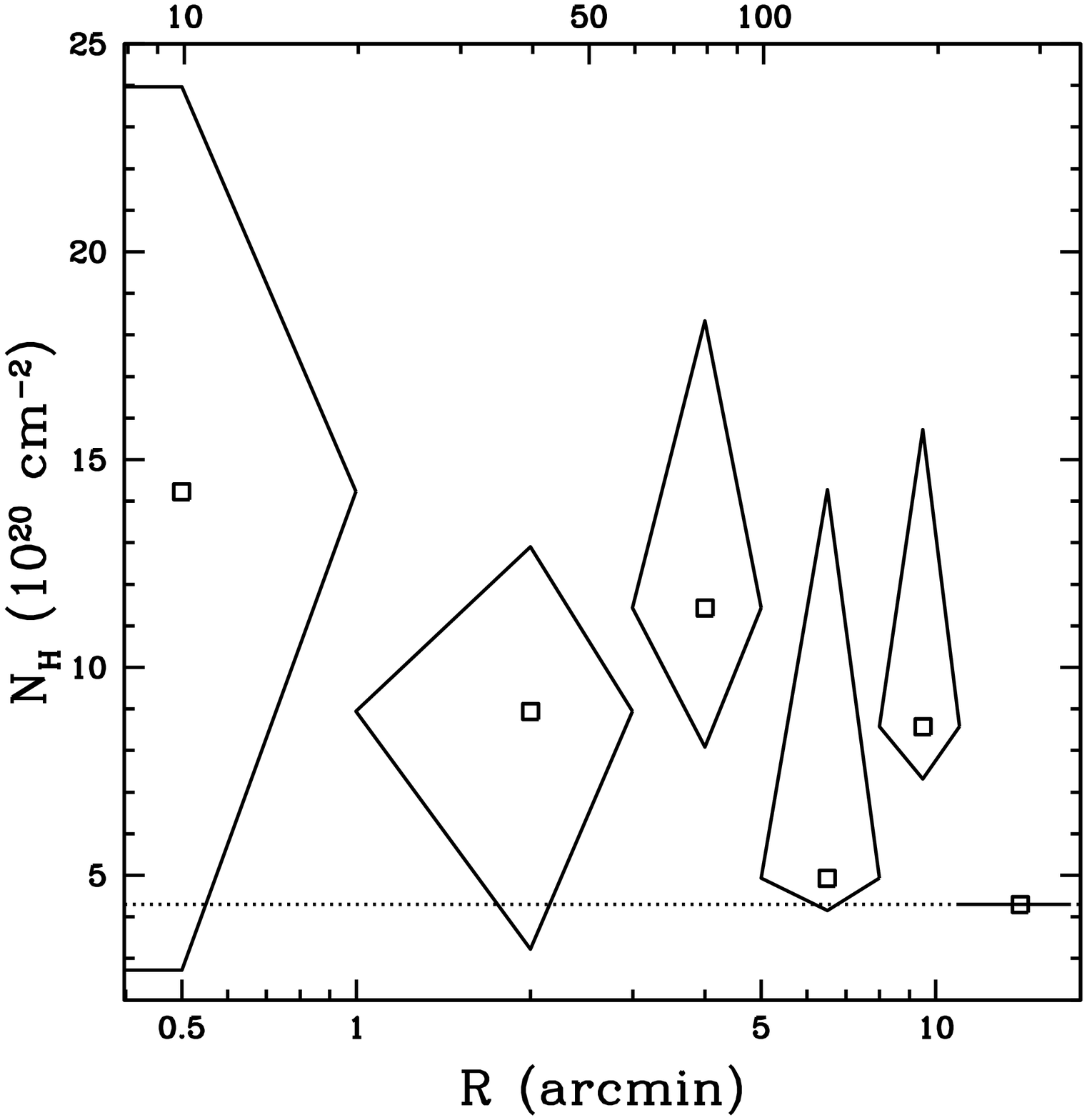,height=0.22\textheight}}
}
\parbox{0.32\textwidth}{
\centerline{\psfig{figure=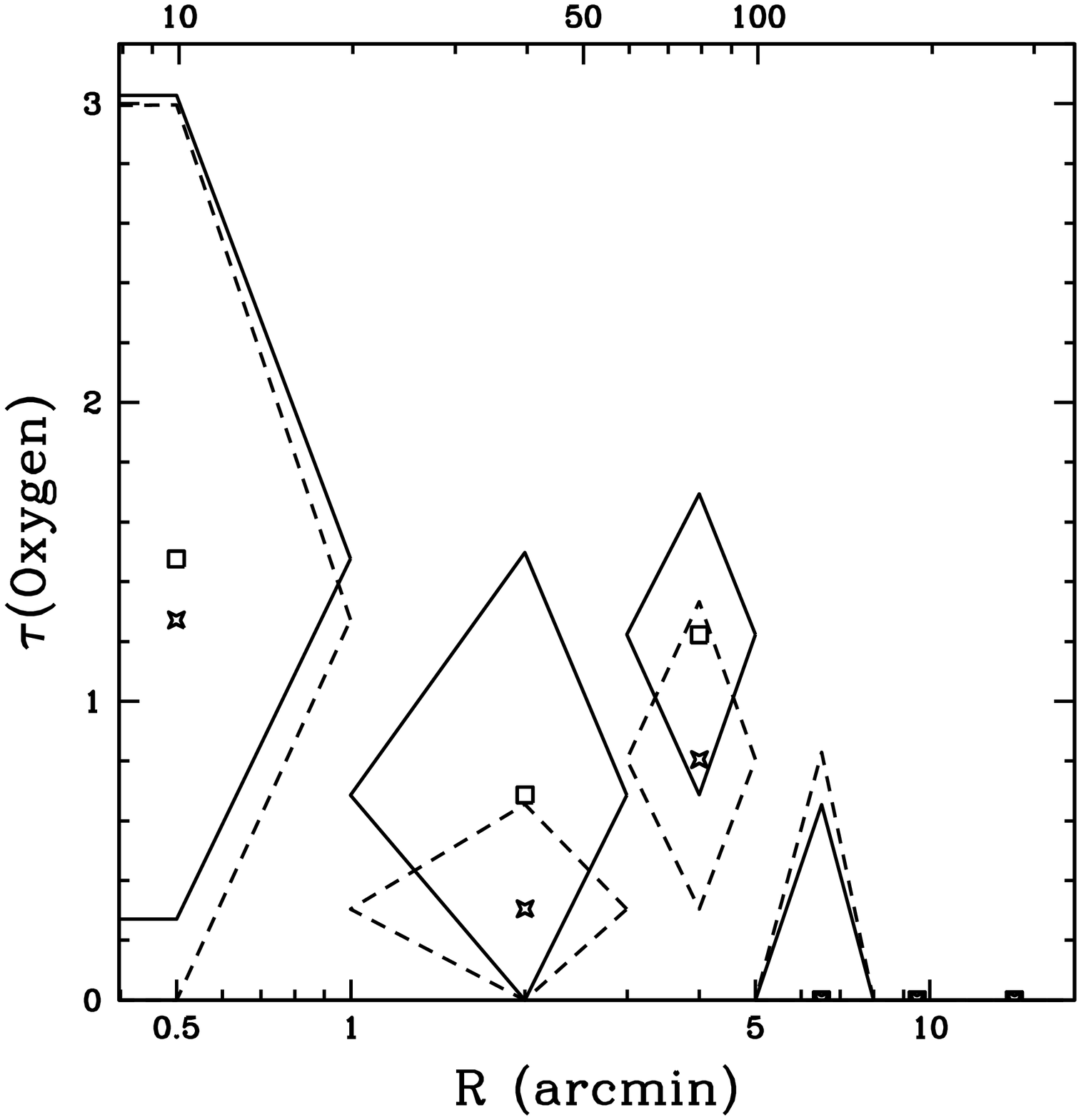,height=0.22\textheight}}
}
\vskip 0.25cm
\centerline{\large\bf NGC 4472} \vskip 0.1cm
\parbox{0.32\textwidth}{
\centerline{\psfig{figure=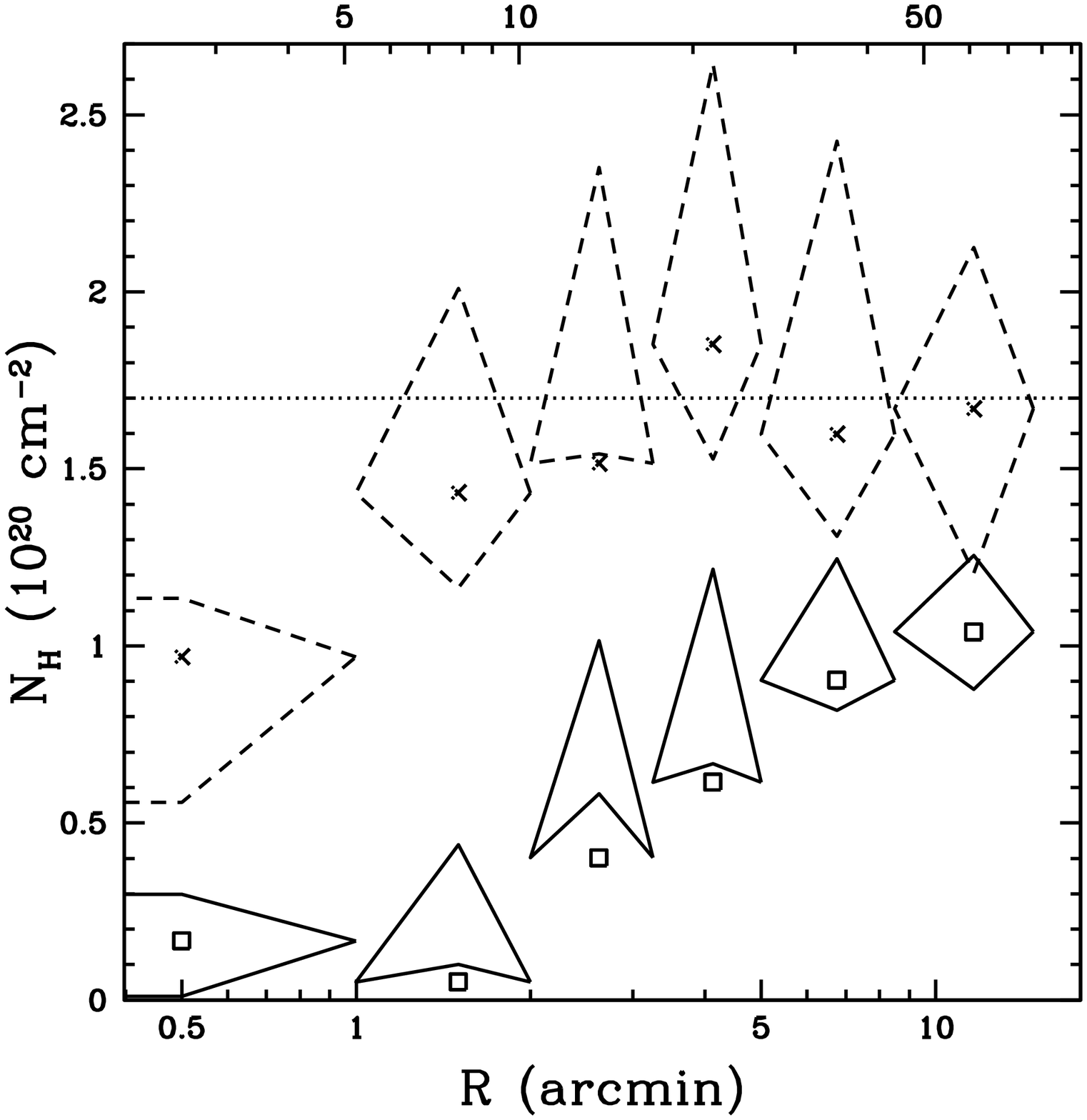,height=0.22\textheight}}
}
\parbox{0.32\textwidth}{
\centerline{\psfig{figure=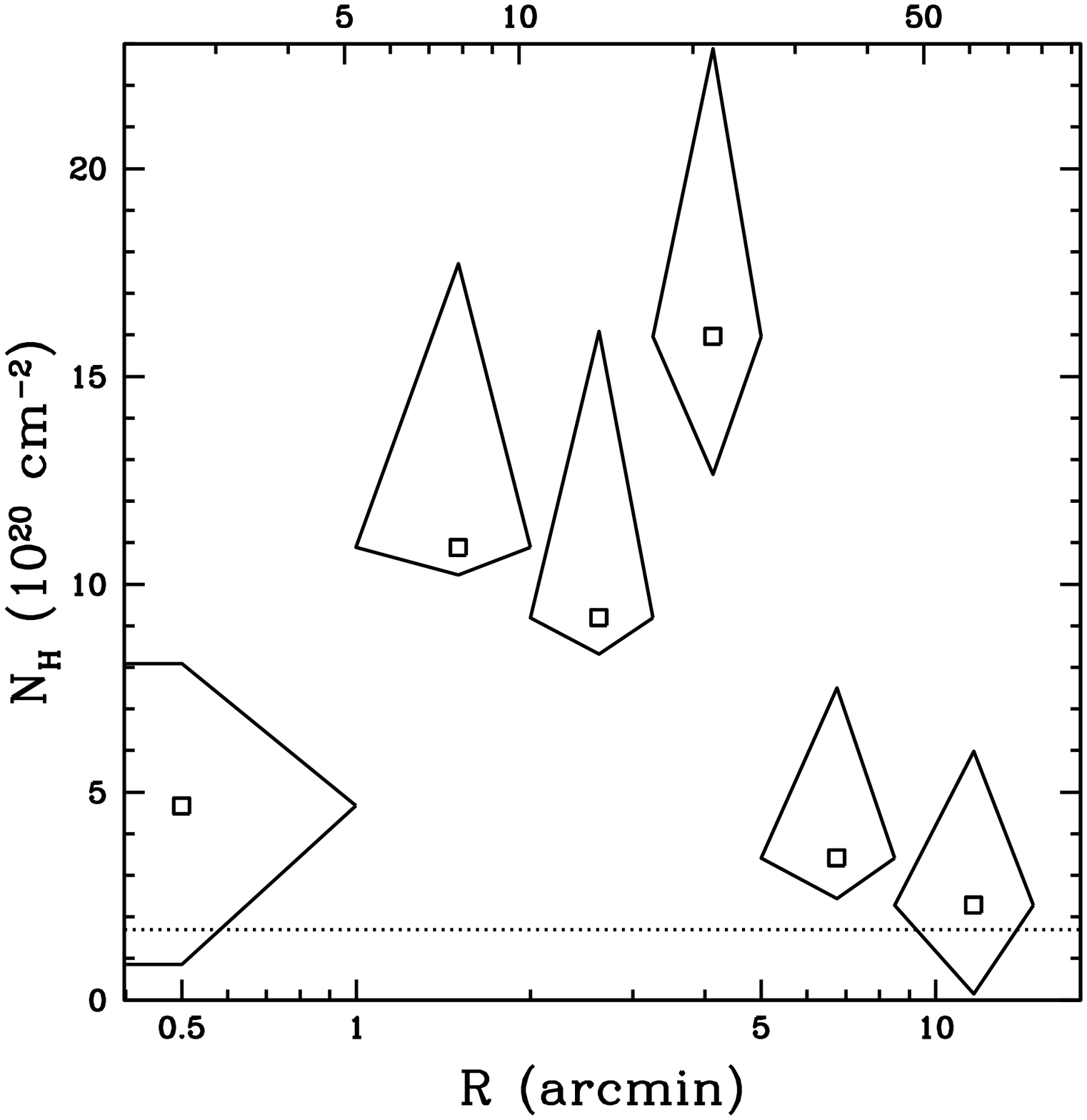,height=0.22\textheight}}
}
\parbox{0.32\textwidth}{
\centerline{\psfig{figure=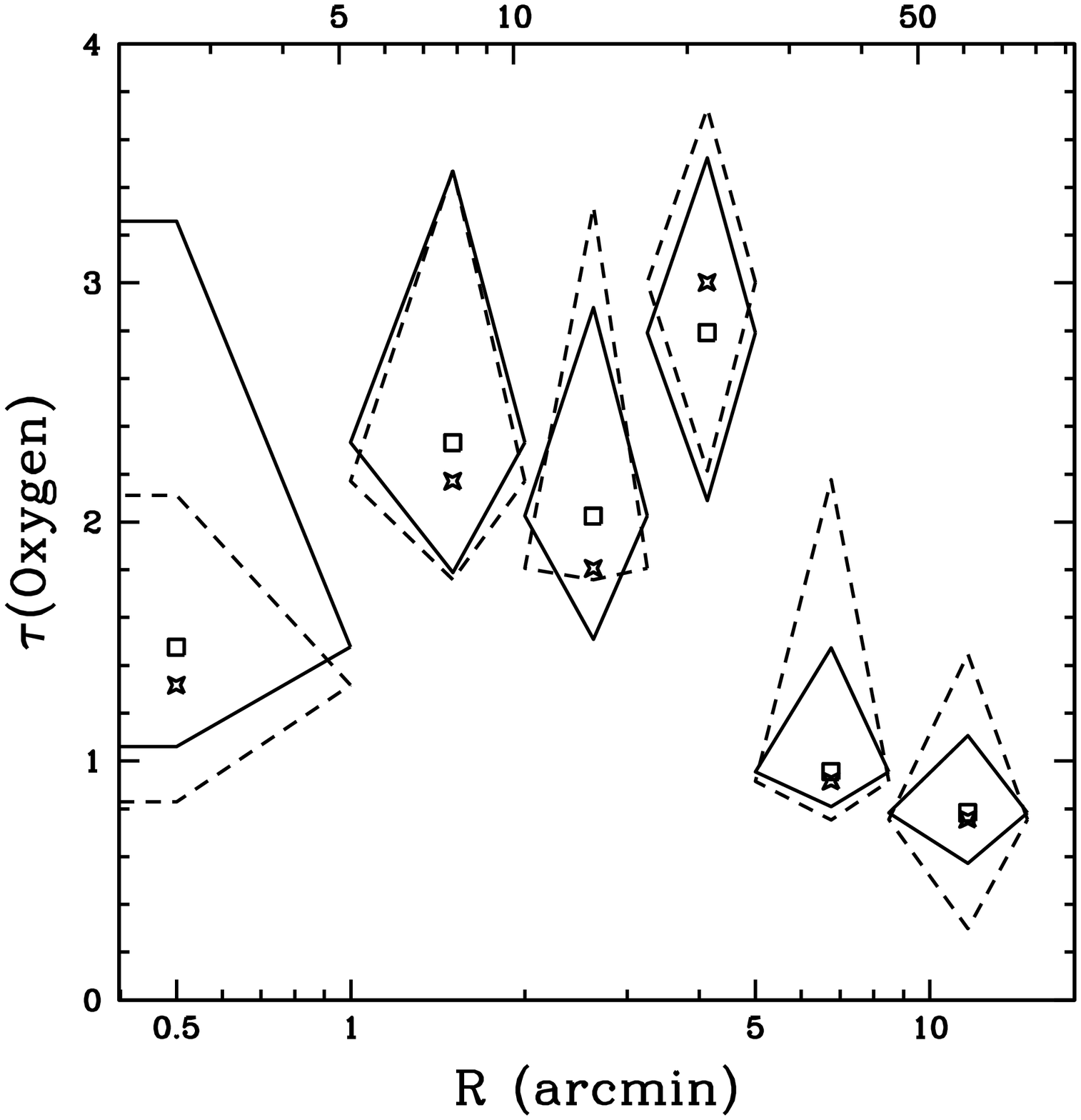,height=0.22\textheight}}
}
\vskip 0.25cm
\centerline{\large\bf NGC 5846} \vskip 0.1cm
\parbox{0.32\textwidth}{
\centerline{\psfig{figure=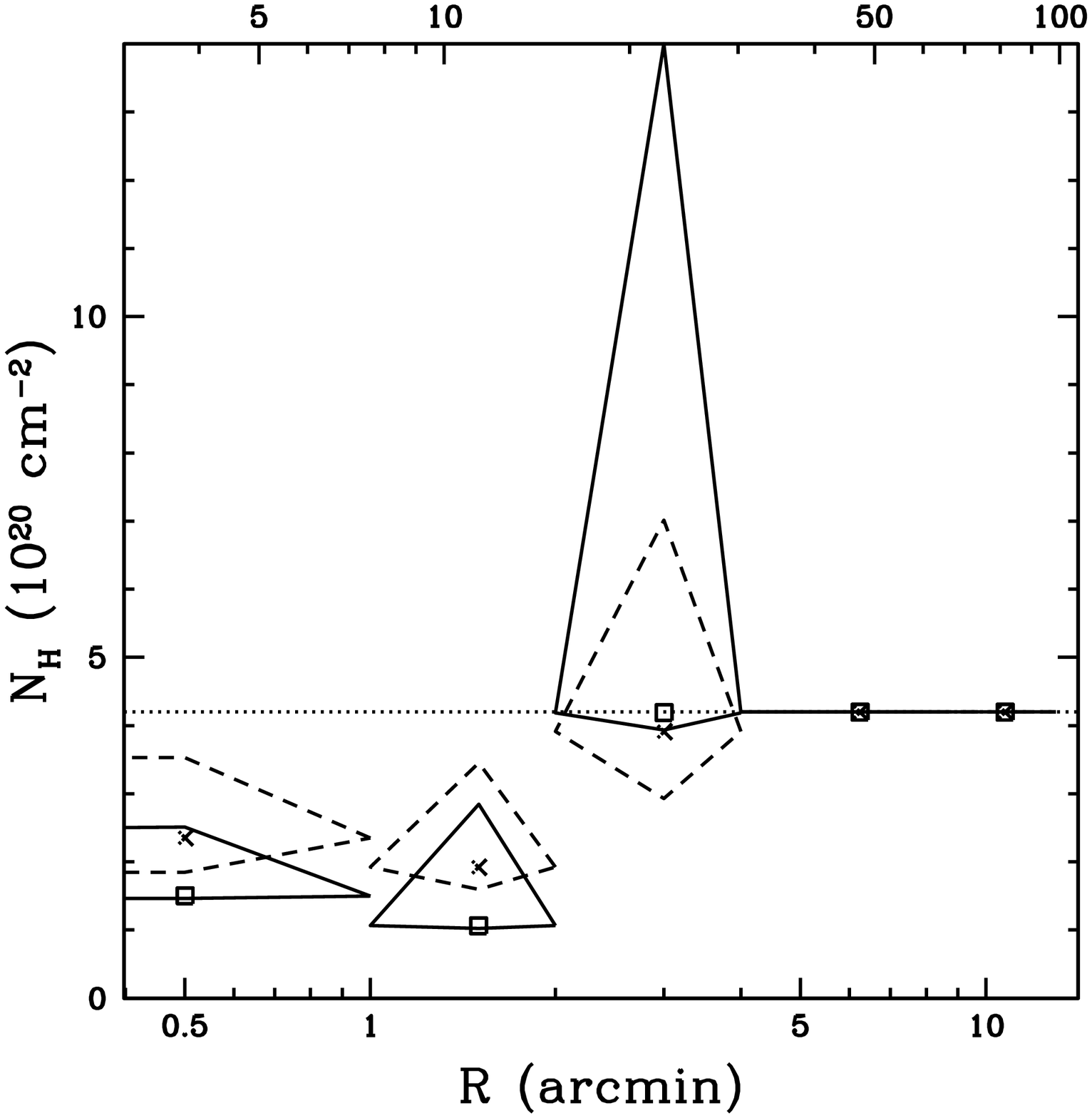,height=0.22\textheight}}
}
\parbox{0.32\textwidth}{
\centerline{\psfig{figure=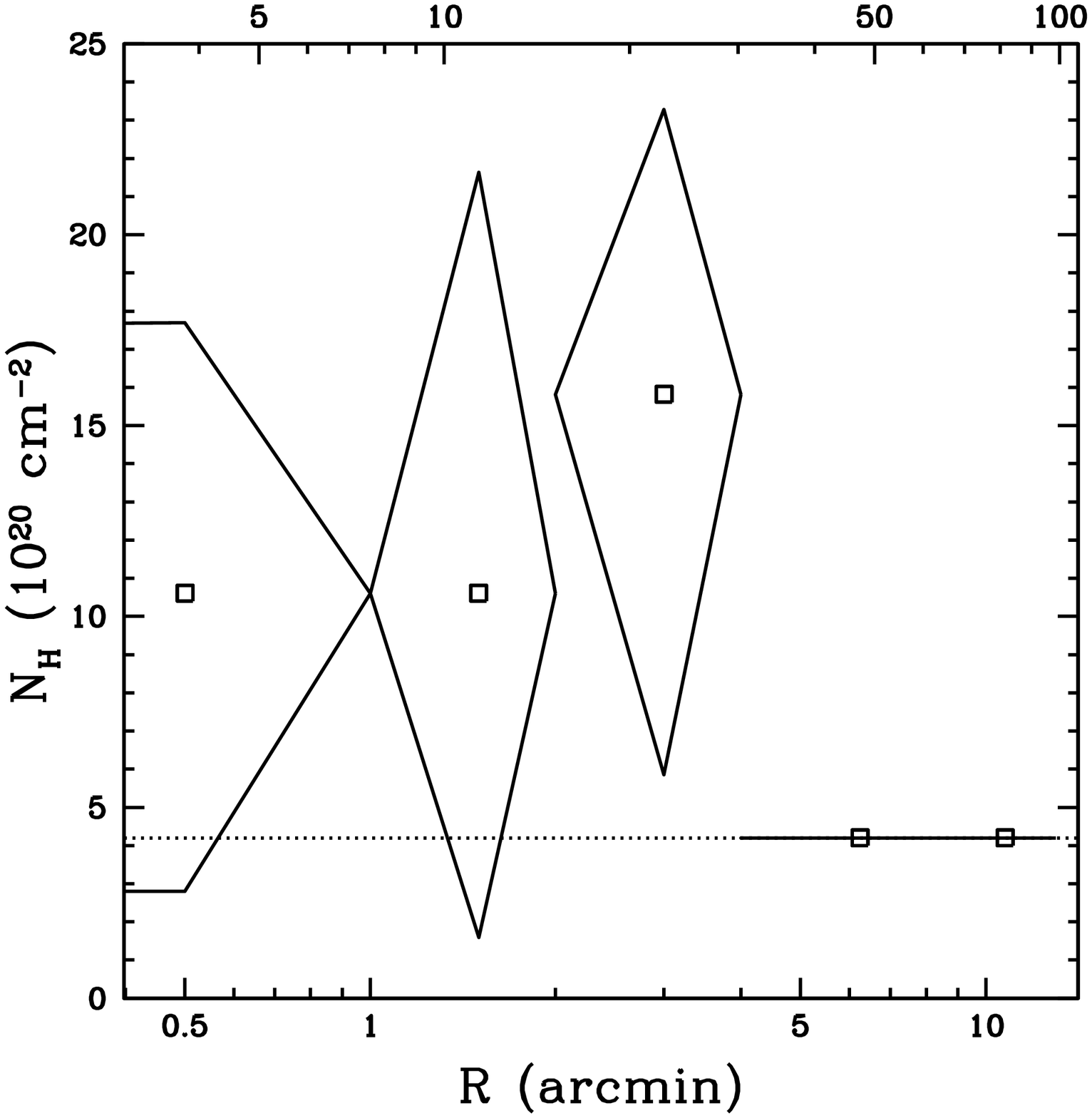,height=0.22\textheight}}
}
\parbox{0.32\textwidth}{
\centerline{\psfig{figure=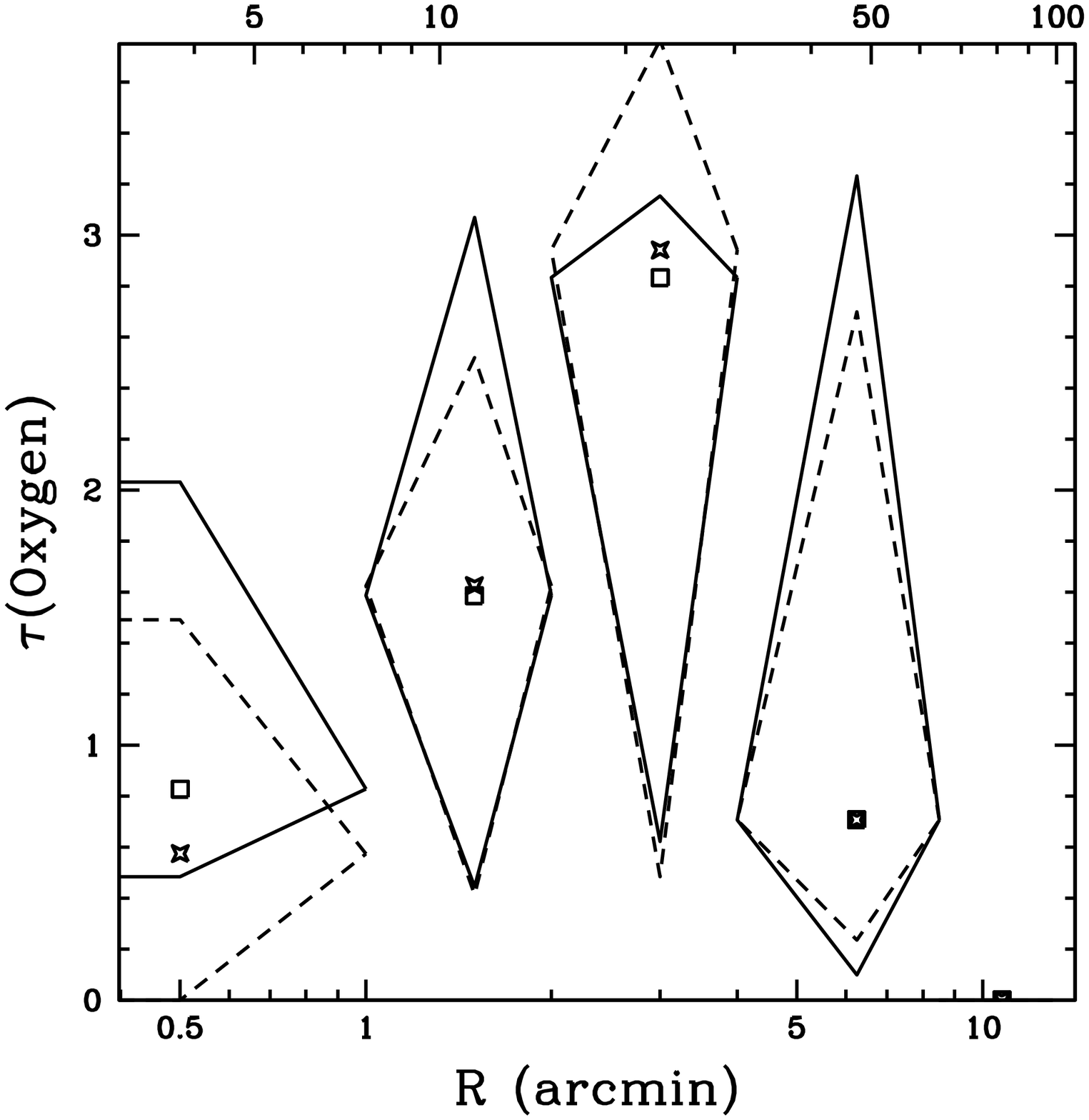,height=0.22\textheight}}
}
\caption{\label{fig.6ann} As Figure \ref{fig.7ann} but for systems
with 5 or 6 annuli.}
\end{figure*}

\begin{figure*}[t]

{\Large\boldmath \Large\bf 
\hskip 1.25cm $\emin = 0.2$ keV \hskip 2.1cm $\emin = 0.5$ keV \hskip 2.2cm
$\emin = 0.2$ keV 
}

\vskip 0.5cm

\centerline{\large\bf NGC 533} \vskip 0.1cm
\parbox{0.32\textwidth}{
\centerline{\psfig{figure=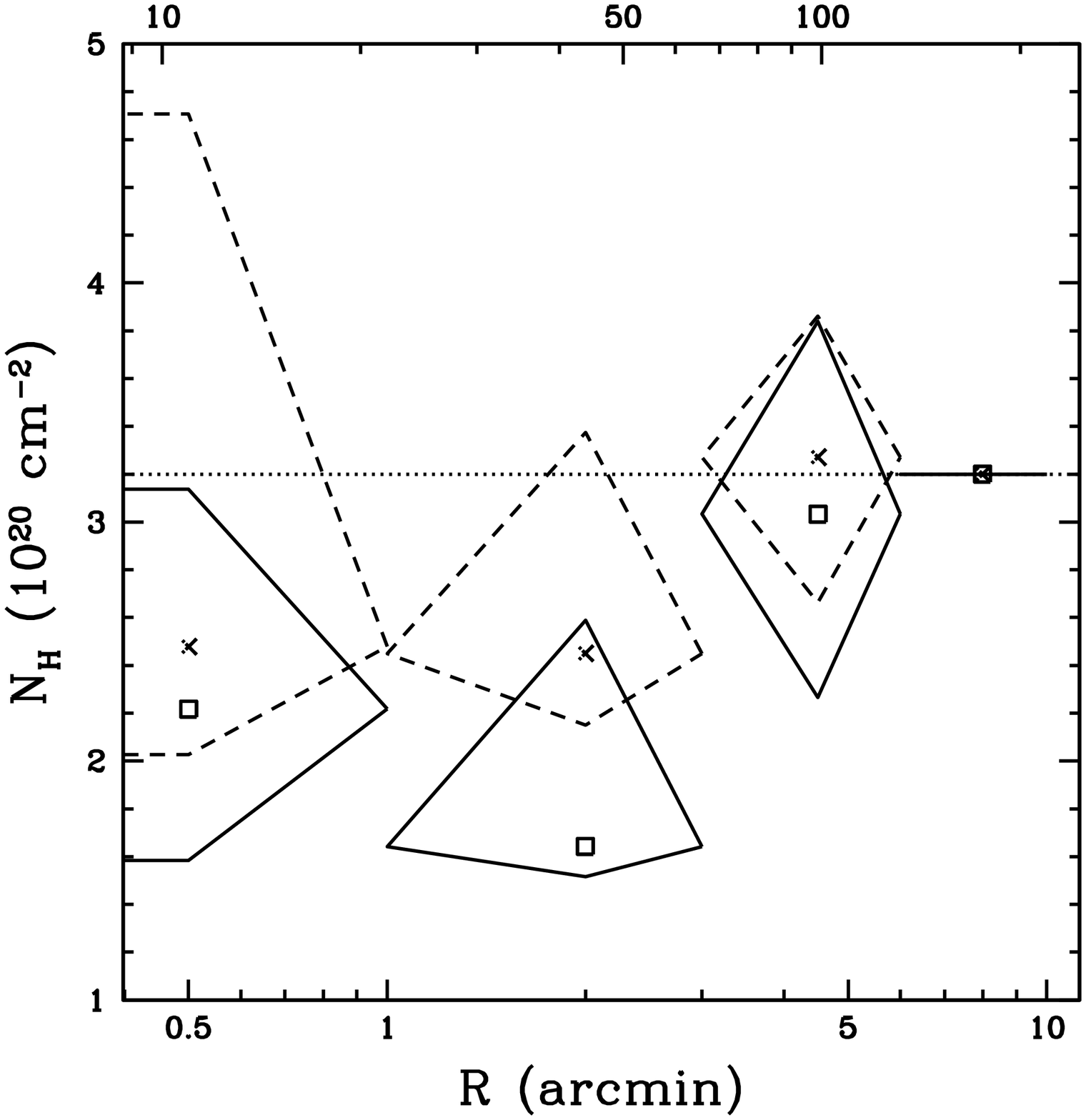,height=0.22\textheight}}
}
\parbox{0.32\textwidth}{
\centerline{\psfig{figure=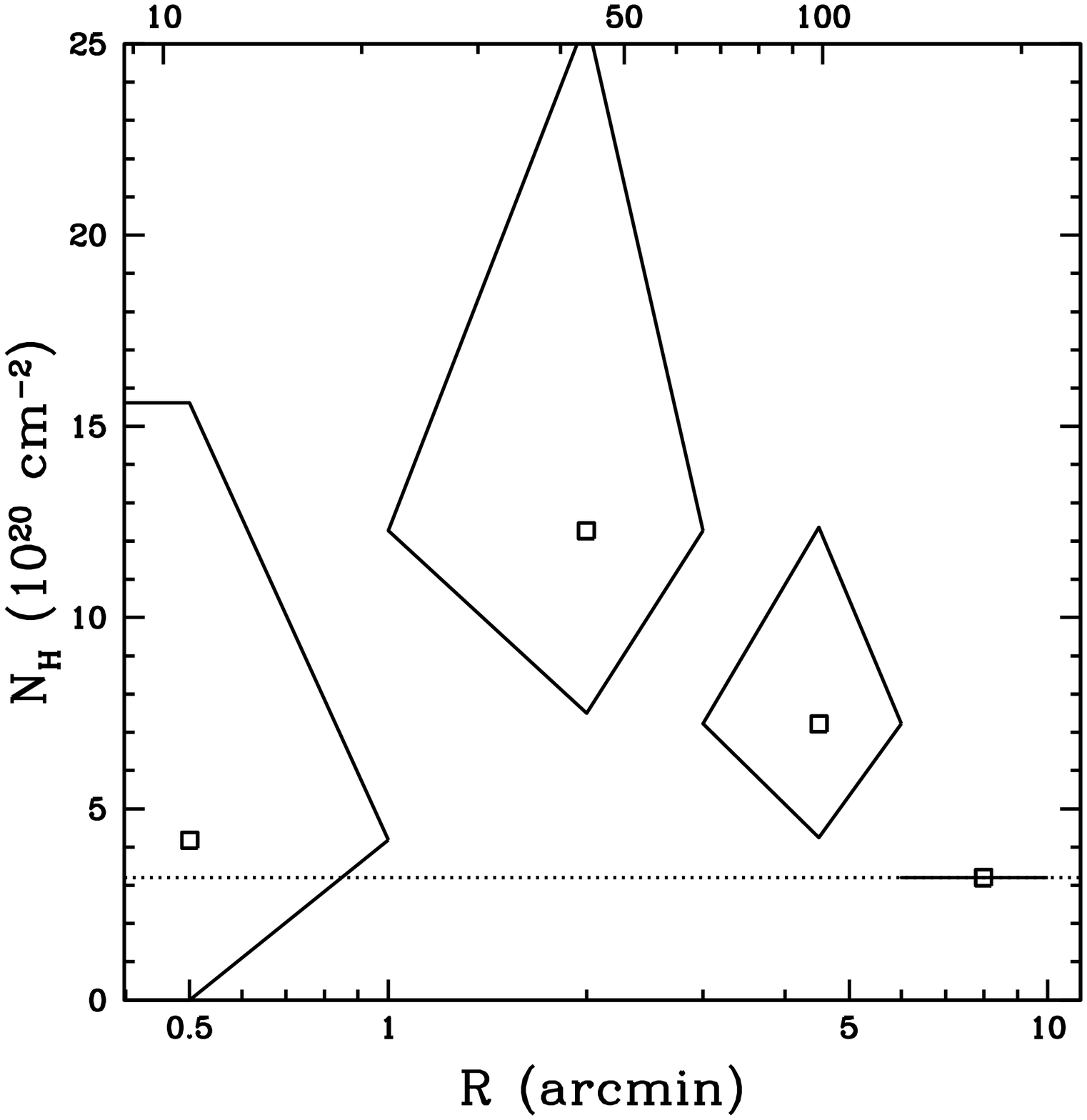,height=0.22\textheight}}
}
\parbox{0.32\textwidth}{
\centerline{\psfig{figure=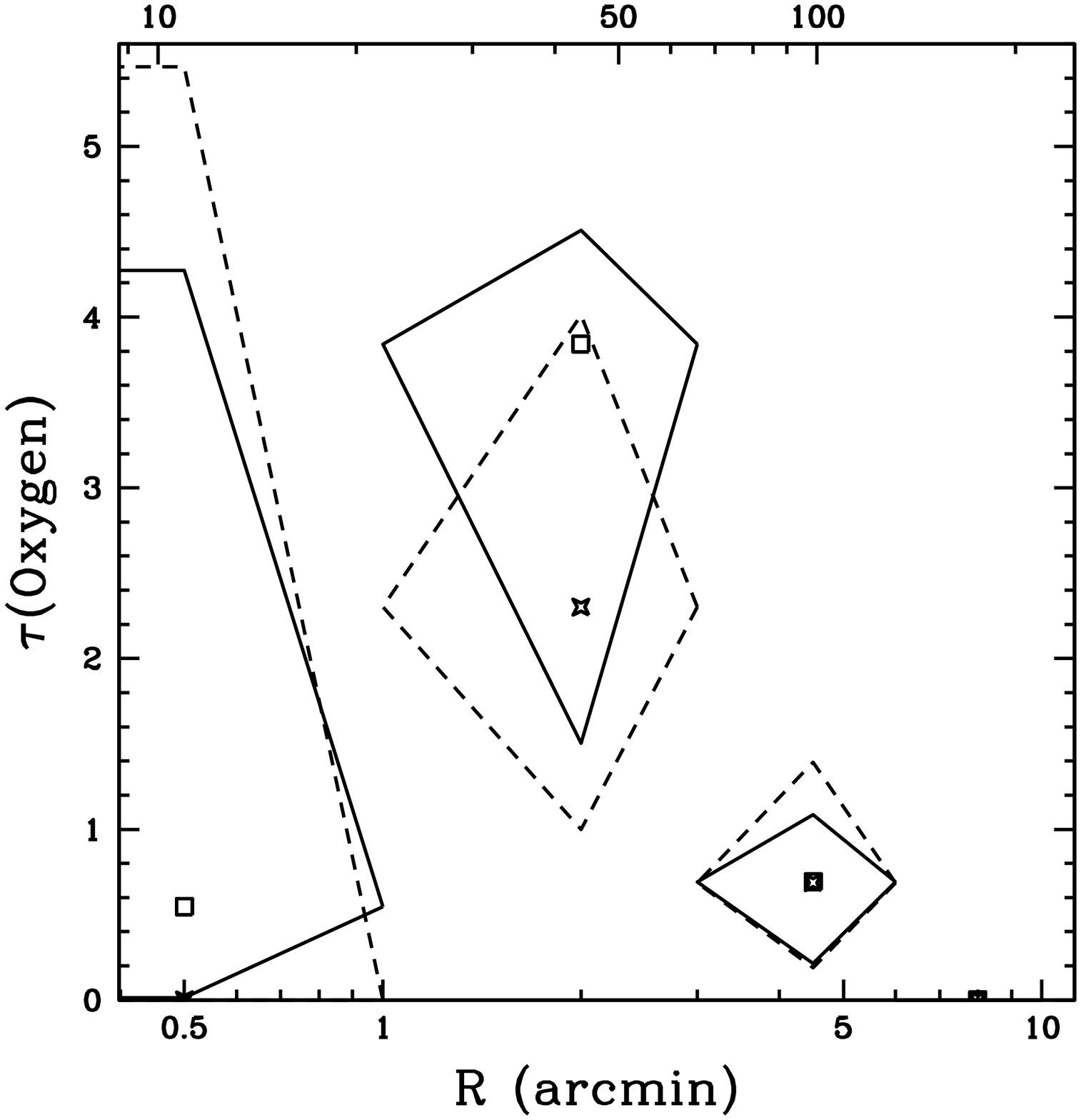,height=0.22\textheight}}
}
\vskip 0.25cm
\centerline{\large\bf NGC 4636} \vskip 0.1cm
\parbox{0.32\textwidth}{
\centerline{\psfig{figure=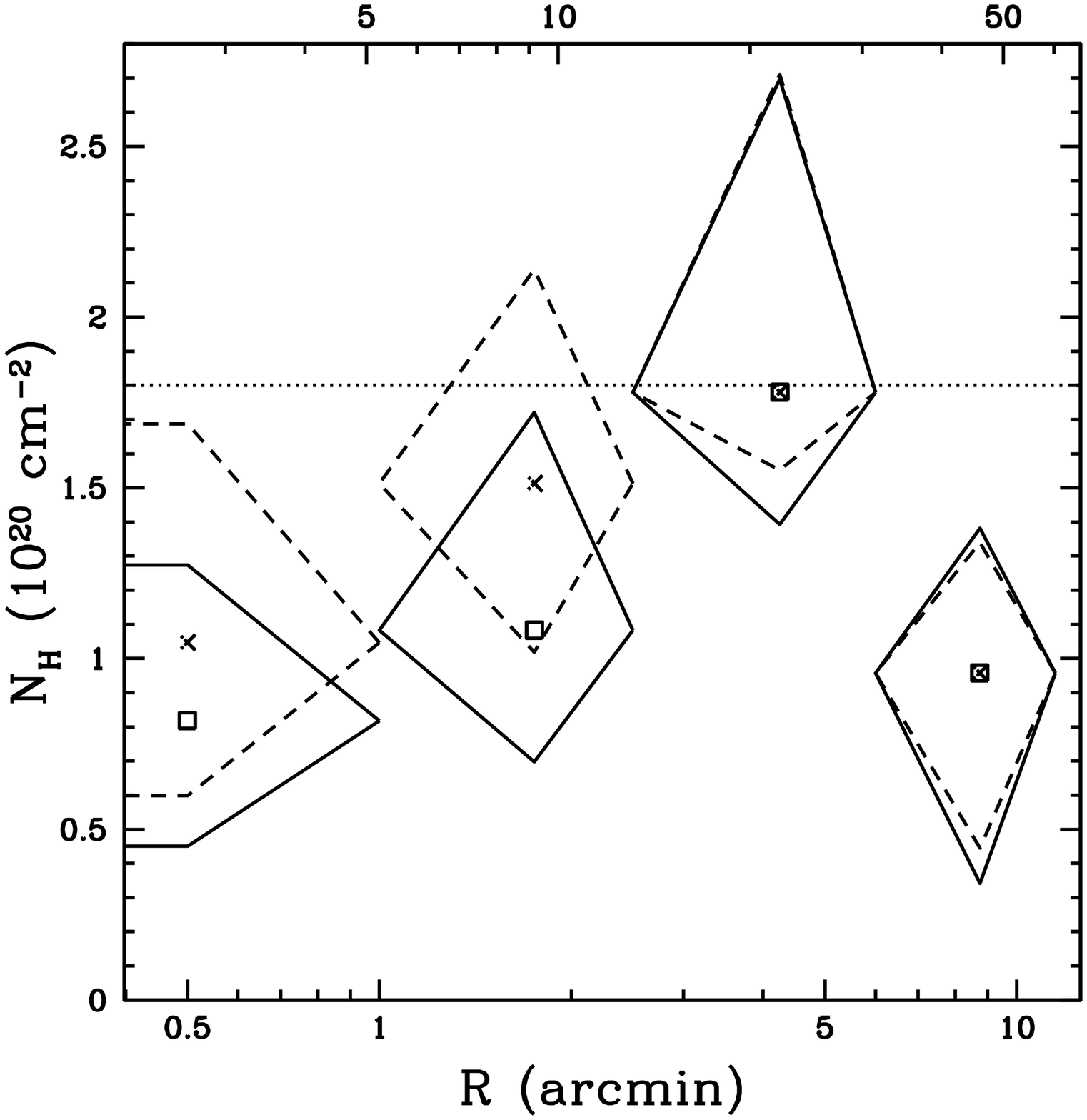,height=0.22\textheight}}
}
\parbox{0.32\textwidth}{
\centerline{\psfig{figure=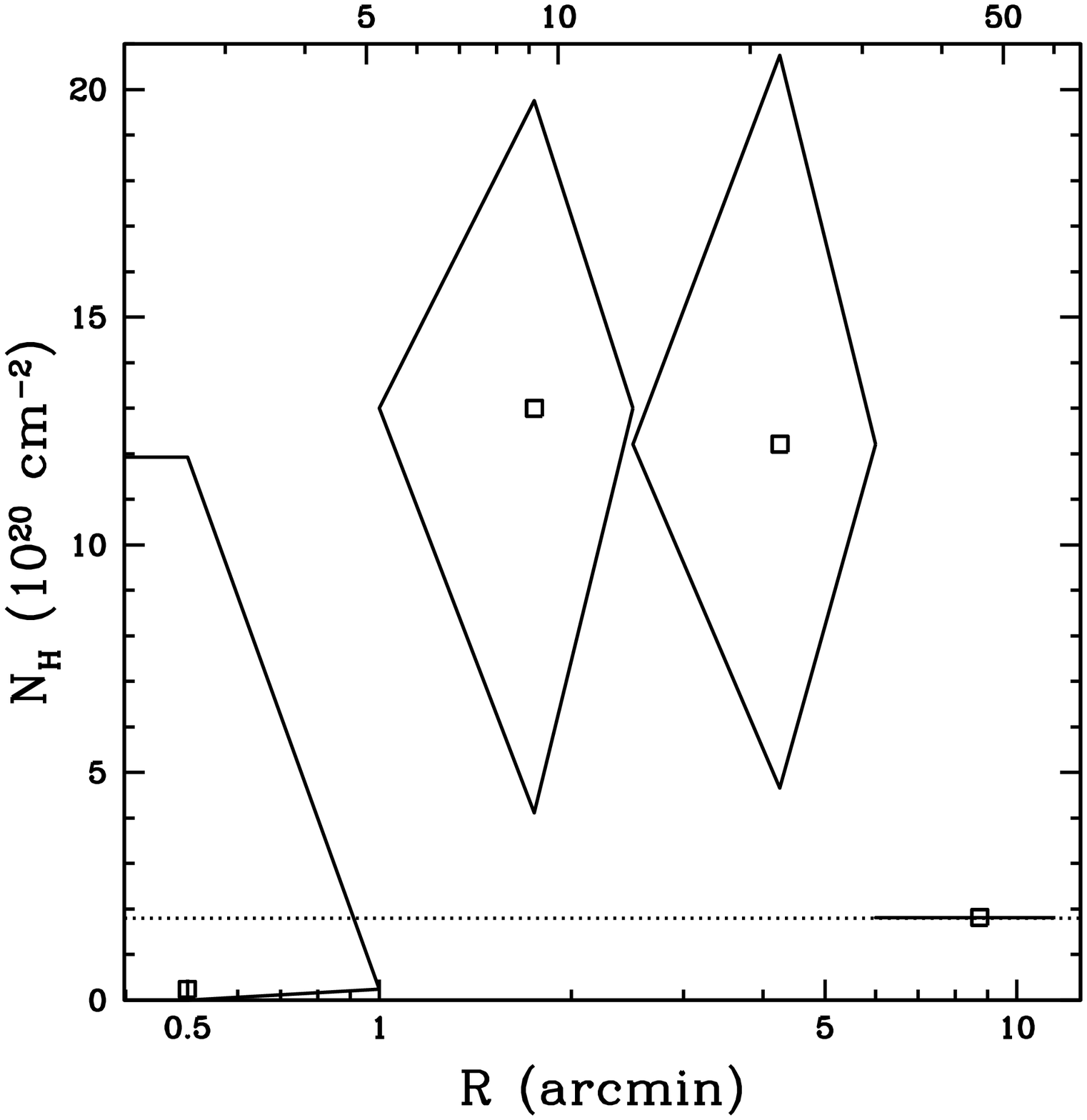,height=0.22\textheight}}
}
\parbox{0.32\textwidth}{
\centerline{\psfig{figure=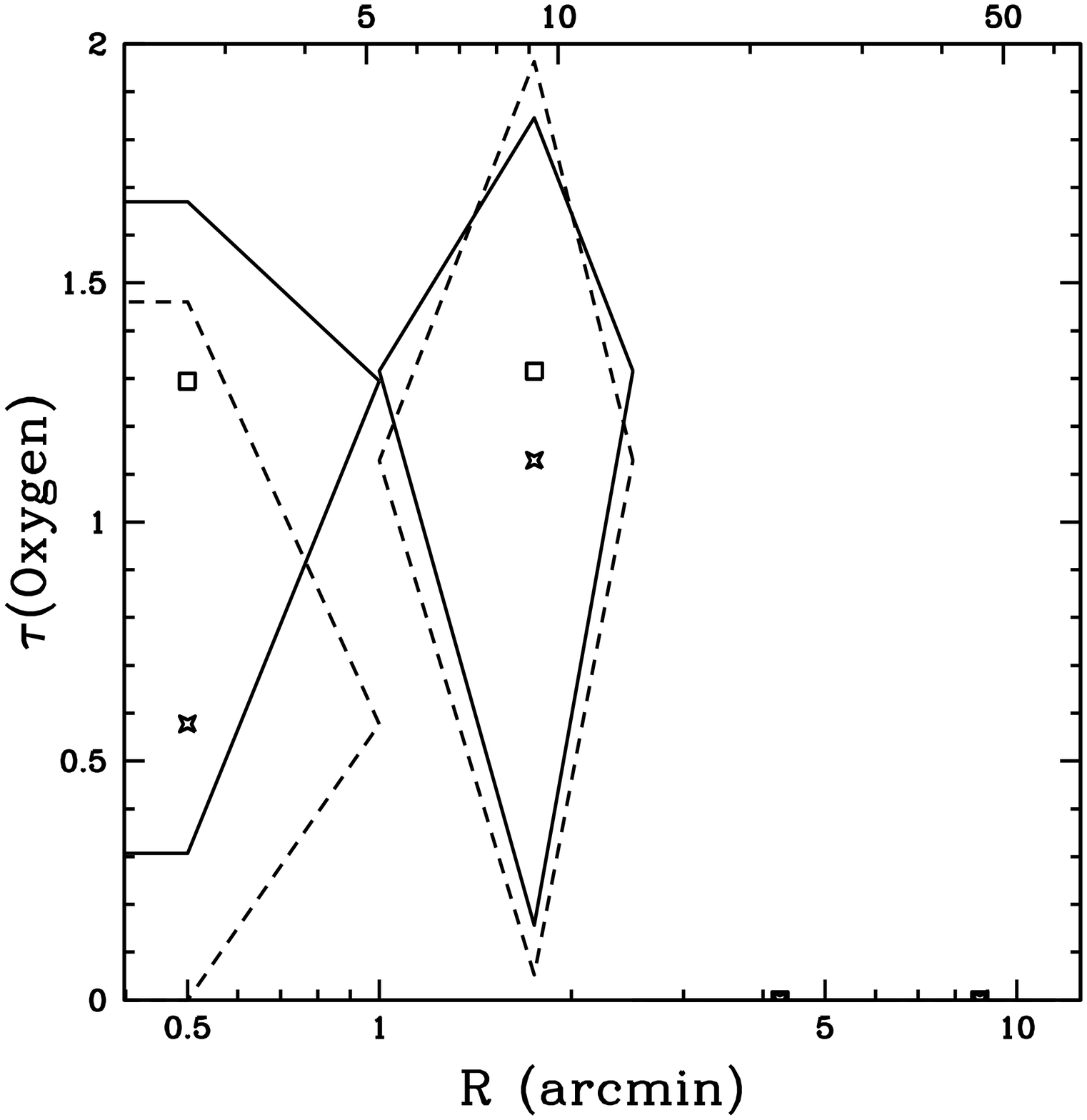,height=0.22\textheight}}
}
\vskip 0.25cm
\centerline{\large\bf HCG 62} \vskip 0.1cm
\parbox{0.32\textwidth}{
\centerline{\psfig{figure=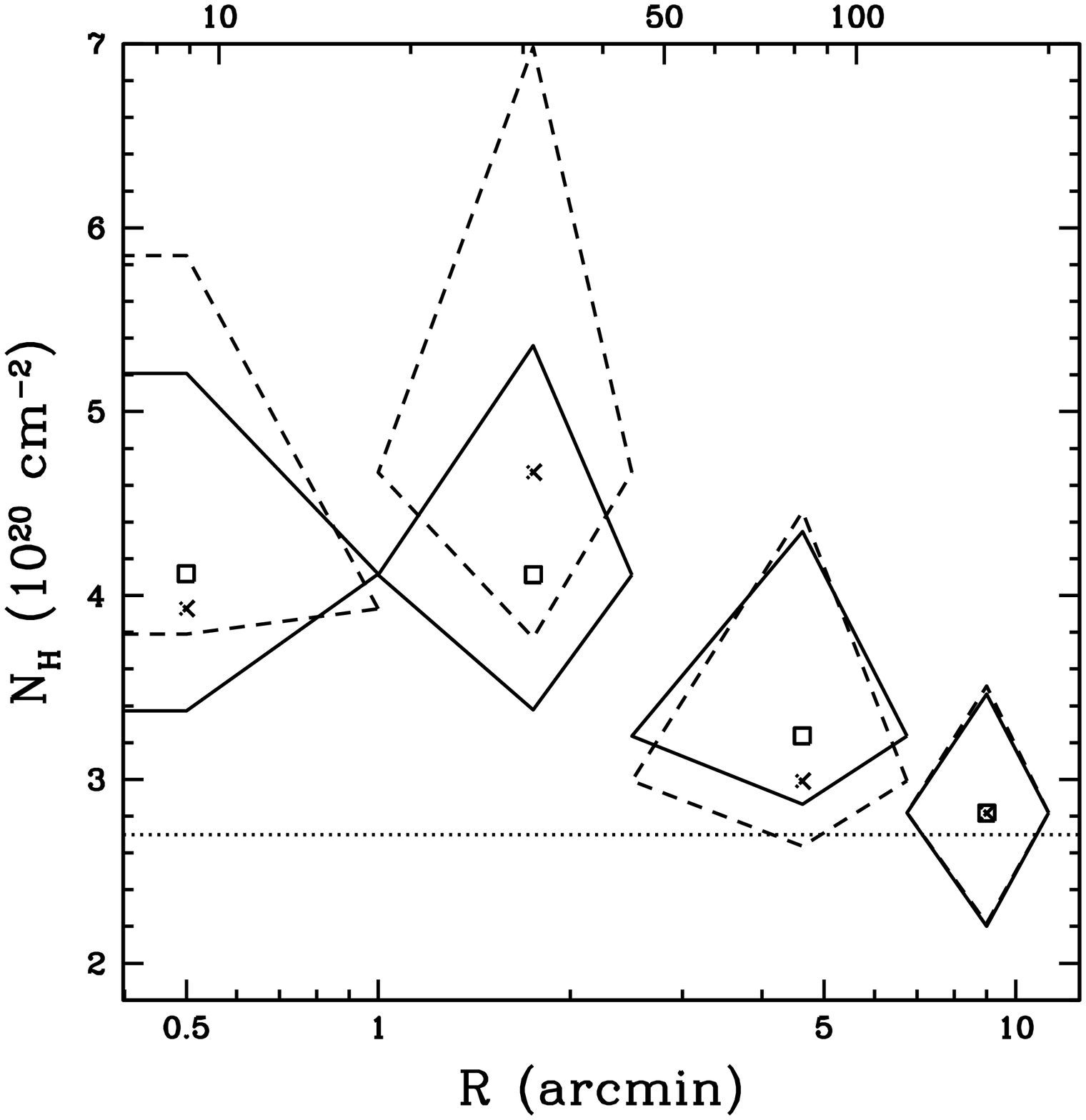,height=0.22\textheight}}
}
\parbox{0.32\textwidth}{
\centerline{\psfig{figure=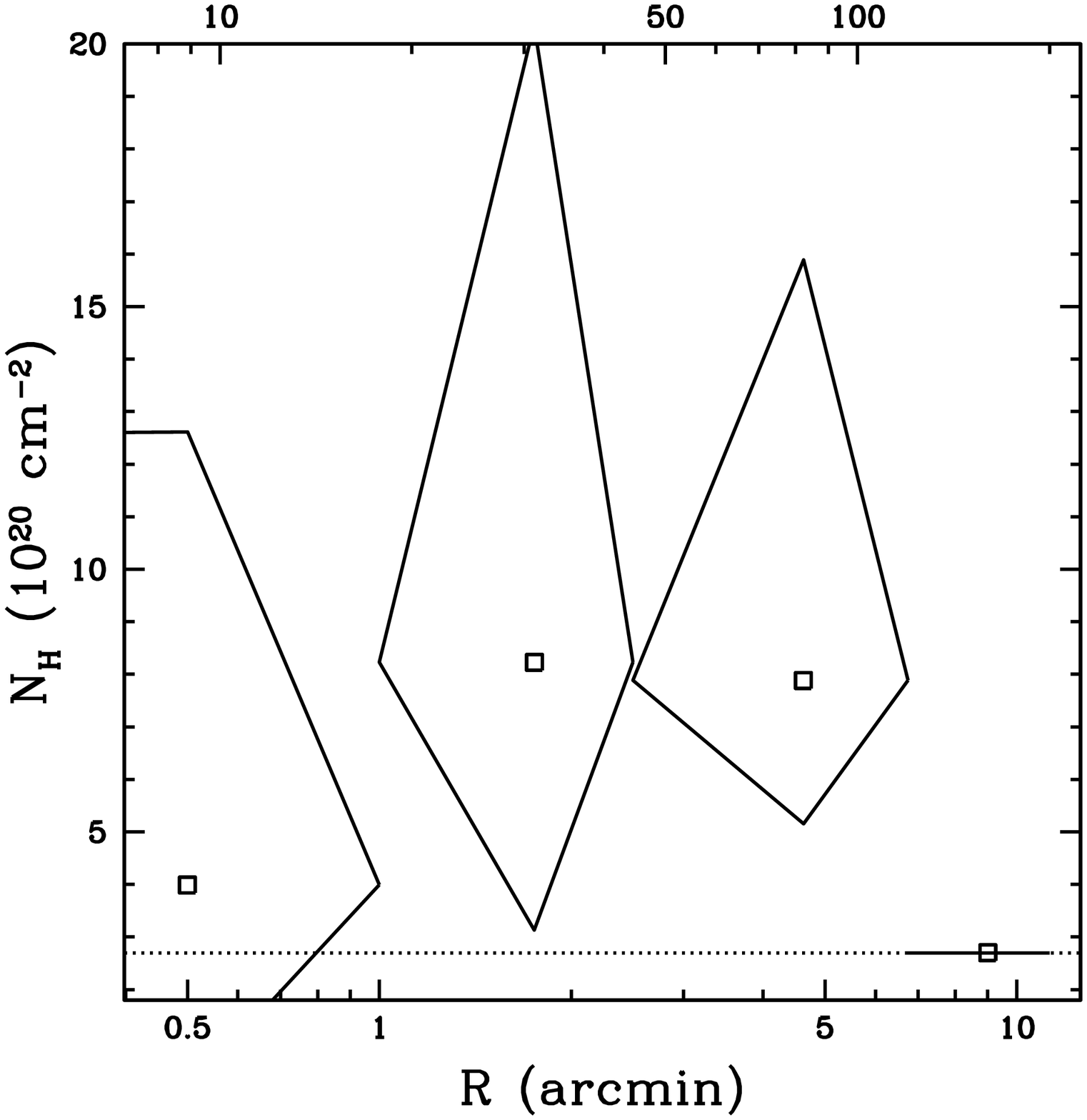,height=0.22\textheight}}
}
\parbox{0.32\textwidth}{
\centerline{\psfig{figure=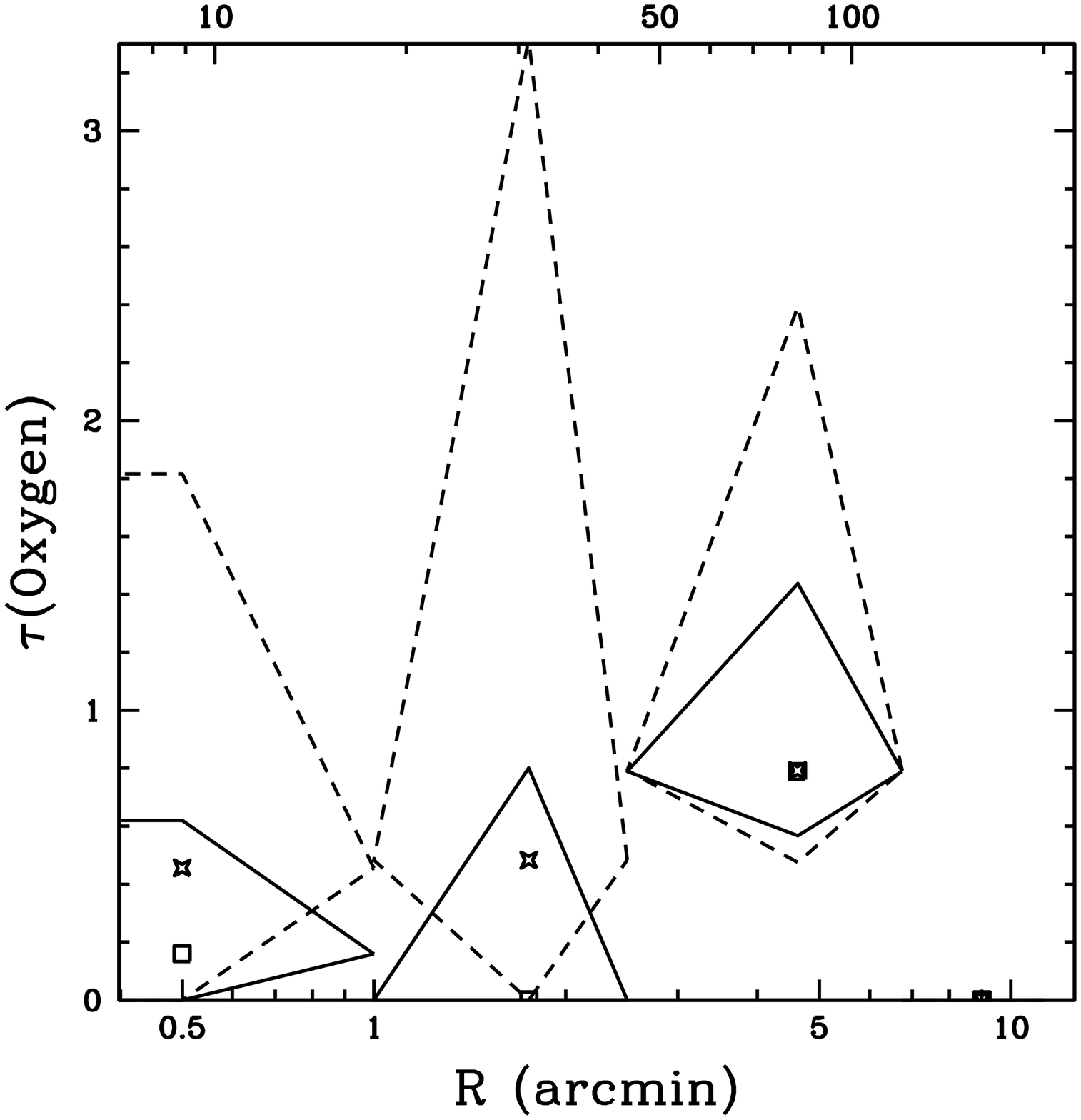,height=0.22\textheight}}
}
\caption{\label{fig.4ann} As Figure \ref{fig.7ann} but for systems
with 4 annuli.}
\end{figure*}

\begin{figure*}[t]

{\Large\boldmath \Large\bf 
\hskip 1.25cm $\emin = 0.2$ keV \hskip 2.1cm $\emin = 0.5$ keV \hskip 2.2cm
$\emin = 0.2$ keV 
}

\vskip 0.5cm

\centerline{\large\bf NGC 4649} \vskip 0.1cm
\parbox{0.32\textwidth}{
\centerline{\psfig{figure=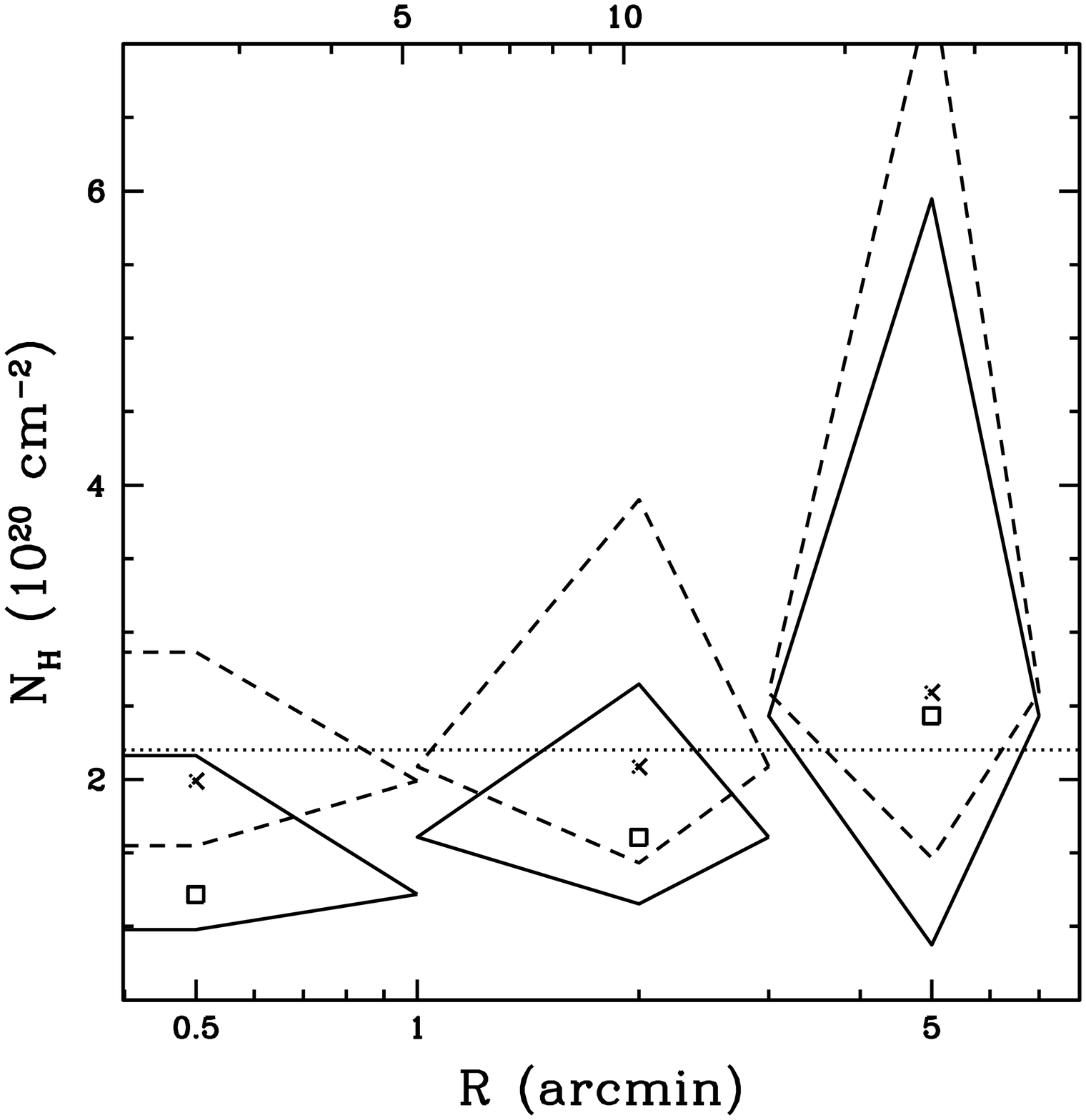,height=0.22\textheight}}
}
\parbox{0.32\textwidth}{
\centerline{\psfig{figure=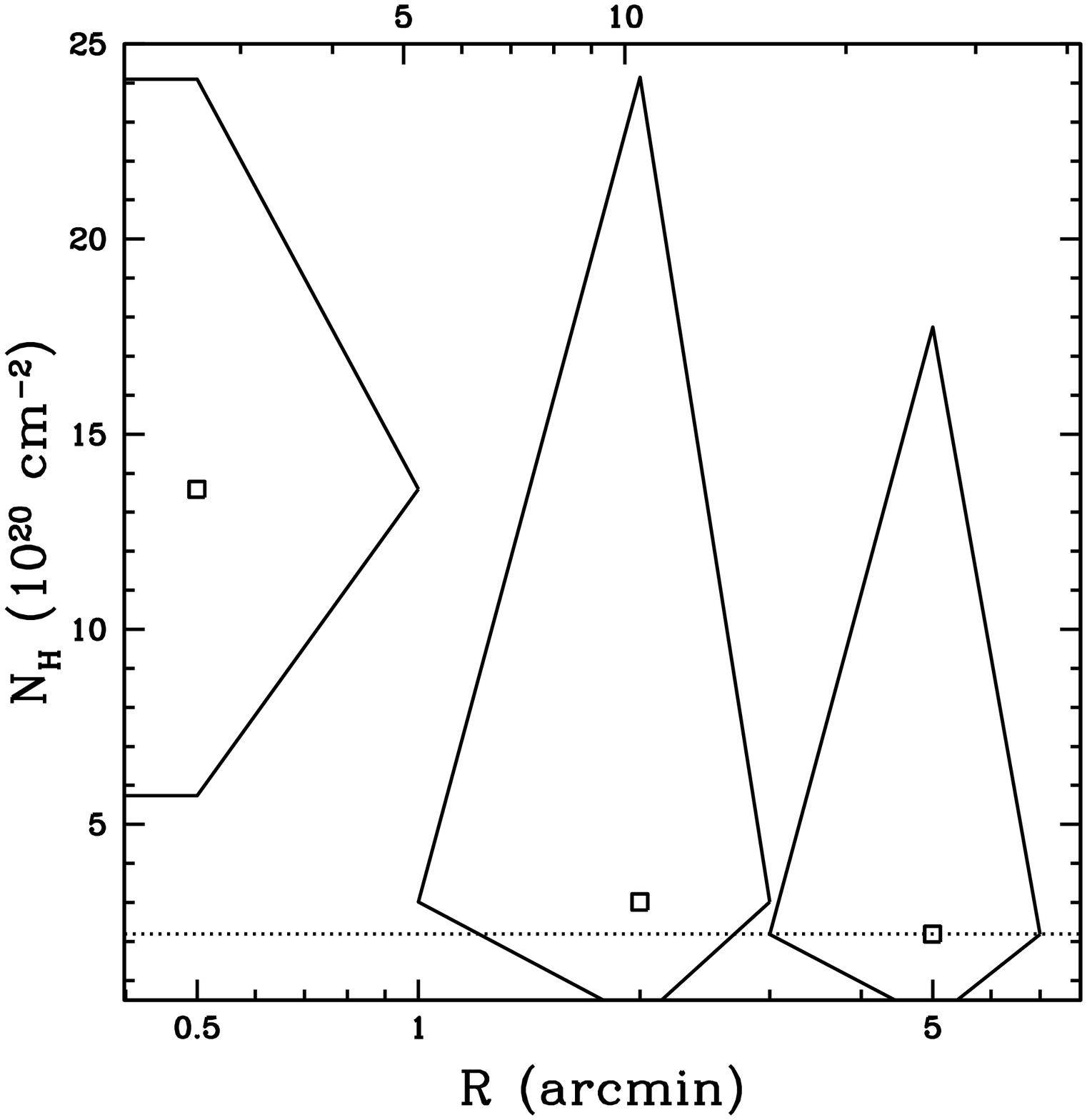,height=0.22\textheight}}
}
\parbox{0.32\textwidth}{
\centerline{\psfig{figure=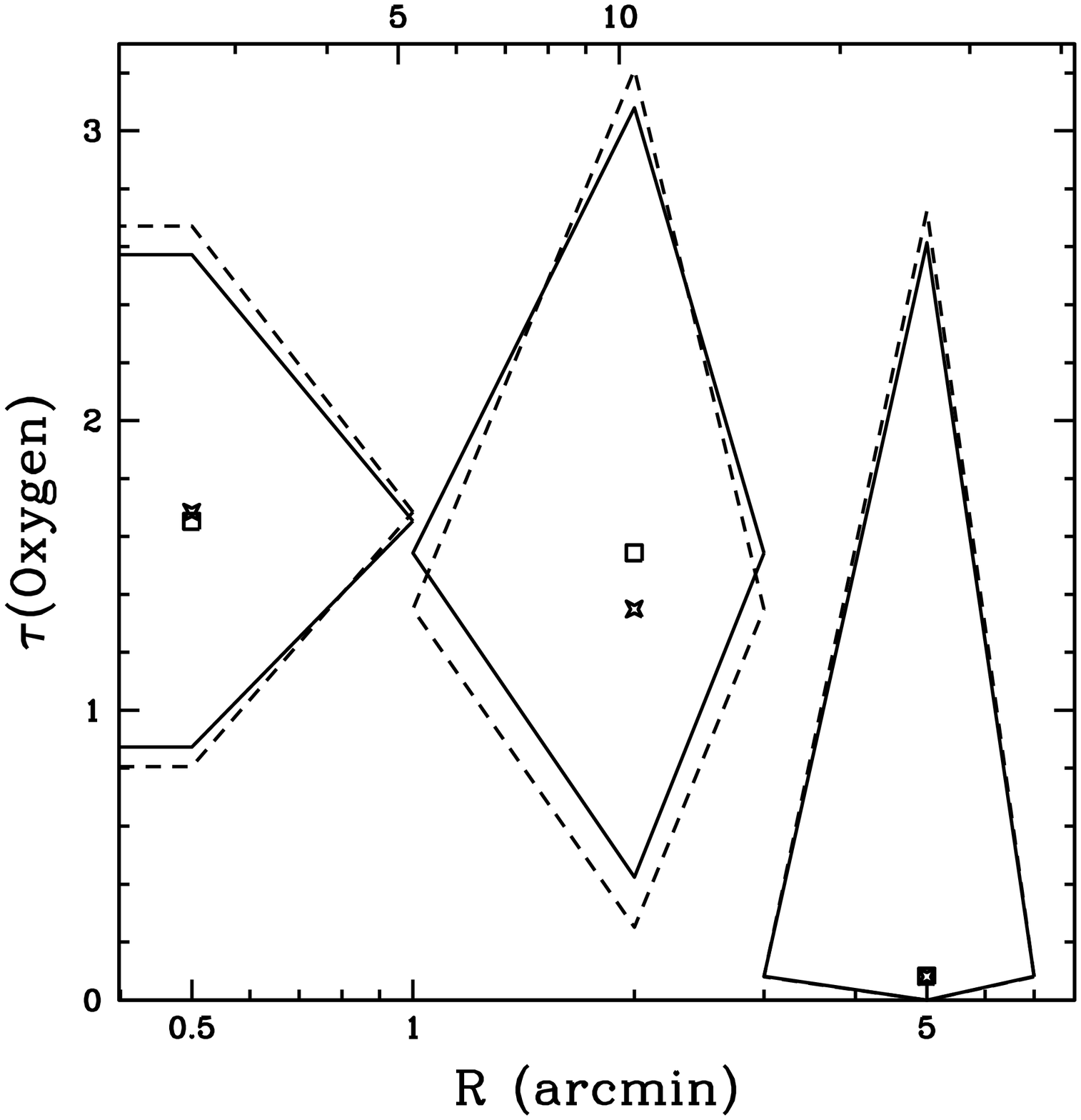,height=0.22\textheight}}}
\caption{\label{fig.3ann} As Figure \ref{fig.7ann} but for the one system
with 3 annuli.}
\end{figure*}

\begin{table*}[t] \footnotesize
\begin{center}
\caption{Quality of Fits and Edge Optical Depth for $r=1\arcmin$\label{tab.fits}}
\begin{tabular}{lc|ccc|cccc|c} \tableline\tableline\\[-7pt]
& & \multicolumn{3}{c}{No Edge} & \multicolumn{4}{c}{Edge} & F Test\\
Name & \nh\, & $\chi^2$ & dof & $P$ & $\chi^2$ & dof & $P$ & $\tau$ & $P(F)$\\ \tableline\\[-7pt]
N507	& Fix	& 36.2	& 47	& 0.87	& 32.2	& 46	& 0.94	& 1.9(0.59) & 2.7E-2\\
	& Free	& 34.0	& 46	& 0.91	& 30.4	& 45	& 0.95	& 1.5(0.49) & 2.6E-2\\
N533 	& Fix	& 29.9 	& 31	& 0.52 	& 29.7	& 30	& 0.48	& 0.55(0.00)& 0.66\\
	& Free	& 28.9	& 30	& 0.52	& 28.9	& 29	& 0.47	& 0.00(0.00)& 1.00\\ 
N1399	& Fix	& 106.9	& 97	& 0.23	& 83.7	& 96	& 0.81	& 3.8(0.89) & 1.3E-6\\
	& Free	& 104.9	& 96	& 0.25	& 83.7	& 95	& 0.79	& 3.3(0.97) & 3.9E-6\\
N2563	& Fix	& 24.0	& 25	& 0.52	& 22.9	& 24	& 0.53	& 1.5(0.00) & 0.29\\
	& Free	& 23.6	& 24	& 0.48	& 22.9	& 23	& 0.47	& 1.3(0.00) & 0.41\\	
N4472	& Fix	& 209.0	& 104	& 4.7E-9 & 158.1 & 103 & 4.0E-4 & 1.5(0.74) & 8.8E-8\\
	& Free 	& 158.9	& 103	& 3.4E-4 & 142.2 & 102 & 5.3E-3 & 1.3(0.25) & 7.9E-4\\
N4636	& Fix	& 109.9	& 52	& 4.9E-6 & 105.8 & 51 & 1.0E-5	& 1.3(0.00) & 0.17\\
	& Free	& 104.7	& 51	& 1.4E-5 & 103.1 & 50 & 1.5E-5 & 0.58(0.00) & 0.38\\
N4649	& Fix	& 67.2	& 61	& 0.27	& 58.1	& 60	& 0.55	& 1.7(0.44) & 3.2E-3\\
	& Free	& 63.4	& 60	& 0.36	& 57.8	& 59	& 0.52	& 1.7(0.00) & 2.0E-2\\
N5044	& Fix	& 191.2	& 118	& 2.3E-5& 156.5 & 117	& 8.7E-3& 1.8(0.59) & 9.2E-10\\
	& Free	& 138.6	& 117	& 8.5E-2& 126.4 & 116	& 0.24	& 1.3(0.45) & 1.1E-3\\
N5846	& Fix	& 84.0	& 47	& 7.4E-4& 78.4	& 46	& 2.1E-3& 0.83(0.00)& 7.6E-2\\
	& Free	& 74.6	& 46	& 4.8E-3& 72.6	& 45	& 5.7E-3& 0.57(0.00)& 0.27\\
H62	& Fix	& 37.8	& 46	& 0.80	& 37.7	& 54	& 0.77	& 0.16(0.00)& 0.73\\
	& Free	& 38.6	& 45	& 0.74	& 37.7	& 44	& 0.74	& 0.46(0.00)& 0.31\\ \tableline
\end{tabular}
\tablecomments{Results of deprojection analysis in the central radial
bin ($R=1\arcmin$). Models where the column density of the standard
foreground absorber with solar abundances is fixed to the Galactic
value are indicated by ``Fix'' under the \nh\, column; variable column
models are indicated by ``Free''.  For the models with an intrinsic
oxygen edge at 0.532 keV (rest frame) we list the best-fitting optical
depth, $\tau$, with its 95\% confidence lower limit in
parentheses. The $\chi^2$ null hypothesis probability is listed in the
column $P$ under the assumption of gaussian random errors (see section
3.4 of PAPER2). The F-Test probability, $P(F)$ (e.g., Bevington 1969),
quantifies the improvement in $\chi^2$ when adding the edge; i.e.,
$P(F)\ll 1$ indicates significant improvement.}
\end{center}
\end{table*}

Following PAPER2 we plot in Figures \ref{fig.7ann}-\ref{fig.3ann} the
radial profiles of the column density and oxygen edge optical depth
obtained from the deprojection analysis according to the number of
annuli for which useful constraints on the parameters were
obtained. This categorizes the systems essentially according to the
S/N of the data.  We refer the reader to PAPER2 for the temperature
and metallicity profiles corresponding to these models.

In several cases the column densities and optical depths could not be
constrained in the outermost annuli. Owing to the nature of the
deprojection method large errors in the outer annuli can significantly
bias the results for nearby inner annuli. Hence, in some systems we
fixed the column densities to their nominal Galactic values or the
edge optical depths to zero in the relevant outer annuli.

\subsection{Foreground Absorber with Solar Abundances}
\label{solar}

We begin by examining the spectral fits using the standard absorption
model of a foreground screen $(z=0)$ with solar abundances.  The left
panels of Figures \ref{fig.7ann}-\ref{fig.3ann} show $\nh(R)$ obtained
from spectral fits over the energy range 0.2-2.2 keV. Our column
density profiles are consistent with those presented in previous
\rosat studies (Forman et al 1993; David et al 1994; Trinchieri et al
1994; Kim \& Fabbiano 1995; Jones et al 1997; Trinchieri et al 1997;
Buote 1999) after accounting for the different plasma codes and solar
abundances used.

The column densities are always within a factor of $\sim 2$ of the
Galactic value (\nhgal), though in most cases \nh\, decreases as $R$
decreases such that $\nh<\nhgal$ at small $R$. The quality of the fits
for four of these systems is also formally poor $(P<0.01)$ in the
central $1\arcmin$ bin (Table \ref{tab.fits}), and for several objects
in the sample the metallicities are very large and very inconsistent
with all \asca studies (see PAPER2).\footnote{Here $P$ represents the
$\chi^2$ null hypothesis probability under the assumption of gaussian
random errors. We discuss the suitability of this approximation for
interpreting goodness of fit in section 3.4 of PAPER2.} Since
$\nh\sim\nhgal$ in the outer radii the observation that $\nh<\nhgal$
at small $R$ indicates that, whatever the origin of the deficit, it
must be intrinsic to the source. (We provide an explanation in \S
\ref{xray}.)

The approximately Galactic columns are wholly inconsistent with the
large excess columns inferred from multitemperature models of the
spatially integrated \asca spectral data of these systems (Buote \&
Fabian 1998; Buote 1999, 2000a; Allen et al 2000a). We now address the
origin of this inconsistency.

\subsection{Effects of Partial Covering}
\label{partial}

It has been suggested that the reason why analyses with the PSPC do
not infer large excess column densities for cluster cooling flows is
that the standard foreground model used above systematically
underestimates the true column intrinsic to the system (Allen \&
Fabian 1996; Sarazin, Wise, \& Markevitch 1998). However, in PAPER1 we
have tested this hypothesis for NGC 1399 and 5044 (and the cluster
A1795) using our deprojection code. We find that the hot gas within
the central $r=1\arcmin$ (3D) cannot be absorbed very differently from
the gas projected from larger radii because their spectral shapes for
energies below $\sim 0.5$ keV are very similar. If we do assume an
absorber with covering factor $f=0.5$ we obtain an excess column
$\Delta\nh=0$ at best fit and $\Delta\nh<\nhgal$ at $>90\%$ confidence
for NGC 1399 and 5044. (Note that the $f=0.5$ model implies a flat
absorbing screen that bisects the source so that the 2D and 3D radii
are equal, and thus the values of $\Delta\nh$ quoted do refer to
quantities within the 3D radius $r=1\arcmin$.)

Entirely analogous results are obtained for the other systems in our
sample. We mention that partial covering models never improve the fits
over the $f=1$ case. The only effect is that somewhat larger columns
are generally allowed; e.g., for $f=0.5$ the implied upper limits for
the excess columns are typically a factor of 20\%-40\% larger than for
$f=1$.

Hence, as in PAPER1 we conclude that models with $f<1$ cannot account
for (1) the large excess columns inferred from \asca, (2) the
sub-Galactic columns and poor fits obtained for several systems in the
central $1\arcmin$, or as we now discuss (3) the sensitivity of \nh\,
to the lower energy boundary of the bandpass.

\subsection{Sensitivity of $\nh(R)$ to Bandpass}
\label{bandpass}

\begin{figure*}[t]
\parbox{0.49\textwidth}{
\centerline{\psfig{figure=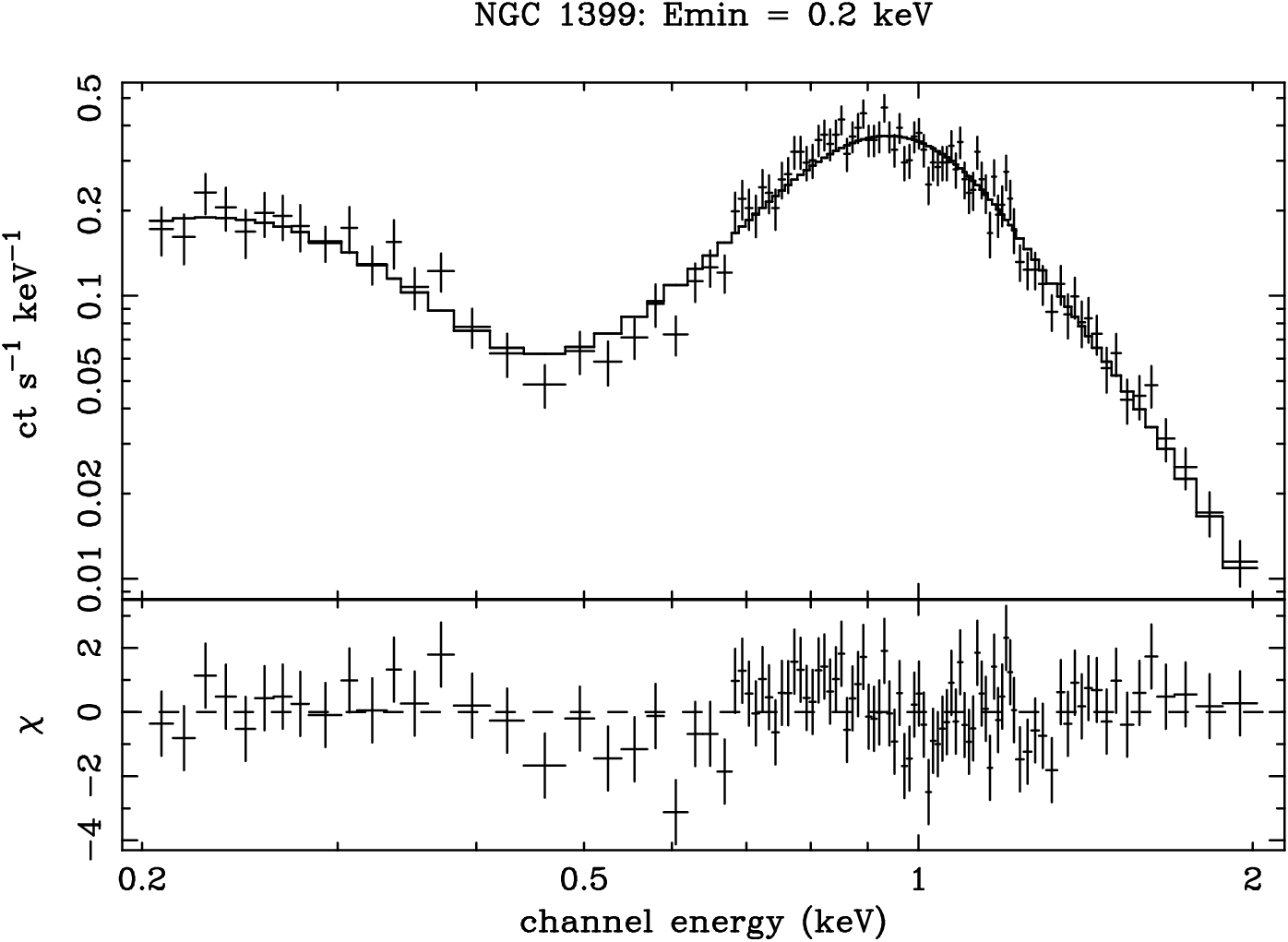,angle=-90,height=0.26\textheight}}}
\parbox{0.49\textwidth}{
\centerline{\psfig{figure=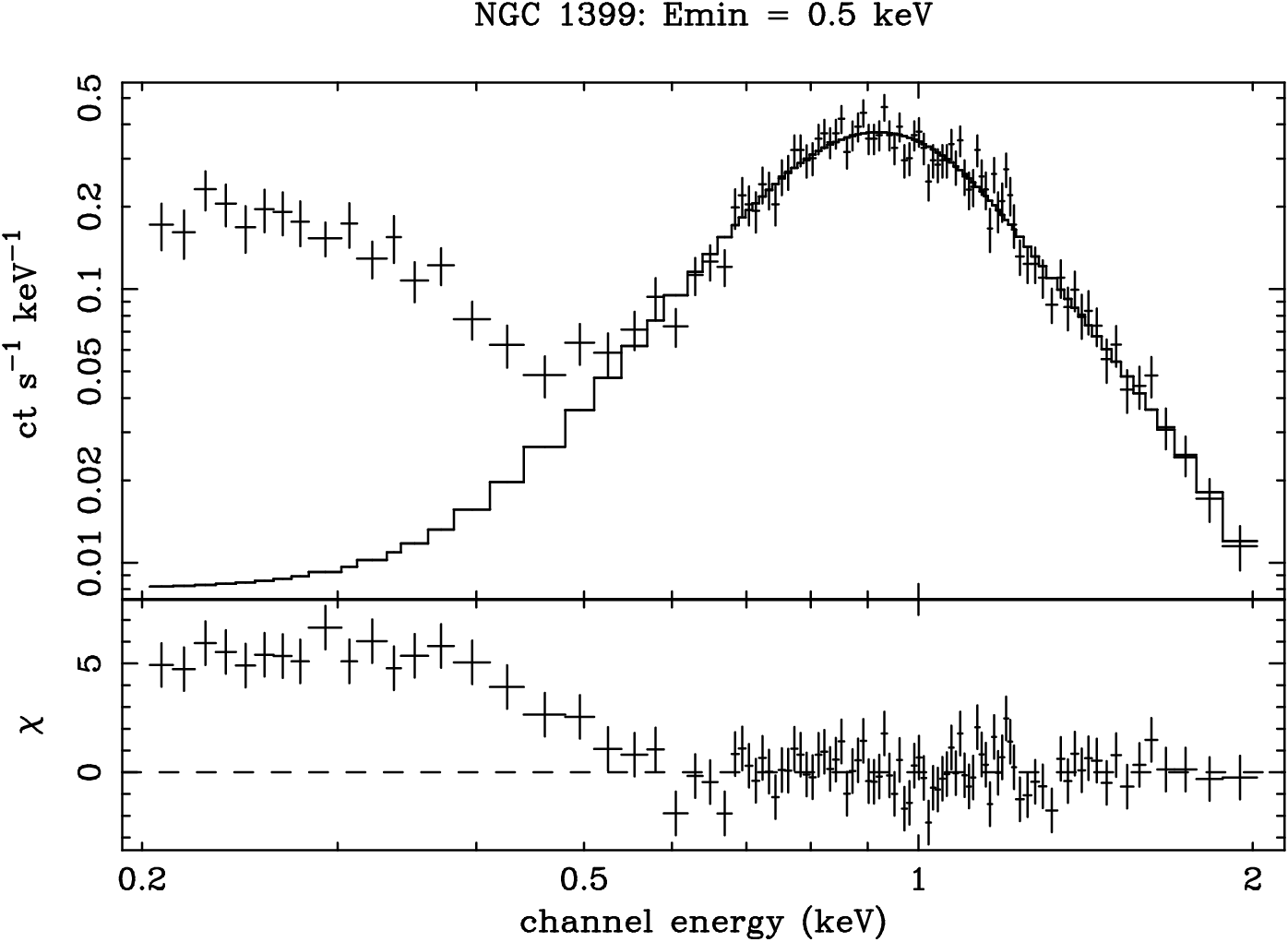,angle=-90,height=0.26\textheight}}}
\caption{\label{fig.emin} (Left) \rosat\, PSPC spectrum within
$R=1\arcmin$ of NGC 1399. Also shown is the best-fitting single-phase
plasma model modified by the standard absorber with $\nh =
\nhgal$. (Right) The best fitting model with variable \nh\, when only
energies above 0.5 keV are included in the fit. In this case the
fitted column density is $\nh\sim 15\nhgal$; i.e., when $\emin = 0.5$
keV the standard absorber model predicts that there should be negligible
emission at lower energies in conflict with the observation.}
\end{figure*}

Thus far we have shown (and confirmed previous results) that the
\rosat PSPC does not indicate the presence of excess absorbing
material arising from cold gas intrinsic to the galaxy or group.  In
Figure \ref{fig.emin} (Left) we display the \rosat PSPC spectrum of
NGC 1399 within the central arcminute along with the best-fitting
plasma model modified by a standard absorber with $\nh = \nhgal$. The
simple model provides a good visual fit to these data as well as a
formally acceptable fit ($\chi^2=107$ for 97 dof and $P=0.23$).

However, the evidence for intrinsic soft X-ray absorption from cold
gas in galaxies, groups, and clusters from the \einstein SSS and the
\asca SIS is obtained with data restricted to energies above $\sim
0.5$ keV because lower energies reside outside the bandpasses of those
instruments. Let us examine what happens to the spectral fits of the
PSPC data of NGC 1399 within the central arcminute if instead we raise
the lower energy limit of the bandpass, \emin, to a value near 0.5 keV
comparable to \asca and \einstein, and we allow \nh\, to be a free
parameter.

We find that when $\emin\approx 0.2-0.3$ keV we obtain fits
essentially as indicated in Figure \ref{fig.emin} (Left) with
$\nh\approx \nhgal$. We see a noticeable change near $\emin=0.4$ keV
when the fitted column density increases to $\nh\approx 2\nhgal$. A
dramatic increase occurs near $\emin=0.5$ keV which we show in Figure
\ref{fig.emin} (Right). The best-fitting model gives $\nh\approx
15\nhgal$ when data below 0.5 keV are excluded from the fits: the
standard absorber model when $\emin=0.5$ keV predicts that there
should be negligible emission at lower energies in clear conflict with
the data for $E\sim 0.2-0.3$ keV. The large column density obtained
when $\emin=0.5$ keV is similar to the large value inferred from \asca
with two-temperature or cooling flow models (see \S 5 in Buote 1999).

For larger values of \emin\, we find that \nh\, does not change
significantly within the uncertainties. The statistical uncertainties
on the fitted spectral parameters increase with increasing \emin\,
since the degrees of freedom decrease as the lower energy data are
ignored.

In the middle panels of Figures \ref{fig.7ann}-\ref{fig.3ann} we plot
$\nh(R)$ for $E_{\rm min}=0.5$ keV.  The character of the \nh\,
profiles for $E_{\rm min}=0.5$ keV is entirely different from the
previous $E_{\rm min}=0.2$ keV case for half of the sample: NGC 507,
1399, 4472, 4649, and 5044. In these systems $\nh(R)$ for $E_{\rm
min}=0.5$ is consistent with the Galactic values in the outermost
annuli and {\it increases} as $R$ decreases until (except for NGC
4472) it reaches a value consistent with a maximum for $R\sim
1\arcmin$. For the other five galaxies the constraints are too poor to
discern a trend, though within the large errors their profiles are
consistent with increasing as $R\rightarrow 0$.

The excess column densities inferred when $E_{\rm min}=0.5$ keV are
most significant for NGC 1399. The lower limits within $R=1\arcmin$
are $8\times 10^{20}$ cm$^{-2}$ and $3\times 10^{20}$ cm$^{-2}$ at
95\% and 99\% confidence respectively which are factors of $\approx 6$
and 3 larger than the Galactic value. Also within $R=1\arcmin$ for NGC
507 we obtain 95\%/99\% lower limits of $8/6\times 10^{20}$ cm$^{-2}$
compared to the adopted Galactic value of $5.2\times 10^{20}$
cm$^{-2}$. The most significant measurement for NGC 5044 is within the
$R=1\arcmin - 2\arcmin$ annulus where we find that $\nh>\nhgal$ at the
95\% confidence level.

In order to obtain measurements of the intrinsic absorption at a
higher significance level we must include the data below 0.5 keV. This
requires a more appropriate model of the intrinsic absorption that
does not conflict with the emission at lower energies which we now
consider.

\subsection{Intrinsic Oxygen Edge}
\label{edge}

Since we find that $\nh(E_{\rm min})\approx constant$ for $E_{\rm
min}\ga 0.5$ keV, the portion of the spectrum responsible for the
excess absorption must be near 0.5 keV. Considering the PSPC
resolution [$\Delta E/E = 0.43 (E/0.93 \rm keV)^{-0.5}$] and effective
area this translates to energies $\sim 0.4-0.7 $ keV. The dominant
spectral features in both absorption and emission over this energy
range are due to oxygen, though ionized carbon and nitrogen can
contribute as well (see \S \ref{temp}). 

Since the PSPC data cannot distinguish between a single edge and
multiple edges, we parameterize the intrinsic absorption with a single
absorption edge at 0.532 keV (rest frame) corresponding to cold atomic
oxygen (\oi). In the right columns of Figures
\ref{fig.7ann}-\ref{fig.3ann} we plot the optical depth profiles,
$\tau(R)$, for the \oi\, edge obtained from fits with $\emin =0.2$
keV.

For every system we find that the shape of $\tau(R)$ is very similar
to that of $\nh(R)$ for $E_{\rm min}=0.5$ for the standard
absorber. Moreover, for those systems where $\nh\gg\nhgal$ for $E_{\rm
min}=0.5$ we find that $\tau(R)\la 0.1$ in the outermost annuli and
increases to $\tau(R)\sim 1$ in the central bin. Therefore, the single
oxygen edge reproduces all of the excess absorption indicated by
$\nh(R)$ for $E_{\rm min}=0.5$ for the standard absorber model.

The fits within the central radial bin are clearly improved for
several systems when the single oxygen edge is added, even though in
some cases the quality of the fit is already judged to be formally
acceptable (null hypothesis $P\gtrsim 0.1$) without the edge.  The
improvement in the $\chi^2$ fit is quantified by the F Test (e.g.,
Bevington 1969), and in Table \ref{tab.fits} for each galaxy and group
we give the F-Test probability, $P(F)$, which compares the fits with
and without the edge; i.e., $P(F)\ll 1$ indicates the edge improves
the fit significantly.

Concentrating on models with the standard absorber with $\nh=\nhgal$
we see in Table \ref{tab.fits} the largest improvements exist for NGC
5044 ($P(F)=9.2\times 10^{-10}$), NGC 4472 ($P(F)=8.8\times 10^{-8}$),
and NGC 1399 ($P(F)=1.3\times 10^{-6}$).  A very significant
improvement is also found for NGC 4649 ($P(F)=3.2\times 10^{-3}$)
while marginal improvements are found for NGC 507 ($P(F)=2.7\times
10^{-2}$) and NGC 5846 ($P(F)=7.6\times 10^{-2}$).  Although only NGC
4472, 5044, and 5846 are indicated to have formally unacceptable fits
in terms of the $\chi^2$ null hypothesis probability for models with
or without the edge, the fact that adding the edge lowers $\chi^2$ by
much more than 1 (i.e., $P(F)\ll 1$) in several cases indicates that
the fits are indeed better with the edge in those systems.

For every system where we found $\nh<\nhgal$ for the standard absorber
(\S \ref{solar}), we find that \nh\, systematically increases when the
oxygen edge is added.  Although the addition of the edge results in
$\nh\sim\nhgal$ for many of these systems we find that in some cases
(most notably NGC 5044) \nh\, is still significantly less than \nhgal,
and the fits in the central bin, though improved, are still formally
unacceptable. Thus, adding the edge does significantly improve the
models, but it apparently is not the only improvement required in some
cases. We discuss another mechanism to improve the fits below in \S
\ref{xray}.

We obtain the best constraints on the edge optical depth for NGC 507,
1399, 4472, 4649, and 5044. Only for these systems is $\tau >0$
significant at the 95\% confidence level in the central bin whether or
not \nh\, is fixed to the Galactic value for the standard absorber
model. In some cases $\tau >0$ is significant at the 99\% level -- we
discuss individual cases below in \S \ref{comments}.

We have also investigated whether these oxygen absorption profiles can
be reproduced by a profile of decreasing oxygen abundance; i.e., the
strong K$\alpha$ lines of \ion{O}{7} and \ion{O}{8} lie between
0.5-0.65 keV. If instead we allow the oxygen abundance in the hot gas
to be a free parameter in the fits we find that typically the
best-fitting oxygen abundance is zero, and the quality of most of the
fits are improved similarly to that found when the oxygen edge is
added. This degeneracy is not surprising owing to the limited energy
resolution of the PSPC.  The notable exception is NGC 1399 where the
fits for $r=1\arcmin$ are only improved to $\chi^2=95.9$ for zero
oxygen abundance as opposed to 83.7 for the edge (both have variable
\nh). However, zero oxygen abundance in the centers is highly unlikely
because of the expected enrichment from the stars in the central
galaxy. And if instead we consider plausible O/Fe ratios to be at
least 1/2 solar in NGC 507, 1399, 4472, 4649, and 5044, then the fits
are not as much improved as when adding the edge.

As mentioned above (and in PAPER1) we find that for most radii the
constraints on the edge energy are not very precise which is why we
fixed the edge energy in our analysis. The best constraints are
available for NGC 1399 and 5044 within the central bin for which we
obtain $0.51^{+0.05}_{-0.05}$ keV and $0.51^{+0.09}_{-0.05}$ keV (90\%
confidence) for the edge energies for the standard absorber models
with variable \nh. (Models with fixed foreground Galactic columns give
similar results; e.g., $0.53^{+0.05}_{-0.03}$ keV for NGC 1399.)

These constraints are consistent with the lower ionization states of
oxygen but not edges from the highest states {\small O\thinspace
\footnotesize\sc\romannumeral 6-\footnotesize\sc\romannumeral 8}.  Due
to the limited resolution we can add additional edges to ``share'' the
$\tau$ obtained for the \oi\, edges, although when using a two-edge
model a significant $\tau$ cannot be obtained for edge energies above
$\sim 0.65$ keV corresponding to $\sim$\ovi.

\subsection{Comments on Individual Systems}
\label{comments}

We elaborate further on the results for individual systems.  When
comparing \nh\, profiles to those obtained from previous \rosat PSPC
studies we implicitly account for any differences in the plasma codes
and solar abundances used.

\subsubsection{Systems with 7 Annuli}

In Figure \ref{fig.7ann} we display the results for the three systems
where the spectral parameters are well determined in seven
annuli. These observations thus generally correspond to the highest
S/N data in our sample.  The evidence for intrinsic oxygen absorption
is strongest for these systems.

\medskip
\noindent {\bf NGC 507:} The oxygen edge optical depth, $\tau(R)$,
falls gradually from the center and remains significantly non-zero out
to $R\sim 4\arcmin$; e.g., in the $R=3\arcmin$-$4.25\arcmin$ annulus
the 95\% confidence lower limits on $\tau$ are 0.39 and 0.24
respectively in models with fixed (Galactic) and variable \nh\, for
the standard absorber. The values of \nh\, for the standard absorber
are consistent with those obtained by \citet{kf} with the PSPC data.

\noindent {\bf NGC 1399:} This system exhibits the most centrally
peaked $\tau$ profile in our sample. Within the central arcminute
$\tau>0.26$ and 0.33 at 99\% confidence respectively for the fixed and
free \nh\, models. At larger radii the non-zero optical depths are
also quite significant; e.g., for $R=2.5\arcmin$-$4\arcmin$
$\tau>0.64$ and 0.50 at 99\% confidence for the fixed and free \nh\,
models. We obtain values of \nh\, for the standard absorber consistent
with the previous PSPC study by \citet{j97}.

\noindent {\bf NGC 5044:} The second radial bin
($R=1\arcmin$-$2\arcmin$) actually has a smaller uncertainty on the
oxygen edge optical depth than the central bin; i.e., the 95\% lower
limits on $\tau$ are 0.87 and 0.79 respectively for the fixed and free
\nh\, models. In fact, the corresponding 99\% lower limits are 0.75
and 0.42 for $R=1\arcmin$-$2\arcmin$ which are larger than the values
for the $R=1\arcmin$ bin. The optical depth remains significantly
non-zero out to the $R=3\arcmin$-$4.5\arcmin$ bin in which we obtain
95\% lower limits on $\tau$ of 0.36 and 0.10 for the fixed and free
\nh\, models. Our values of \nh\, are consistent with those obtained
from the PSPC data by \citet{djfd}.

\subsubsection{Systems with 5-6 Annuli}

In Figure \ref{fig.6ann} we display the results for the three systems
where the spectral parameters are well determined in 5-6 annuli. Only
for NGC 4472 is the intrinsic oxygen absorption clearly significant.

\medskip
\noindent {\bf NGC 2563:} The uncertainties on $\tau(R)$ are large and
consistent with zero at the center. However, the shapes of the error
regions, especially the large uncertainty at the center, are
consistent with the same type of increasing $\tau$ profile with
decreasing $R$ found for the systems with seven annuli.

\noindent {\bf NGC 4472:} The optical depth is consistent with
$\tau\sim 2$ for $R\la 5\arcmin$ and then decreases rapidly at larger
radii. Unlike the other systems with evidence for intrinsic absorption
$\tau$ is most significant away from the central bin; i.e., for
$R=3.25\arcmin$-$5\arcmin$ the 95\% confidence lower limits on $\tau$
are 1.62 and 1.75 respectively for the fixed and free \nh\,
models. (The corresponding 99\% lower limits are 1.27 and 1.46.)
Interestingly, $R\sim 5\arcmin$ corresponds to the region where
\citet{is} identified holes in the X-ray emission from visual
examination of the PSPC image, and thus the large oxygen edge optical
depths could be related to these holes. The sub-Galactic columns
obtained for the standard absorber with variable \nh\, are consistent
with those obtained by \citet{forman}.

\noindent {\bf NGC 5846:} The uncertainties on $\tau(R)$ are large,
and although a substantial amount of intrinsic absorption is allowed
by the data, no excess absorption is required.

\subsubsection{Systems with 4 Annuli}

In Figure \ref{fig.4ann} we display the results for the three systems
where the spectral parameters are well determined in 4 annuli. The
uncertainties on $\tau(R)$ are large for each of these systems and
thus no intrinsic oxygen absorption is clearly required by the data.

\subsubsection{Systems with 3 Annuli}

In Figure \ref{fig.3ann} we display the results for the one system
where the spectral parameters are well determined in only 3 annuli,
NGC 4649. In contrast to the systems with 4 annuli we find evidence
for significant intrinsic oxygen absorption in the central bin for NGC
4649.

\medskip
\noindent {\bf NGC 4649:} In the central radial bin the 90\% lower
limits on $\tau$ are 0.55 and 0.20 respectively for for the fixed and
free \nh\, models, although only the 95\% lower limit for the model
with fixed \nh\, is significantly larger than zero (Table
\ref{tab.fits}). Our values of \nh\, for the standard absorber are
consistent with those obtained from the PSPC data by \citet{tfk}.

\subsection{Multiphase Models}
\label{multi}

\subsubsection{Simple Cooling Flow}
\label{cf}

In the inhomogeneous cooling flow scenario the hot gas is expected to
emit over a continuous range of temperatures in regions where the
cooling time is less than the age of the system.  We consider a simple
model of a cooling flow where the hot gas cools at constant pressure
from some upper temperature, $T_{\rm max}$ (e.g., Johnstone et al
1992).  The differential emission measure of the cooling gas is
proportional to $\dot{M}/\Lambda(T)$, where $\dot{M}$ is the mass
deposition rate of gas cooling out of the flow, and $\Lambda(T)$ is
the cooling function of the gas (in our case, the MEKAL plasma code).

Since the gas is assumed to be cooling from some upper temperature
$T_{\rm max}$, the cooling flow model requires that there be a
reservoir of hot gas emitting at temperature $T_{\rm max}$ but is not
cooling out of the flow. Consequently, our cooling flow model actually
consists of two components, CF+1T, where ``CF'' is the emission from
the cooling gas and ``1T'' is emission from the hot ambient gas. We
set $T_{\rm max}$ of the CF component equal to the temperature of the
1T component, and both components are modified by the same
photoelectric absorption.  This simple model of a cooling flow is
appropriate for the low energy resolution of the \rosat PSPC and has
the advantage of being well studied, relatively easy to compute, and a
good fit to the \asca data of many elliptical galaxies and groups
(e.g., Buote \& Fabian 1998; Buote 1999, 2000a).

When fitting this cooling flow model to the \rosat spectra of the
galaxies and groups in our sample we find that if only absorption from
the standard cold absorber model with solar abundances is included
then we obtain results identical to the single-phase models; i.e.,
the CF component is clearly suppressed by the fits.  Since the CF
model includes temperature components below $\sim 0.5$ keV it has
stronger \ion{O}{7} and \ion{O}{8} lines than the single-phase models
which, as shown above, already predict too much emission from these
lines. It is thus not surprising that only when an intrinsic oxygen
absorption edge is included in the fits can we obtain a significant
contribution from a CF component.

Even when adding the oxygen edge the cooling flow models do not
improve the fits perceptively in any case. And since the magnitude of
$\tau$ for the oxygen edge is degenerate with $\mdot$ of the cooling
flow component, the constraints on both parameters are quite
uncertain.  For NGC 5044, which has the best constraints, adding the
CF component improves $\chi^2$ from 156.5 to 154.1 for 116 dof within
the central arcminute (i.e., marginal improvement). The oxygen edge
optical depths are consistent with those obtained from the
single-phase analysis.  Only within $r=2\arcmin$ is there an
indication of significant cooling with $\mdot\sim 12\msun \rm yr^{-1}$
which is very consistent with the \rosat results of \citet{djfd}
within the same radius. (Our mass deposition rates have large
statistical uncertainties because of the inclusion of the intrinsic
oxygen edge.)

Thus, simple cooling flow models give results that are entirely
consistent with the single-phase models.

\subsubsection{Two Hot Phases}

A two-temperature model (2T) is a more flexible multiphase emission
model than the constant-pressure cooling flow and can very accurately
mimic a cooling flow spectrum over typical X-ray energies (Buote,
Canizares, \& Fabian 1999). If we restrict the temperatures of the 2T
model to lie between $\sim 0.5-2$ keV appropriate for hot gas near the
virial temperatures of these galaxies and groups, then we obtain
results equivalent to the cooling flow models above; e.g., (1) the
extra temperature component is only allowed if the intrinsic oxygen
edge included, and (2) results are entirely consistent with the
single-phase models. Since the temperatures of the 2T models are
fitted separately, constraints on the 2T models are even poorer than
the cooling flows because of the extra free parameter.

\subsubsection{Two-Phase Medium: Warm and Hot Gas}
\label{xray}

Recall that when fitted with only a standard absorber with solar
abundances that $\nh(R)$ {\it decreases} as $R$ decreases such that
\nh\, is {\it less} then \nhgal\, near $R=0$ for most of the galaxies
and groups (\S \ref{solar}). This trend implies the existence of
excess soft X-ray emission above that produced by the hot gas, the
signature of which is most evident at the centers of these galaxies
and groups.  Although the sub-Galactic values of \nh\, are only
marginally significant in many cases, they are highly significant for
NGC 4472 and NGC 5044.

If the excess soft X-rays near the centers represent emission from
coronal gas, then the temperatures must be $\la 0.1$ keV (i.e.,
distinctly cooler than the hot gas phase). For NGC 5044, which has the
most significant sub-Galactic column densities in the central bins, we
find that when adding an extra temperature component to the
single-phase model modified only by the standard absorber with
$\nh=\nhgal$ (i.e., no intrinsic oxygen edge), that the fits are
improved very similarly to the single-phase models when \nh\, is
allowed to take a value significantly less than \nhgal.  For example,
in the central bin the fit improves from $\chi^2=191$ (118 dof) for
the 1T model to $\chi^2=133$ (116 dof) for the 2T model (both models
with $\nh=\nhgal$) very similar to the result obtained for the 1T
model with $\nh < \nhgal$ (i.e., $\chi^2=139$ for 117 dof -- ``Free''
in Table \ref{tab.fits}). Essentially the same large improvement is
found for the second radial bin ($R=1\arcmin-2\arcmin$). Significant
but smaller improvement in the fits ($\Delta\chi^2\sim 10$) is also
seen in in the third and fourth bins (i.e. $R=2\arcmin-4.5\arcmin$).

The inferred temperature of the soft component in the central bin is
$T=0.05^{+0.03}_{-0.03}$ keV ($6^{+4}_{-4}\times 10^5$ K) at 90\%
confidence consistent with values obtained for other radii; i.e.,
consistent with ``warm'' gas rather than ``hot'' gas near the virial
temperatures of the galaxies and groups.  Gas at these warm
temperatures is not optically thin to photons with energies near 0.5
keV (e.g., Krolik \& Kallman 1984), and thus this warm gas may be
responsible for the intrinsic oxygen absorption inferred in \S
\ref{edge}.  However, since this warm gas is apparently partially
photoionized by the hot gas (and perhaps by itself) a proper
calculation of the emission from the warm gas must consider radiative
transfer effects which is beyond the scope of this paper. (We shall
continue to use the optically thin models in this paper.) We discuss
the properties of this warm gas in more detail in \S \ref{warm}.

The improvement obtained in the fit when adding a component of warm
gas to the single-phase model ($\nh=\nhgal$) is somewhat larger than
that obtained when adding the single oxygen edge (i.e., $\chi^2=133$
for 116 dof for the 2T model versus $\chi^2=156.5$ for 117 dof for the
edge -- Table \ref{tab.fits}).  Unfortunately, due to the limited
energy resolution of the PSPC it is very difficult to obtain
simultaneous constraints on both absorption and emission models for
both the warm and hot gas.  

The most reliable constraints on the warm gas component are possible
for NGC 5044 because for 1T models we find that \nh\, of the standard
absorber is well below \nhgal\, at small $R$ whether or not an
intrinsic oxygen edge is included (Figure \ref{fig.7ann}).  For a 2T
model with an intrinsic oxygen edge we find that the temperature of
the warm component in the central bin is $T=0.06^{+0.03}_{-0.04}$ keV
($7^{+4}_{-5}\times 10^5$ K) at 90\% confidence consistent with that
obtained for the 2T model without the oxygen edge. The oxygen edge
optical depth is not very well constrained, $\tau=0.8^{+0.4}_{-0.6},
1.2^{+0.8}_{-0.7}$ (90\% confidence) in the inner two bins
respectively, which is about half the best-fitting value and near the
95\% lower limit of $\tau$ obtained without including the emission
from the warm gas (Table \ref{tab.fits}).  The weakened constraint on
the oxygen edge optical depth for the 2T model of NGC 5044 reflects
the relatively small improvement in the fit ($\chi^2$ of 127 for 115
dof) over the 2T model without an edge.  For the other galaxies, most
notably NGC 1399, the constraints on the oxygen edge are not weakened
nearly as much.

Emission from such a warm gas component is consistent with, though not
as clearly required by, the PSPC data for the other systems. Only for
NGC 4472 is clearly significant improvement found when adding both the
warm gas component and the oxygen edge. Temperatures ($T\sim
(5-10)\times 10^5$ K) and the reduction in edge optical depth are
obtained similarly as found for NGC 5044.  Future observations with
better energy resolution are required to definitively confirm and
measure the emission and absorption properties of the warm gas in all
of these systems.

\subsection{Caveats}
\label{caveats}

\noindent {\it (i) Calibration:} The gain of the PSPC is well
calibrated and in particular no significant calibration problems near
0.5 keV have been reported.\footnote{See
http://heasarc.gsfc.nasa.gov/docs/rosat.} The large values of
$\tau\sim 1$ obtained in the central regions imply that the absorption
significantly affects a large energy range comparable to the energy
resolution of the PSPC; e.g., the \oi\, edge absorbs 25\% of the flux
at 0.8 keV for $\tau=1$; i.e. possible residual calibration errors
near 0.5 keV where the effective area is changing rapidly (e.g.,
Figure 1 of Snowden et al 1994) can not explain the need for
absorption at higher energies. Moreover, the shape of $\tau(R)$ is not
the same for each system as would be expected if calibration were
responsible for the intrinsic oxygen absorption found in half of the
sample; e.g., $\tau(R)$ for NGC 1399 is much more centrally peaked
than for NGC 4472 or the other systems.  The agreement between the
absorption inferred by the PSPC and \asca mentioned below in \S
\ref{asca} further argues against a systematic error intrinsic to the
PSPC being responsible for the measured oxygen absorption.

\noindent {\it (ii) Galactic Columns:} All of the objects in our
sample reside at high Galactic latitude, and thus the Galactic
absorption should be fairly uniform over the relevant
$5\arcmin$-$10\arcmin$ scales. Errors in the assumed Galactic columns
\citep{dl} should only affect the baseline value and not the variation
with radius.

\noindent {\it (iii) Background:} Errors in the background level
affect most seriously the lowest energies ($\la 0.4$ keV) which are
also most sensitive to the column density of the standard absorber
model. Hence, measurements of \nh\, at the largest radii (which have
lowest S/N) are the most affected by background errors.  The hydrogen
column densities measured in the outer radii (see Figures
\ref{fig.7ann}-\ref{fig.3ann}) are very similar to and are usually
consistent with the assumed Galactic values within the estimated
$1\sigma$ errors which attests to the accuracy of our background
estimates. The intrinsic oxygen optical depths in the central radial
bins are very insensitive to the background level.

\section{Comparison to \asca}
\label{asca}

The intrinsic oxygen absorption indicated by the PSPC data in half of
our sample is most significant within the central $1\arcmin-2\arcmin$
which is similar to the width of the \asca PSF. In addition, since the
\asca SIS is limited to $E>0.5$ keV, and the efficiency near 0.5 keV
is severely limited due to instrumental oxygen absorption, it cannot
be expected that \asca can distinguish an oxygen edge from a standard
absorber with solar abundances. However, it is instructive to examine
the consistency between results obtained from spatially resolved
(PSPC) and single-aperture (\asca) methods.  As mentioned in PAPER1
the results obtained when adding an oxygen edge to the \asca data of
NGC 1399 and 5044 are consistent with those obtained with the
PSPC. Since similar results are found for other objects in our sample
showing intrinsic absorption, we focus on NGC 1399 for illustration.

Previously we \citep{b99} have fitted two-temperature models to the
accumulated {\sl ASCA} SIS and GIS data within $R\approx 5\arcmin$ of
NGC 1399 and obtained $N_{\rm H}^{\rm c}=49_{-9}^{+6}\times 10^{20}$
cm$^{-2}$ (90\% confidence) for the standard absorber on the cooler
temperature component.  Using the meteoritic solar abundances (see \S
4.1.3 of PAPER2) slightly modifies this result to $N_{\rm H}^{\rm
c}=40_{-7}^{+6}\times 10^{20}$ cm$^{-2}$ which is about 30 times
larger than the Galactic value ($N_{\rm H}^{\rm c}=1.3\times 10^{20}$
cm$^{-2}$). Comparing this result to those obtained from the PSPC for
$E_{\rm min}=0.5$ keV (Fig \ref{fig.7ann}) we see that the \asca
column density is qualitatively similar to the value of $N_{\rm
H}\approx 20\times 10^{20}$ obtained within $R=1\arcmin$ and is
consistent with the total column within $R\approx 5\arcmin$.  If
instead the columns of the standard absorber are fixed to Galactic on
both components, and an intrinsic \oi\, edge is added to the cooler
component, then we obtain $\tau=6.0^{+0.7}_{-0.7}$ (90\% confidence)
for the \asca data.  Again this \asca result is similar to $\tau\sim
4$ in the central PSPC bin and is very consistent with the value of
$\tau=5.7$ obtained from adding up the best-fitting values obtained
with the PSPC within $R=5\arcmin$.  Therefore, the oxygen edge
provides as good or better description of the excess absorption
inferred from multitemperature models of {\sl ASCA} data within the
central few arcminutes as a standard cold absorber with solar
abundances, and yields optical depths that are consistent with those
obtained with the PSPC data.

Although a similar consistency is achieved for NGC 4472 the
interpretation of the absorption of the two-component model of the
single-aperture \asca data is not so straightforward because the
absorption is not obviously concentrated at the center (i.e. on the
cooler temperature component). In fact, a spatially uniform absorber
within $R=5\arcmin$ is probably a better description of the PSPC data.
If the column densities on both temperature components are tied
together for the two-temperature model of the \asca data the quality
of the fit is slightly worse ($\Delta\chi^2=8$ for 361 dof), but
$\nh\approx 11\times 10^{20}$ cm$^{-2}$ which is a fair representation
of the average \nh\, profile obtained from the PSPC for $E_{\rm
min}=0.5$ keV (Fig \ref{fig.6ann}). Similar agreement is found for the
\oi\, edge when applied to both temperature components.

\section{Warm Ionized Gas in Cooling Flows}
\label{warm}

We consider now in some detail the properties of the intrinsic
absorber and their implications. Initially we focus our attention on
the physical state of the absorber.

\subsection{Why Not Dust?}
\label{dust}

Models of dust grains indicate that dust can give rise to significant
amounts of absorption between 0.1-1 keV (e.g., Laor \& Draine
1993). In principle such grains could explain the intrinsic absorption
we have inferred for energies above $\sim 0.5$ keV in the PSPC data of
half of the galaxies and groups in our study as well as the excess
absorption detected for energies above $\sim 0.5$ keV in the \asca
data of bright galaxies, groups, and clusters. 

However, dust should also heavily absorb X-rays with energies between
$0.1-0.4$ keV (e.g., Laor \& Draine 1993) which is inconsistent with
the \rosat data for the 10 galaxies and groups in our study and, e.g.,
the large sample of \rosat clusters studied by \citet{af}.  To evade
this constraint (i.e., no absorption below 0.4 keV) an unconventional
model for the formation of dust grains has to be postulated; e.g.,
dust grains condense out of a medium in which helium remains ionized
\citep{am}.  Even if we consider the (rather unlikely) possibility
that dust does not absorb X-rays below $\sim 0.4$ keV, we still have
that dust cannot account for the excess X-ray emission at those
energies implied by the sub-Galactic column densities (\S \ref{solar}
and \ref{xray}).

Other strong arguments against the existence of large quantities of
oxygen-absorbing dust in cooling flows have been made in the past.  At
the centers of cooling flows where the gas density is largest and the
soft X-ray absorption is most significant dust should not be present
in large quantities because the grains are rapidly destroyed by
sputtering by the hot gas (e.g., Tsai \& Mathews 1995; Voit \& Donahue
1995).  A different argument due to \citet{vd} considers that the
transient heating of the grains by X-rays from the hot gas prevents CO
from fully condensing onto dust grains. Consequently, significant
amounts of oxygen would remain in molecular gas which would produce CO
emission from rotational lines that are inconsistent with the
generally negligible CO detections in cooling flows (e.g., Bregman,
Hogg, \& Roberts 1992; O'Dea et al 1994).

\subsection{Temperature of the Warm Ionized Gas}
\label{temp}

Unlike dust all of the key features of the observed soft X-ray
absorption and emission can be accounted for if the absorber is
(primarily) collisionally ionized gas. The lack of intrinsic
absorption observed for energies between $\sim 0.2-0.4$ keV in the
\rosat data requires that H and He be completely ionized implying a
gas temperature $T\ga 1.0\times 10^5$ K (e.g., Sutherland \& Dopita
1993). In order to have significant absorption at 0.5 keV the
temperature cannot be larger than $\approx 1\times 10^6$ K (see Figure
2 of Krolik \& Kallman 1984). This {\it absorbing} temperature range
($T=10^{5-6}$ K) is entirely consistent with that inferred from the
gas {\it emission} to explain the sub-Galactic column densities
especially for NGC 5044 (\S \ref{xray}). This consistency of
absorption and emission properties lends strong support to the idea
that the absorber is warm ionized gas.

At these temperatures carbon and nitrogen are not completely ionized
(e.g., Nahar \& Pradhan 1997; Sutherland \& Dopita 1993), and thus
these elements will also contribute to the soft X-ray absorption. For
nitrogen the states {\small N\thinspace \footnotesize\sc\romannumeral
4-\footnotesize\sc\romannumeral 6} are significant, and their edge
energies span 0.46-0.55 keV (rest frame) which are consistent with the
edge energy range determined from the PSPC data. Since $\rm N/O =
1/8.51$ (assuming solar abundance ratios), and the threshold cross
sections for absorption are similar for N and O, only $\sim 12\%$ of
the optical depth we have measured in each system likely arises from
ionized nitrogen.

The ionization fraction for carbon changes rapidly near $T=10^5$ K with
\cv\, dominating for temperatures above this value and up to $T\approx
10^6$ K. The edge energy for \cv\, is 0.39 keV, and since $\rm C/O
\approx 0.5$ (assuming solar abundance ratios) and the threshold cross
sections of C and O are similar, the optical depth of \cv\, is about
half that expected from a dominant ionized state of oxygen. However,
the strong instrumental carbon absorption leaves the PSPC with
essentially no effective area over energies 0.28 keV to $\sim 0.4$ keV
(e.g., see Figure 1 of Snowden et al 1994), and thus it is only
possible to detect intrinsic absorption from the \cv\, edge for
energies above $\sim 0.4$ keV even considering the smearing due to the
limited energy resolution. This is entirely consistent with the
variation of \nh\, with $E_{\rm min}$ described in \S \ref{bandpass}.

Considering the energy resolution and effective area curve of the
PSPC, the 0.532 keV edge that we have used to parameterize the
intrinsic absorption is likely an average of the $\sim 40\%$
contribution from the \cv\, edge (0.39 keV) and {\small N\thinspace
\footnotesize\sc\romannumeral 4-\footnotesize\sc\romannumeral 6} edges
(0.46-0.55 keV) with a $\sim 60\%$ contribution from ionized oxygen
states. Although as discussed in \S \ref{edge} the PSPC data cannot
distinguish between multi-edge models, when using a more realistic
absorber model consisting of \cv\, and \nvi\, and an oxygen edge, we
are able to obtain a significant optical depth for the oxygen edge for
energies as high as $\sim 0.75$ keV consistent with \ovii\, (0.74
keV). However, to insure that oxygen produces at least as much
absorption as the C and N edges, the PSPC data also require a
contribution from edges around $\sim 0.6-0.65$ keV corresponding to
edges from {\small O\thinspace \footnotesize\sc\romannumeral
4-\footnotesize\sc\romannumeral 6}.

Consideration of these maximum allowed ionization states for oxygen
indicates that the maximum temperature of the warm gas is more like
$T\approx 5\times 10^5$ K if the gas is isothermal and collisionally
ionized (e.g., Nahar 1999; Sutherland \& Dopita 1993). The absorption
signature of this warm gas is not one dominant feature near 0.5 keV
but is rather a relatively broad trough over energies 0.4 to $\sim
0.8$ keV for total optical depths of unity (see Figure 1 of Krolik \&
Kallman 1984).

(We mention that the edge energies we have quoted are from Daltabuit
\& Cox (1972) though Gould \& Jung (1991) argue that the edge energy
for \oi\, is $\sim 10$ eV higher. Such differences may be relevant for
modeling future high resolution spectra but are unimportant for our
present discussion with the PSPC data.)

Hence, a proper absorber model needs to consider several edges from
different ionization states of oxygen as well as edges from ions of C
and N. Since the warm gas absorbs photons from the hot gas the
assumption of collisional ionization equilibrium is also not strictly
valid nor is it clear that the warm gas is fully optically thin to its
own radiation as was assumed in \S \ref{xray} for
convenience. Consequently, the single-edge oxygen absorber that we
have used throughout this paper (and PAPER1) has been intended
primarily as a phenomenological tool to establish the existence and to
study the gross properties of the absorber which is appropriate for
the low spectral resolution afforded by the PSPC data. If our basic
results are confirmed with the substantially higher quality data from
\chandra and \xmm, then it will be appropriate to expend the effort to
construct rigorous models of the warm absorber accounting for many
edges and possible radiative transfer effects that are not currently
available in \xspec.

\subsection{Absorber Masses vs Mass Drop-Out}
\label{mass}

\begin{table*}[t] \footnotesize
\begin{center}
\caption{Absorber Masses and Implied Accumulation Timescales\label{tab.mass}}
\begin{tabular}{lccccc|cccc} \tableline\tableline\\[-7pt]
& & \multicolumn{4}{c}{Central Bin} & \multicolumn{4}{c}{Total}\\
& \mdot & R & \mabs & \tacc & \mhot & R & \mabs & \tacc & \mhot\\
Name & (\msun yr$^{-1}$) & (kpc) & ($10^{10}\msun$) & ($10^{10}$ yr) &
($10^{10}\msun$) & (kpc) & ($10^{10}\msun$) & ($10^{10}$ yr) & ($10^{10}\msun$)\\ \tableline\\[-7pt] 
N507 & $18.50_{- 3.30}^{+ 3.50}$ &  20.0 & $ 5.86_{- 4.02}^{+ 5.85}$ & $ 0.32_{- 0.23}^{+ 0.45}$ & $ 0.430_{- 0.069}^{+ 0.221}$ & 200.0 & $125.67_{-92.61}^{+599.92}$ & $ 6.79_{- 5.29}^{+40.94}$ & $43.68_{- 9.41}^{+12.31}$\\
N1399 & $ 1.62_{- 0.18}^{+ 0.20}$ &   5.2 & $ 0.81_{- 0.62}^{+ 0.10}$ & $ 0.50_{- 0.39}^{+ 0.13}$ & $ 0.019_{- 0.001}^{+ 0.012}$ &  90.3 & $ 9.41_{- 7.10}^{+23.11}$ & $ 5.81_{- 4.54}^{+16.78}$ & $ 5.55_{- 0.63}^{+ 0.51}$\\ 
N4472 & $ 2.54_{- 0.37}^{+ 0.34}$ &   5.2 & $ 0.32_{- 0.16}^{+ 0.54}$ & $ 0.12_{- 0.07}^{+ 0.27}$ & $ 0.031_{- 0.018}^{+ 0.002}$ &  77.2 & $47.26_{-24.67}^{+39.53}$ & $18.61_{-10.76}^{+21.39}$ & $ 3.65_{- 0.33}^{+ 0.36}$\\ 
N4649 & $ 2.20_{- 0.60}^{+ 0.90}$ &   5.2 & $ 0.35_{- 0.26}^{+ 0.41}$ & $ 0.16_{- 0.13}^{+ 0.32}$ & $ 0.026_{- 0.014}^{+ 0.004}$ &  36.7 & $ 3.70_{- 3.61}^{+60.64}$ & $ 1.68_{- 1.65}^{+38.53}$ & $ 0.56_{- 0.25}^{+ 0.19}$\\
N5044 & $41.32_{- 5.42}^{+ 4.59}$ &  11.0 & $ 1.69_{- 1.13}^{+ 1.17}$ & $ 0.04_{- 0.03}^{+ 0.04}$ & $ 0.266_{- 0.109}^{+ 0.052}$ & 143.5 & $26.52_{-17.87}^{+38.17}$ & $ 0.64_{- 0.45}^{+ 1.16}$ & $22.08_{- 4.13}^{+ 2.91}$\\ \tableline
\end{tabular}
\tablecomments{Total cooling flow mass deposition rates (\mdot) and
90\% confidence limits inferred from \asca data (see text). The quoted
errors on \mabs\, refer to 95\% confidence limits on $\tau$ (``Fix''
in Table \ref{tab.fits}) and assume solar abundances. The accumulation
timescale, $\tacc = \mabs/\mdot$, reflects both the 95\% uncertainties
on \mabs\, and the 90\% errors on \mdot. The mass of hot gas, \mhot,
and the 95\% errors computed within the 3D radius are also
given. These values of \mabs\, probably are over-estimates (see
beginning of \S \ref{mass}).}
\end{center}
\end{table*}

As suggested in PAPER1 this warm ionized absorber might be the gas
that has dropped out of the putative cooling flow during the lifetime
of the galaxy or group and thus could provide the confirmation of the
inhomogeneous cooling flow scenario that has been suggested to operate
in massive elliptical galaxies, groups, and clusters (e.g., Fabian
1994). To make this connection between our measurements of oxygen
absorption and the cooling flow scenario we first estimate the mass of
absorbing material implied by the measured optical depths. This mass
is then compared to the cooling flow mass deposition rate inferred
from \asca data.

Before computing the absorber masses some caveats must be
discussed. First, although the simple constant-pressure cooling flow
model discussed in \S \ref{cf} can describe the X-ray emission of the
hot gas just as well as the single-phase model, it does not describe
the excess 0.2-0.4 keV X-rays presumably arising from the emission of
the warm absorbing gas (\S \ref{xray}). Although at this time we
cannot exclude the possibility that with a more rigorous treatment of
the absorption and emission properties of the warm gas (see end of
previous section) that the simple cooling flow model could be
compatible with the PSPC data between 0.2-0.4 keV, it is quite
possible that an important modification of the simple cooling flow
model is required (see \S \ref{theory}). As a convenient benchmark for
comparison to most previous studies we shall consider here the mass
deposition rates predicted by the constant-pressure cooling flow
models.

Second, the oxygen edge optical depths predicted by the single-edge
models without including the emission from the warm gas certainly are
overestimates. Recall that when including the warm gas emission in NGC
5044 that $\tau$ reduces to a value near the 95\% lower limit of the
result obtained without including the warm gas emission (\S
\ref{xray}).  Of equal importance, when including additional edges the
inferred optical depth for each edge decreases, and we expect several
edges from different ionization states of oxygen, carbon, and nitrogen
to contribute (\S \ref{temp}). Since the absorption of an edge is not
linear in the energy of the edge (i.e.,
$A_0(E)=\exp[-\tau_0(E/E_0)^{-3}]$ and $A_1(E)A_2(E) \ne A_{1+2}(E)$
if the edge energies $E_1 \ne E_2$), by spreading multiple edges over
a large energy range one can produce the observed absorption with
smaller total optical depth than can be achieved with a single edge.
Consequently, the single-edge optical depths obtained in \S \ref{edge}
should be considered upper limits.

Let us now estimate the amount of absorbing material implied by these
absorption measurements with these caveats in mind, and in particular
with the assumption that the inferred absorber masses are most likely
over-estimates. Assuming the optical depths refer to the \oi\, edge,
then the measured values of $\tau$ imply a hydrogen column density
(assuming $\sigma=5.5\times 10^{-19}$ cm$^{2}$ at threshold) and thus
a mass within a projected radius, $R$,
\begin{eqnarray}
\mabs(<R) = \hspace{5.7cm} \nonumber \\
(7.8\times 10^9) (\tau) \left( {R\over 10\rm kpc}\right)^2
\left( { \rm O/H \over 8.51\times 10^{-4}}\right)^{-1}\msun ,
\hspace{0.2cm} \label{eqn.mabs} 
\end{eqnarray}
where $\tau$ is the optical depth of the \oi\, edge and O/H is the
oxygen abundance of the absorber. Although the projected mass is
larger than the mass within the 3D radius $r=R$, the value of $\tau$
in equation (\ref{eqn.mabs}) slightly underestimates the 3D value as
discussed in (\S \ref{partial}); i.e. these projection effects
approximately cancel. The metallicity of the hot gas in the central
bins for the objects in our sample are larger than solar (PAPER2), and
thus we expect the same for the absorber. Since the oxygen abundance
is uncertain we shall quote results assuming O is solar and recognize
that \mabs\, could be overestimated by a factor of 2-3 in the central
bin. The expected contribution from carbon and nitrogen (\S
\ref{warm}) to $\tau$ also reduces \mabs\, by another $\sim 40\%$.

In Table \ref{tab.mass} we give \mabs\, for NGC 507, 1399, 4472, 4649,
and 5044 in both the central bin $(R=1\arcmin)$ and the total mass
interior to the largest bin investigated; the edge optical depths used
refer to the single-phase models since the cooling flow models give
entirely consistent values (\S \ref{cf}). The mass deposition rates,
\mdot, listed in the second column are determined from the accumulated
\asca data within radii of $r\sim 3\arcmin-5\arcmin$. (The \asca
spectra place much tighter constraints on the total \mdot\, than do
the \rosat spectral data.)  The results for NGC 1399, 4472, and 5044
are taken from \citet{b99}. For NGC 507 and 4649 we re-analyzed the
data sets as prepared in \citet{bf} and fitted cooling flow models
analogously to that done in \citet{b99}. That is, the spectra were
fitted with (1) a constant pressure cooling flow component, (2) an
isothermal component representing the ambient gas, and (3) for NGC
4649 an extra high-temperature bremsstrahlung component. Since the
cooling flow model assumes constant pressure and neglects the
gravitational work done on the cooling gas, the value of \mdot\, is an
upper limit. This over-estimate is typically $\la 30\%$ (e.g., the
agreement of different cooling flow models in Allen et al 2000b).

Let us focus on \mabs\, and the accumulation time,
$\tacc=\mabs/\mdot$, within the central bin $(R=1\arcmin)$ where the
measured optical depths are most significant and the fits are most
clearly improved when the edge is added; i.e. we consider the results
for the central bin to be most reliable. Assuming an age of the
universe, $\tage=1.3\times 10^{10}$yr, examination of Table
\ref{tab.mass} reveals that within the central bin $\tacc\sim
(0.1-0.5)\tage$ using the best-fitting values or a 95\% upper limit of
$\tacc\sim (0.3-0.6)\tage$; NGC 5044 actually has smaller values, but
if the second bin is included (which has intrinsic absorption just as
significant as the inner bin) then \tacc\, is consistent with the
values quoted above. These accumulation timescales are a sizeable
fraction of \tage, and thus \mabs\, within the central bin(s) can
account for most, if not all, of the mass deposited by the cooling
flow over the lifetime of the flow; the exact value depends on
precisely when the cooling flow begins and whether \mdot\, varies with
time.

The total absorbing masses have large errors within the 95\%
confidence limits. Although the best-fitting values for \tacc\, are
typically larger than \tage, the 95\% lower limits are $\la\tage$ for
all but NGC 4472. We reiterate that we consider these values at the
largest radius to be less secure than the central bin(s) because the
fits do not clearly require the addition of the oxygen edge outside
the inner 1-2 bins. Nevertheless, the result for NGC 4472 is striking
and deserves comment. Clearly the approximation of a spherically
symmetric, relaxed cooling flow is invalid for $R\ga 3\arcmin$ because
the isophotal distortions suggest a strong interaction with the
surrounding Virgo gas \citep{is} and thus the estimate of \mdot\,
unlikely applies at larger radii. If the large value of \mabs\, at
large radius is confirmed then another mechanism must have produced
the warm gas in NGC 4472.

Hence, within the central 1-2 bins ($R\sim 10-20$ kpc) where the model
constraints are most secure we conclude that the absorbing mass
inferred from the oxygen edges can explain most (and perhaps all) of
the mass deposited by a cooling flow over the age of the system. If
the edges also apply at larger radii (as we have assumed), then all of
the deposited mass (except for NGC 4472) can be easily explained by
the inferred absorbing mass. These qualitative conclusions still apply
if the oxygen edge optical depths are really closer to their 95\%
lower limits as discussed near the top of this section.

We mention that systems without strong cooling flows will not have had
sufficient time to accumulate the $\sim 10^9\msun$ within $R\sim
1\arcmin$ to produce detectable soft X-ray emission. Thus, the ``Very
Soft Components'' found in galaxies with low ratios of X-ray to
optical luminosity unlikely arise from warm gas deposited in a cooling
flow and instead probably reflect the collective emission from X-ray
binaries (e.g., Irwin \& Bregman 1999). 

\subsection{Constraints from the Optical and FUV}
\label{optical}

Since collisionally ionized gas at temperatures of $10^{5-6}$ K emits
many strong lines at optical and ultraviolet wavelengths (e.g.,
Pistinner \& Sarazin 1994; Voit \& Donahue 1995), we consider whether
the large amounts of absorbing material implied by the intrinsic X-ray
absorption (Table \ref{tab.mass}) violate published constraints on
line emission in the optical and UV spectral regions.  The best
published constraints available in the optical are for H$\alpha$ from
studies of extended ionized gas in the centers of elliptical galaxies
(e.g., Trinchieri \& di Alighieri 1991; Goudfrooij et al 1994;
Macchetto et al 1996). In most cases the emission line gas is only
detected within $r\la 20\arcsec$ which is significantly smaller than
the central $1\arcmin$ used in our analysis.

The object in our sample where H$\alpha$ has been detected out to the
largest angular radius is NGC 5044. \citet{macc96} measure $\rm
F(H\alpha)= 1.4\times 10^{-13}$ \ergcms within $R=0.5\arcmin$. We can
estimate the temperature at which the H$\alpha$ emission implied by
\mabs\, (Table \ref{tab.mass}) within $R=1.0\arcmin$ equals the
observed flux. We take the predicted H$\alpha$ line intensity at peak
temperature from Pistinner \& Sarazin and the temperature dependence
of $\sim T^{-2.5}$ from inspection of Figure 7 of Voit \&
Donahue. After accounting for the different region sizes we find that
the required temperature is $\approx 1.5\times 10^5$ K using the
best-fitting \mabs, although when using the 95\% lower limit on
\mabs\, we find that $T\approx 0.5\times 10^5$ K.

Similar results hold for NGC 1399, 4472, and 4649 although the
comparison is less certain because of the larger aperture
corrections. If no aperture correction is made for NGC 5044 then the
implied temperatures rise to $T\approx 2.5\times 10^5$ K at best fit
and $T\approx 1.7\times 10^5$ K at the 95\% lower limit. If we consider
also that \mabs\, in Table \ref{tab.mass} is over-estimated because
the oxygen abundance is larger than solar ($\sim 1.5\solar$ -- see
PAPER2) and carbon and nitrogen contribute $\sim 40\%$ to the measured
optical depths (\S \ref{mass}), we obtain $T\approx 1.8\times 10^5$ K
at best fit and $T\approx 1.2\times 10^5$ K at the 95\% lower limit
(again without aperture correction). Therefore, the published
constraints on H$\alpha$ are satisfied if $T\ga 2.0\times 10^5$ K.

Stronger lines from warm gas are expected to appear in the UV, but
Hopkins Ultraviolet Telescope (HUT) observations detected no
significant emission lines in NGC 1399, 4472, and 4649 (Ferguson et al
1991; Brown, Ferguson, \& Davidsen 1995). It is unfortunate that the
strongest emission lines for temperatures $T\sim (2-3)\times 10^5$ K
are \ov (1218\AA), \ovi (1034\AA), and \nv (1240\AA) which appear to
be lost in the background geocoronal emission (e.g., Figure 1 of
Ferguson et al 1991, though see below for \ion{O}{6}). However, the
lines \civ (1549\AA), \oiv (1401\AA), and \neiv (1602\AA) are also
strong and uncontaminated by geocoronal emission.

To determine the gas temperature at which, e.g., the \oiv (1401\AA)
flux would not violate the published UV constraints we estimate that
the \oiv\, flux would have to be less than $\sim 10\%$ of the
continuum considering the error bars on the spectrum of NGC 1399
(Figure 2 of Ferguson et al 1991). This limit corresponds to a flux of
$\sim 1.4\times 10^{-12}$ \ergcms. We take the predicted \oiv\, line
intensity at peak temperature from Pistinner \& Sarazin and assume the
temperature gradient above the peak falls similarly to that displayed
for \civ\, in Figure 7 of Voit \& Donahue. Using the best-fitting
\mabs\, (and accounting for the smaller HUT aperture) we find that a
temperature of at least $3\times 10^5$ K is required, though the 95\%
lower limits on \mabs (which are probably more realistic -- \S
\ref{xray}) coupled with the oxygen abundance and C/N issues as above
indicate the limit is more conservatively $\sim 2\times 10^5$
K. Similar limits are obtained for the other lines and for the HUT
spectra of NGC 4472 and 4649 \citep{bfd}.

Our procedure of requiring the line fluxes to be less than 10\% of the
continuum may result in limits that are too restrictive. \citet{dhf}
have estimated $2\sigma$ upper limits on the \ovi (1034\AA) intensity
from M87 which has a HUT spectrum very similar to NGC 1399. Their
$2\sigma$ upper limit on the \ovi\, flux within a $1\arcmin$ circle is
$\sim 1\times 10^{-10}$ \ergcms. If the warm gas has a temperature of
$3.2\times 10^5$ K corresponding to the peak temperature for \ovi,
then the observed limit is comparable to the \ovi\, emission expected
from the warm gas of NGC 1399 when using the 95\% lower limit on
\mabs. Hence, if the HUT results for M87 apply to NGC 1399 (as they
appear to), then the predicted \ovi\, emission agrees with the limits,
especially if the temperature is not precisely at the peak temperature
for \ovi.

\subsection{Theoretical Issues}
\label{theory}
 
Although the hypothesis of warm, mostly collisionally ionized, gas
apparently can explain the X-ray observations and the matter deposited
by the cooling flows, this model has serious theoretical difficulties
which must be overcome before it can be considered a viable model:

\begin{enumerate}
\item $T_{\rm warm} < T_{\rm virial}$. The temperature of the warm gas
($\la 0.1$ keV) is less than the virial temperatures of the halos
($\sim 1$ keV) implying the gas is not thermally supported. How does
the warm gas support itself in the gravitational fields of these
systems?
\item $t_{\rm cool}^{\rm warm\, gas} \ll t_{\rm cool}^{\rm hot\,
gas}$. The cooling time of the warm gas is very short, even more so
than the hot gas in the central regions. How are large quantities of
gas maintained at these temperatures?
\end{enumerate} 

Most likely these problems can only be solved with a substantial
modification of the standard cooling flow scenario. Clearly an
additional energy source and very possibly the important role of
magnetic fields will have to be considered. A small number of models
have been proposed throughout the years which consider these issues,
though they are not well developed and do not at present make detailed
predictions regarding the issues (1) and (2) above. These models have
generally been designed to completely suppress cooling flows, and thus
will have to be modified to allow some cooling of the hot gas. We now
briefly review some of the candidate models.

\citet{binney} and \citet{co} have proposed feedback from the central
black hole as a promising means of inhibiting the cooling of hot gas
in the central regions of galactic cooling flows. In this model
whenever the black hole accretes a sufficient amount of gas to
stimulate nuclear activity, the accompanying radiation stimulated by
the accretion heats up the hot gas and prevents further cooling. This
is supposed to be a cyclical process such that the AGN phase is
sufficiently rare to be consistent with the lack of nuclear activity
in most cooling flows.  If the AGN feedback energy does not completely
suppress cooling of the hot gas but instead merely prevents most of
the gas from cooling down to temperatures below $10^{5-6}$ K, this
might be able to explain issues (1) and (2). We also note that most of
the cooling gas is not expected to reach the central black hole in
standard models \citep{bmbh}.

Another energy source proposed to exist in the centers of cooling
flows is that from the reconnection of tangled magnetic field lines.
It has been known for some time that small seed magnetic fields can be
amplified within a cooling flow to produce sizeable fields at the
centers (e.g., Soker \& Sarazin 1990; Lesch \& Bender 1990; Moss \&
Shukurov 1996; Mathews \& Brighenti 1997). \citet{zsr} show that
reconnection energy can significantly reduce the cooling rates and may
account for significant amounts of warm ($T=10^{5-6}$ K) gas.
Similarly, \citet{nm} propose a two-phase model to recycle warm and
hot gas along magnetic flux loops also resulting in much lower rates
of mass deposition.

Both the AGN and magnetic field reconnection models would explain the
warm gas phase as a non-equilibrium configuration where the warm gas
is continuously diluted by heating processes and replenished by
cooling from the hot phase. If we consider the (unlikely) possibility
that the warm gas is a long-lived equilibrium phase (see, e.g., Fabian
1996) then magnetic pressure is probably the most viable non-thermal
process which can support the gas. \citet{mb} describe a possible
equilibrium model of the warm gas as the the outer envelopes of low
mass stars forming in a cooling flow.

The details of how the magnetic field would support the gas are
uncertain. \citet{dft} suggest that the cool gas blobs would be
anchored to the hot gas by the magnetic fields, and thus the pressure
support would actually come from the hot gas. However, inspection of
Table \ref{tab.mass} reveals that $\mabs\sim (1-10)\mhot$ within the
central bins indicating that the hot gas could not support the cool
gas. (The 95\% lower limits on \mabs are considered here.) At larger
radii it may be possible for the hot gas to support the cool gas.

Whatever the details of the magnetic support the condition of
hydrostatic equilibrium requires that $B^2\sim 6\mabs GM_{\rm
grav}r^{-4}$. Using the values for \mabs\, within the central
$1\arcmin$ bin quoted in Table \ref{tab.mass} and the gravitating
masses obtained from previous \rosat studies (David et al 1995; Kim \&
Fabbiano 1995; Rangarajan et al 1995; Irwin \& Sarazin 1996) we find
that $B\sim 100\mu\rm G$ is required within a 5 kpc radius.

Estimates of the magnetic field strengths from radio polarization
analyses of these galaxies are lacking, although there exist estimates
using minimum energy arguments for NGC 1399 \citep{kbe}, NGC 4472
\citep{ek}, and NGC 4649 \citep{sw} which give consistent results:
$B\ga 50-100\mu\rm G$ at the centers and $B\ga (5-10)\mu\rm G$ at
$r\sim 0.5\arcmin$. Assuming $B\sim r^{-1.2}$ \citep{mb97} then these
observations imply $\langle B\rangle\sim 5\mu\rm G$ when averaged over
a $1\arcmin$ circle. Since the observations only set lower limits the
expected $\langle B\rangle\sim 100\mu\rm G$ fields are consistent with
the observations. 

Interestingly, the need for $B\sim 100\mu\rm G$ in cooling flows has
been suggested by \citet{bm97} on entirely different grounds. In their
analysis of the gravitating mass distributions of NGC 4472, 4636, and
4649 \citet{bm97} find in every case that the gravitating mass
determined from the X-ray analysis falls below that estimated from
stellar dynamics for $r\la (\rm few) kpc$. If $B\sim 100\mu\rm G$
within the centers of these systems then the X-ray and stellar
dynamical masses agree. (A similar result holds for NGC 1399 as well
-- W. Mathews 2000, private communication.)

Finally, we mention it may be useful to consider a model for the warm
gas that does not originate from cooling out of the hot phase. One
such possibility is offered by the material continuously ejected by
the stars. This material is injected into the ISM with low energy and
is shock-heated up to the virial temperature of the halo. Although
theoretical arguments suggest that the heating is very rapid
\citep{bill}, more detailed calculations are required to rule out the
possibility that in fact the transition is gradual and could give rise
to an observable phase of warm gas consistent with the X-ray
observations.

\section{Conclusions}
\label{conc}

From deprojection analysis of the \rosat PSPC data of 10 cooling flow
galaxies and groups with low Galactic columns we have detected oxygen
absorption at the $2\sigma/3\sigma$ level intrinsic to the central
$\sim 1\arcmin$ in half of the sample: NGC 507, 1399, 4472, 4649, and
5044.  The data for the other systems are insufficient to place
interesting constraints on the absorption profile but are consistent
with substantial absorption.  We modeled the oxygen absorption as a
single edge (rest frame $E=0.532$ keV) which produces the necessary
absorption in both the PSPC and \asca data for $E\ga 0.5$ keV without
violating the PSPC constraints over $0.2$ $\sim 0.4$ keV for which no
significant excess absorption is indicated. Assuming the absorber is
collisionally ionized gas we infer a temperature of $10^{5-6}$ K from
consideration of the possible edge energies consistent with the PSPC
data. 

The intrinsic oxygen absorption reconciles the longstanding problem of
why negligible column densities for a foreground absorber with solar
abundances were inferred from \rosat data whereas large columns were
obtained from \asca and other instruments with bandpasses above $\sim
0.5$ keV. Moreover, since the absorption is confined to energies above
$\sim 0.5$ keV there is no need for large columns of cold H which are
known to be very inconsistent with the negligible atomic and molecular
H measured in galactic and cluster cooling flows (e.g., Bregman et al
1992; O'Dea et al 1994).

In most of the galaxies and groups we have found that single-phase and
cooling flow models cannot explain the X-ray emission in the soft
(0.2-0.4 keV) energy channels of the \rosat PSPC data (\S
\ref{xray}). That is, when \nh\, of the standard absorber model is
freely fitted it is found that $\nh<\nhgal$ in the central bins of
most systems, with NGC 4472 and NGC 5044 showing the most significant
soft excesses. If we model this soft emission as coronal gas we obtain
temperatures $10^{5-6}$ K in excellent agreement with those inferred
from the energy ranges of the absorption edges.

Hence, the sub-Galactic column densities are consistent with a direct
detection of the emission from the intrinsic absorbing gas. The
agreement between the temperatures inferred from the emission and
absorbing properties of the warm gas lends strong support to the
ionized gas model. In contrast, dust can not explain the excess soft
X-ray emission. (Other problems exist with the dust hypothesis -- see
\S \ref{dust}.)

Our simple estimates of the amount of absorbing matter implied by our
single-edge absorption measurements are consistent with the total
amount of matter expected to have been deposited by a cooling flow (\S
\ref{mass}). With the arrival of higher quality data from \chandra and
\xmm more accurate estimates should be made which account for a range
of absorbing edges and possible radiative transfer effects in the warm
gas.

We have examined the theoretical difficulties associated with
attributing the absorption to warm ionized gas and have discussed some
candidate models that may be able to account for these
problems. Future detailed calculations are required to assess the
viability of these models (\S \ref{theory}).

Fortunately on the observational front it will be very easy to verify
the intrinsic oxygen absorption with new \chandra and \xmm data.  The
\xmm (EPIC) and \chandra (ACIS-S) CCDs both extend down to 0.1-0.2 keV
and have substantially better energy resolution than the
PSPC. Observations with these instruments can easily test our
prediction for warm gas in both absorption and emission. The grating
spectrometers of \xmm and \chandra have even better energy resolution
(but smaller effective area) and, in principle, might detect
individual edges.

We emphasize that the absorption signature of the warm gas is expected
to be a relatively broad trough over energies $0.4\sim 0.8$ keV, and
thus future \chandra and \xmm observations will not see a single sharp
feature. The most straightforward means to confirm our results will be
to reproduce the sensitivity of \nh\, to $E_{\rm min}$ for a standard
absorber model (\S \ref{bandpass}). To obtain the properties of the
absorber (e.g., temperature and abundances) a model for the soft X-ray
opacity such as that described by \citet{kk} must be compared to the
new data. In so doing the emission from the warm gas must also be
accounted for (\S \ref{xray}), and thus it is very important that the
detector bandpass extend down to $\sim 0.1$ keV which it does for the
\chandra and \xmm CCDs.

As discussed in \S \ref{optical} optical and FUV constraints imply a
lower limit of $T\sim 2\times 10^5$ K. It is possible that precise
measurements of \oiii (5007\AA) could refine this limit, but since
this line peaks at $T\sim 0.8\times 10^5$ K its emissivity is already
falling rapidly at $T\sim 2\times 10^5$ K. Future high resolution FUV
spectroscopy of the \ion{O}{6} (1034\AA) line (peak temperature
$3.2\times 10^5$ K) with, e.g., FUSE may be able to place additional
interesting constraints on the warm gas if its temperature does not
exceed  $T\approx 5\times 10^5$ K.

It should be remembered that to infer the properties of the warm gas
from X-ray observations the absorption and emission spectrum arising
from warm gas must be disentangled from Galactic absorption and the
emission from hot plasma. Since (if confirmed) the warm gas almost
certainly represents the mass deposited by an inhomogeneous cooling
flow (\S \ref{mass}), the hot gas at each radius should also emit over
a range of temperatures. Hence, the thermodynamic state of the X-ray
emitting plasma appears to be very complex in the central regions of
the (X-ray) brightest galaxies and groups, and the analogous results
for A1795 presented in PAPER1 suggest the same applies for galaxy
clusters.

\acknowledgements

I thank W. Mathews for fruitful discussions and the anonymous referee
for detailed comments. Support for this work was provided by NASA
through Chandra Fellowship grant PF8-10001 awarded by the Chandra
Science Center, which is operated by the Smithsonian Astrophysical
Observatory for NASA under contract NAS8-39073.


\begin{thebibliography}{}
\bibitem[Allen \& Fabian(1997)]{af} Allen, S. W., \& Fabian,
A. C. 1997, \mnras, 286, 583 
\bibitem[Allen et al(2000a)]{swa_agn} Allen, S. W., Di Matteo, T., \&
Fabian, A. C., 2000a, \mnras, in press (astro-ph/9910188)
\bibitem[Allen et al(2000b)]{swa_nearby} Allen, S. W., Fabian, A. C.,
Johnstone, R. M., \& Nulsen, P. E. J., 2000b, \mnras, submitted
(astro-ph/9910188) 
\bibitem[Arabadjis \& Bregman(1999)]{ab} Arabadjis, J. S., \&
Bregman, J. N., 1999, \apj, 514, 607
\bibitem[Arnaud \& Mushotzky(1998)]{am} Arnaud, K. A., \& Mushotzky,
R. F., 1998, \apj, 501, 119
\bibitem[Baluci\'{n}ska-Church \& McCammon(1992)]{phabs}
Baluci\'{n}ska-Church, M., \& McCammon, D., 1992, \apj, 400, 699
\bibitem[Bevington(1969)]{bevington} Bevington, P. R., 1969, Data Reduction
and Error Analysis for the Physical Sciences, (New York: McGraw-Hill)
\bibitem[Binney(1996)]{binney} Binney, J. J., 1996, in Gravitational
Dynamics Proc. 36th Herstmonceux conf., ed O. Lahav (Cambridge:
Cambridge Univeristy Press), 89
\bibitem[Bregman et al(1992)]{bhr} Bregman, J. N., Hogg, D. E., \&
Roberts, M. S., 1992, \apj, 387, 484
\bibitem[Briel \& Henry(1996)]{bh} Briel, U. G., \& Henry, J. P.,
1996, \apj, 472, 131 
\bibitem[Brighenti \& Mathews(1997)]{bm97} Brighenti, F., \& Mathews,
W. G., 1997, \apjl, 486, L83
\bibitem[Brighenti \& Mathews(1998)]{bmcircum} Brighenti F., \&
Mathews W. G., 1998, \apj, 495, 239
\bibitem[Brighenti \& Mathews(1999)]{bmbh} Brighenti F., \&
Mathews W. G., 1999, \apjl, 527, L89
\bibitem[Brown et al(1995)]{bfd} Brown, T. M., Ferguson, H. C.,
Davidsen, A. F., 1995, \apjl, 454, L15
\bibitem[Buote(1999)]{b99} Buote, D. A., 1999, \mnras, 309, 695
\bibitem[Buote(2000a)]{b00a} Buote, D. A., 2000a, \mnras, 311, 176
\bibitem[Buote(2000b)]{b00b} Buote, D. A., 2000b, \apjl, 532, L113 (PAPER1)
\bibitem[Buote(2000c)]{b00c} Buote, D. A., 2000c, \apj, in press
(PAPER2) (astro-ph/0001329)
\bibitem[Buote \& Fabian(1998)]{bf} Buote, D. A., \& Fabian,
A. C. 1998, \mnras, 296, 977
\bibitem[Buote et al(1999)]{bcf} Buote, D. A., Canizares, C. R., \&
Fabian, A. C. 1999, \mnras, 310, 483
\bibitem[Ciotti \& Ostriker(2000)]{co} Ciotti, L., \& Ostriker, J. P.,
2000, \apj, submitted (astro-ph/9912064)
\bibitem[Daines et al(1994)]{dft} Daines, S. J., Fabian, A. C., \&
Thomas, P. A., 1994, \mnras, 268, 1060
\bibitem[Daltabuit \& Cox(1972)]{dalt} Daltabuit, E., \& Cox, D. P., 1972,
\apj, 177, 855
\bibitem[David et al(1994)]{djfd} David, L. P., Jones, C., Forman,
W., Daines, S., 1994, \apj, 428, 544
\bibitem[Dickey \& Lockman(1990)]{dl}
Dickey J. M., Lockman F. J., 1990, \araa, 28, 215
\bibitem[Dixon et al(1996)]{dhf} Dixon, W., Hurwitz, M., \& Ferguson, H. C.,
1996, \apjl, 469, L77
\bibitem[Ekers \& Kotanyi(1977)]{ek} Ekers, R. D., \& Kotanyi, C. G.,
1977, \aap, 67, 47
\bibitem[Fabian (1994)]{acf} 
Fabian A. C., 1994, \araa, 32, 277
\bibitem[Fabian (1994)]{acf96} 
Fabian A. C., 1996, Science, 271, 1244
\bibitem[Fabian et al(1994)]{fabt} Fabian, A. C., Arnaud, K. A.,
Bautz, M. W., \& Tawara, Y., 1994, \apj, 436, L63
\bibitem[Ferguson et al(1991)]{ferg} Ferguson, H. C., et al, 1991,
\apj, \apjl, 382, L69
\bibitem[Forman et al(1993)]{forman} Forman W., Jones C., David L.,
Franx M., Makishima K., \& Ohashi T., 1993, \apjl, 418, L55
\bibitem[Goudfrooij et al(1994)]{goud} Goudfrooij, P., Hansen, L.,
Jorgensen, H. E., \& Norgaard-Nielsen, H. U., 1994, \aaps, 105, 341
\bibitem[Gould \& Young(1991)]{gy} Gould, R. J., \& Jung, Y.-D., 1991,
\apj, 373, 271
\bibitem[Ikebe et al(1997)]{ikebe97}
Ikebe Y., et al., 1997, \apj, 481, 660
\bibitem[Irwin \& Bregman(1999)]{ib} Irwin, J. A., \& Bregman, J. N.,
1999, \apj, 527, 125
\bibitem[Irwin \& Sarazin(1996)]{is} Irwin, J. A., \& Sarazin, C. L.,
1996, \apj, 471, 663
\bibitem[Johnstone et al(1992)]{rjcool} Johnstone R. M., Fabian A. C.,
Edge A. C., \& Thomas P. A., 1992, \mnras, 255, 431
\bibitem[Jones et al(1997)]{j97} Jones, C., Stern, C., Forman, W.,
Breen, J., David, L., Tucker, W., \& Franx, M., 1997, \apj, 482, 143
\bibitem[Killeen et al(1988)]{kbe} Killeen, N. E. B., Bicknell, G. V.,
\& Ekers, R. D., 1988, \apj, 325, 180
\bibitem[Kim \& Fabbiano(1995)]{kf} Kim, D.-W., \& Fabbiano, G., 1995,
\apj, 441, 182
\bibitem[Krolik \& Kallman(1984)]{kk} Krolik, J. H., \& Kallman,
T. R., 1984, \apj, 286, 366
\bibitem[Laor \& Draine(1993)]{ld} Laor, A., \& Draine, B. T., 1993,
\apj, 402, 441
\bibitem[Lesch \& Bender(1990)]{lb} Lesch, H., \& Bender, R., 1990,
\aap, 233, 417
\bibitem[Lester et al(1995)]{l95} Lester, D. F., Zink, E. C.,
Doppmann, G. W., Gaffney, N. I., Harvey, P. M., Smith, B. J., \&
Malkan, M., 1995, \apj, 439, 185
\bibitem[Macchetto et al(1996)]{macc96} Macchetto, F., Pastoriza, M.,
Caon, N., Sparks, W. B., Giavalisco, M., Bender, R., \& Capaccioli,
M., 1996, \aaps, 120, 463
\bibitem[Mathews(1990)]{bill} Mathews, W. G., 1990, \apj, 354, 468
\bibitem[Mathews \& Brighenti (1997)]{mb97} Mathews, W. G., \&
Brighenti, F., 1997, \apj, 488, 595
\bibitem[Mathews \& Brighenti (1999)]{mb} Mathews, W. G., \&
Brighenti, F., 1999, \apj, 526, 114
\bibitem[Morrison \& McCammon(1983)]{wabs} Morrison, R., \& McCammon, D.,
1983, \apj, 270, 119
\bibitem[Moss \& Shukurov(1996)]{ms} Moss, D., \& Shukurov, A., 1996,
\mnras, 279, 229
\bibitem[Nahar(1999)]{n99} Nahar, S. N., 1999, \apjs, 120, 131 
\bibitem[Nahar \& Pradhan(1997)]{np97} Nahar, S. N., \& Pradhan,
A. K., 1997, \apjs, 111, 339
\bibitem[Norman \& Meiksen(1996)]{nm} Norman, C., \& Meiksen, A.,
1996, \apj, 468, 97
\bibitem[O'Dea et al(1994)]{odea} O'Dea, C. P., Baum, S. A., Maloney,
P. R., Tacconi, L. J., Sparks, W. B., 1994, \apj, 422, 467
\bibitem[Pistinner \& Sarazin(1994)]{pist} Pistinner, S., Sarazin,
C. L., 1994, \apj, 433, 577
\bibitem[Rangarajan et al(1995)]{vijay} Rangarajan, F. V. N., Fabian,
A. C., Forman, W. R., \& Jones, C., 1995, \mnras, 272, 665
\bibitem[Sarazin et al (1998)]{swm} Sarazin, C. L., Wise, M. W.,
Markevitch, M. L., 1998, \apj, 498, 606
\bibitem[Snowden et al(1994)]{snow2} Snowden, S. L., McCammon, D.,
Burrows, D. N., \& Mendenhall, J. A., 1994, \apj, 424, 714
\bibitem[Soker \& Sarazin(1990)]{ss} Soker, N., \& Sarazin, C. L.,
1990, \apj, 348, 73
\bibitem[Stanger \& Warwick(1986)]{sw} Stanger, V. J., \& Warwick,
R. S., 1986, \mnras, 220, 363
\bibitem[Sutherland \& Dopita(1993)]{sd} Sutherland, R. S., \& Dopita,
R. S., 1993, \apjs, 88, 253
\bibitem[Thomas et al(1987)]{tfn}
Thomas P. A., Fabian A. C., \& Nulsen P. E. J., 1987, \mnras, 228, 973
\bibitem[Trinchieri \& di Serego Alighieri(1991)]{ta} Trinchieri, G.,
\& di Serego Alighieri, S., 1991, \aj, 101, 1647
\bibitem[Trinchieri et al(1997)]{tfk}
Trinchieri G., Fabbiano G., Kim D.-W., 1997, A\&A, 318, 361
\bibitem[Trinchieri et al(1994)]{trin} Trinchieri G., Kim D.-W.,
Fabbiano G., \& Canizares C., 1994, \apj, 428, 555
\bibitem[Tsai \& Mathews(1995)]{tm} Tsai, J. C., \& Mathews, W. G.,
1995, \apj, 448, 84
\bibitem[Tsai \& Mathews(1996)]{tm96} Tsai, J. C., \& Mathews, W. G.,
1996, \apj, 468, 571
\bibitem[Voit \& Donahue(1995)]{vd} Voit, G. M., \& Donahue, M., 1995,
\apj, 452, 164
\bibitem[White et al(1991)]{daw91} White, D. A., Fabian A. C.,
Johnstone R. M., Musthotzky, R. F., \& Arnaud, K. A., 1991, \mnras,
252, 72
\bibitem[Xu et al(1998)]{xu} 
Xu H., Makishima K., Fukazawa Y., Ikebe Y., Kikuchi K., Ohashi T., \&
Tamura T., 1998, \apj, 500, 738
\bibitem[Zoabi et al(1998)]{zsr} Zoabi, E., Soker, N., \& Regev, O.,
1998, \mnras, 296, 579 
\end{thebibliography}
\end{document}